\documentclass{article}

\usepackage{arxiv}

\usepackage[utf8]{inputenc} 
\usepackage[T1]{fontenc}    
\usepackage{hyperref}       
\usepackage{url}            
\usepackage{booktabs}       
\usepackage{amsfonts}       
\usepackage{nicefrac}       
\usepackage{microtype}      
\usepackage{lipsum}		
\usepackage{graphicx}
\usepackage{doi}

\usepackage{threeparttable} 
\usepackage{caption}
\usepackage{subcaption}
\usepackage{siunitx}

\usepackage[shortlabels]{enumitem}

\usepackage{float}
\usepackage{mathtools}
\DeclarePairedDelimiter{\abs}{\lvert}{\rvert}
\usepackage{makecell}

\title{The Contribution of Human Body Capacitance/Body-Area Electric Field To Individual and Collaborative Activity Recognition}

\author{\hspace{1mm}Sizhen Bian $^{1}$, Vitor Fortes Rey $^{2}$, Siyu Yuan$^{3}$, and Paul Lukowicz $^{4}$ \\
	$^{1,2,4}$Department of Computer Science\\
        $^{3}$Department of Electrical Engineering and Information Technology\\
	Technical University Kaiserslautern\\
	Kaiserslautern, Germany 67663 \\
	\texttt{sbian@rhrk.uni-kl.de} \\
}

\date{}

\renewcommand{\shorttitle}{The Contribution of Human Body Capacitance/Body-Area Electric Field ...}

\hypersetup{
pdftitle={A template for the arxiv style},
pdfsubject={q-bio.NC, q-bio.QM},
pdfauthor={David S.~Hippocampus, Elias D.~Striatum},
pdfkeywords={First keyword, Second keyword, More},
}

\begin{document}
\maketitle

\begin{abstract}
The current dominated sensing modalityin sensor-based human activity tracking is Inertial Measurement Unit ($IMUs$), which combines accelerometer and gyroscope for accurate motion monitoring of separate body parts. This work presented an alternative wearable motion-sensing approach: inferring motion information of various body parts from the human body capacitance ($HBC$, also commonly defined as body-area electric field).  While being less robust in tracking the posture and trajectory, $HBC$ has two properties that make it an attractive complement to $IMU$. First, the deployment of the sensing node on the being tracked body part is not a requirement for $HBC$ sensing approach. Thus, for example, a wrist-worn $HBC$ sensor can be used to track and recognize leg based exercises. Furthermore, $HBC$ is sensitive to the body's interaction with its surroundings, including both touching and being in the immediate proximity of people and objects. In this paper, we first described the sensing principle for $HBC$, our sensor architecture and implementation, and methods for evaluating the signal. We then presented two case studies demonstrating the usefulness of $HBC$ as a complement/alternative to $IMUs$, an individual exercises experiment, and a collaborative TV-Wall assembling and disassembling experiment. 
In the first case, we explored the exercise recognition and repetition counting of seven machine-free leg-only exercises and eleven general gym workouts with the signal source of $HBC$ and $IMU$. The $HBC$ sensing shows significant advantages over the $IMU$ signals in classification(0.89 vs 0.78 in F-score) and counting(0.982 vs 0.938 in accuracy) of the leg-only exercises. For the general gym workouts, $HBC$ only shows improvement for workout recognition accuracy to certain workouts like adductor where legs alone complete the movement.  Although the $HBC$ didn't present competitive contribution in general gym workouts recognition compared with $IMU$, it supplies better results over the $IMU$ for workouts counting(0.800 vs. 0.756 when wearing the sensors on the wrist).
In the second case, we tried to recognize actions related to manipulating objects and physical collaboration between users by using a wrist-worn $HBC$ sensing unit. In particular, activities that challenge the $IMU$ such as carrying an object alone, carrying it jointly with another agent, or carrying nothing just walking. Using both the accelerometer and the capacitive sensor, we detected collaboration between the users with 0.69 F-score when receiving data from a single user and 0.78 when receiving data from both users. The capacitive sensor can improve the recognition of collaborative activities with an F-score over a single wrist accelerometer approach by 16\%. The resources described in this paper, including hardware, firmware, datasets, and algorithms are available from public repository to promote other researchers for further exploration.
\end{abstract}


\section{Introduction and Related Work}\label{Section_1}

\subsection{Dominance of $IMU$ and Introduction of $HBC$}
Wearable inertial motion sensors ($IMUs$) are currently probably the most important sensing modality for sensor-based human activity recognition. This is firstly due to the fact that motion of body parts is a defining characteristics of many human activities. Secondly, there is broad availability in unobtrusive form factor including all sorts of consumer electronic products. $IMUs$ based tracking application increasingly shows attraction to both academic and industry researchers, aiming to get a better understanding of the individual's or group's behaviour\cite{park2003enhancing, caporusso2019comparative, ward2017detecting}, and to supply positive behavioral support\cite{patel2015wearable, naslund2016wearable}. Advances in sensor technique \cite{bian2022state} (like algorithms integrated $IMUs$ enabling plug-and-play functionality) enables a broader range of practical application, like sport\cite{holzemann2018using,kaewkannate2016comparison}, healthcare\cite{alzahrani2016wearable,hansel2015challenges}, clinic\cite{parker2018interplay}, gaming\cite{wang2018intelligent}, working\cite{cha2018towards} etc. Table \ref{Wearable_devices} lists some latest wearables used in industries. A rich set of sensors were integrated into those wearables, and the inertial measurement unit is the most widely embedded one, which contributes the motion detection functionality of those wearables alone as the a motion signal source. 

For certainty, $IMU$ plays the main or even the unique role in wearables that detect motion\cite{uddin2015wearable,magalhaes2015wearable,kozlovszky2018imu}. And in the wearable field, its hardly to find a competitive sensor-based sensing modality other than the $IMU$. In this paper we will introduce our investigation on an alternative wearable motion sensing modality which shows significant potential in motion related human activities recognition: human body capacitance($HBC$).

\begin{figure}[!b]
\begin{minipage}[t]{0.5\linewidth}
\centering
\includegraphics[width=0.7\textwidth,height=3.0cm]{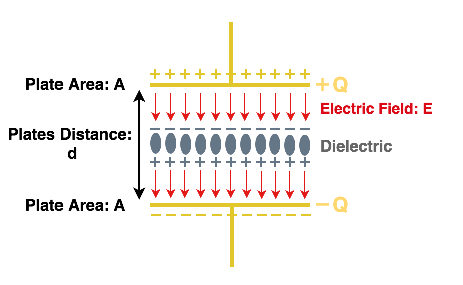}
\caption{Basic structure of a capacitor}
\label{Capacitor_Model}
\end{minipage}
\quad
\begin{minipage}[t]{0.5\linewidth}
\centering
\includegraphics[width=0.7\textwidth,height=3.0cm]{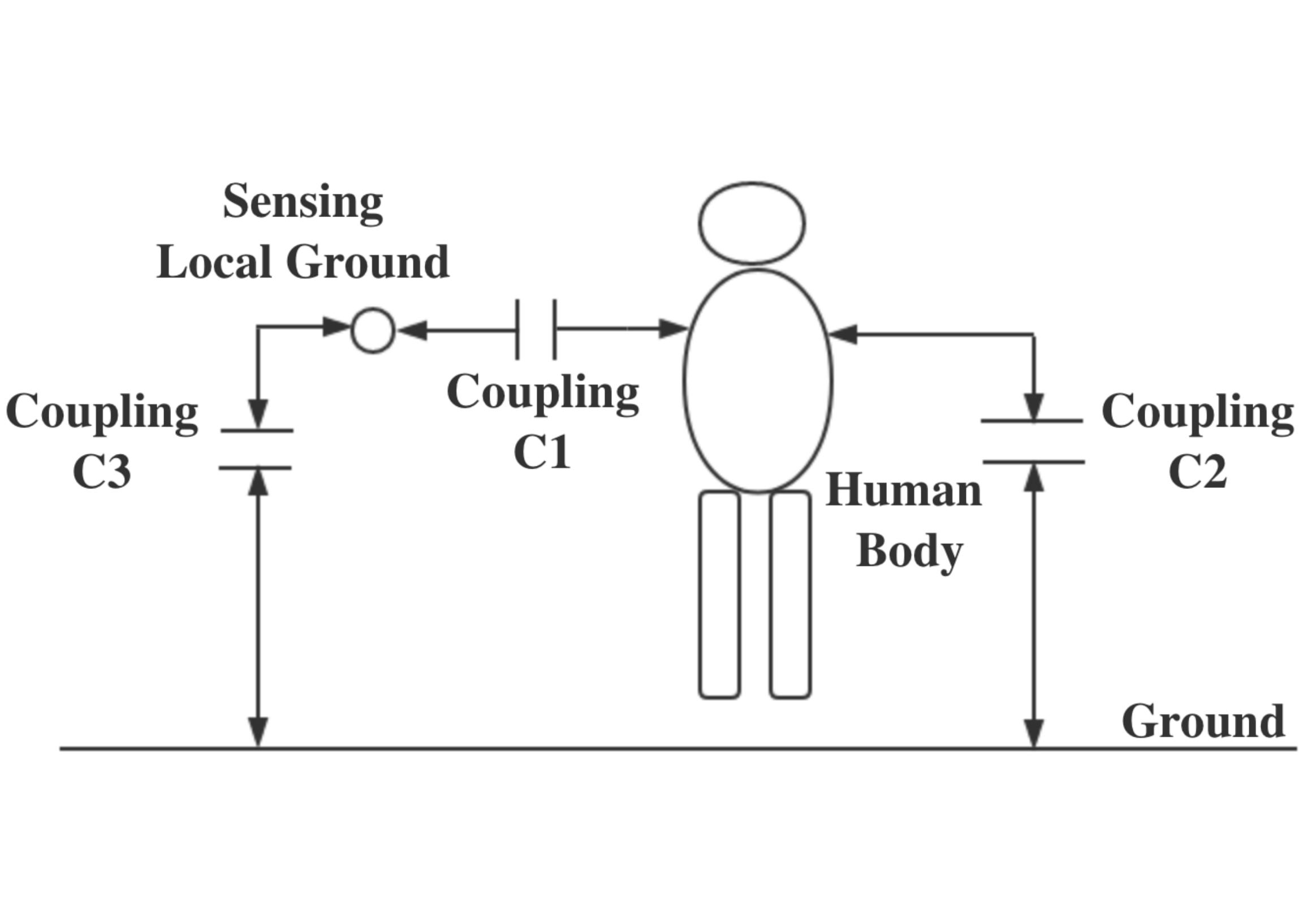}
\caption{Capacitive coupling among ground, human body and sensing hardware's local ground}
\label{Capacitive Model}
\end{minipage}
\end{figure}

Human body, the subject for body motion tracking tasks, has good electric features because of its ideal conductivity\cite{presta1983measurement,cochran1986total}. We tried to exploit the physiological signals, like the skin conductance, EMG, for the body motion detection, and the $Human$ $Body$ $Capacitance(HBC)$ showed significant potential. Unlike the other physiological features, $HBC$ is a feature that interacts with the surrounding, especially the ground. Being insulated by the wearing, the body acts as a pure conductive plate, the ground as another one, forming a natural capacitor, as depicted in Figure \ref{Capacitor_Model} and \ref{Capacitive Model}. $HBC$ is a physiological property that indicates the body's ability to store electrons. Studies\cite{aliau2012fast, aliau2013novel, buller2006measurement, sizhen2021systematic} show that the value of body capacitance is not a constant value but varies around 100 pF. Postures \cite{fujiwara2002numerical}, garment \cite{jonassen1998human}, skin stated\cite{goad2016ambient,egawa2002effect}, etc, are all potential influence factors to the $HBC$. $HBC$-related works include communication\cite{cohn2012humantenna}, cooperation detection\cite{bian2019wrist} as well as motion detection\cite{kan2015social, cheng2010active, sizhen2021capacitive}. Although those works are utilizing the conductivity property of human body, the body-environment relationship is still less understood and explored.

\begin{table}[H]
\centering
\caption{Wearable Devices from Industry}
\label{Wearable_devices}
\begin{tabular}{ p{1.6cm} p{0.4cm} p{1.3cm} p{2.8cm} p{0.8cm} p{1.3 cm} p{3.5cm}}
\toprule
wearables & IMU & Heart Rate Monitor & other motion related Sensors & Body Place & Main Application & motion detection scale \\ 
\midrule
Fitbit versa & x & x & GPS,ambient light sensor,NFC, etc & wrist & fitness & Swim, Run, Walk, Bike, Yoga, etc aerobic workouts\\

Apple Watch 4 & x & x &  ambient light sensor,GPS,NFC,Blood oxygenation/Sugar, etc & wrist & life assistant, fitness & Swim, Run, Walk, Bike, Yoga, etc aerobic workouts  \\ 
MiBand 4 & x & x & Ambient light sensor,NFC& wrist& life assistant, fitness &  Treadmill, Swim, Run, Walk, Bike\\ 
HexoSkin & x & x & cardiac/breath sensor & body& fitness & cardiac and respiratory activity \\
Google Glass & x &  & camera, telephoto, light, microphone & head & life assistant  & head direction, head motion \\
Vuzik Blade & x &  & camera, microphone & head & life assistant & head direction, head motion \\
\bottomrule
\end{tabular}
\end{table}

\subsection{Related work of individual activity}

\begin{figure}[H]
\centering
\includegraphics[width=0.90\columnwidth,height=4.6cm]{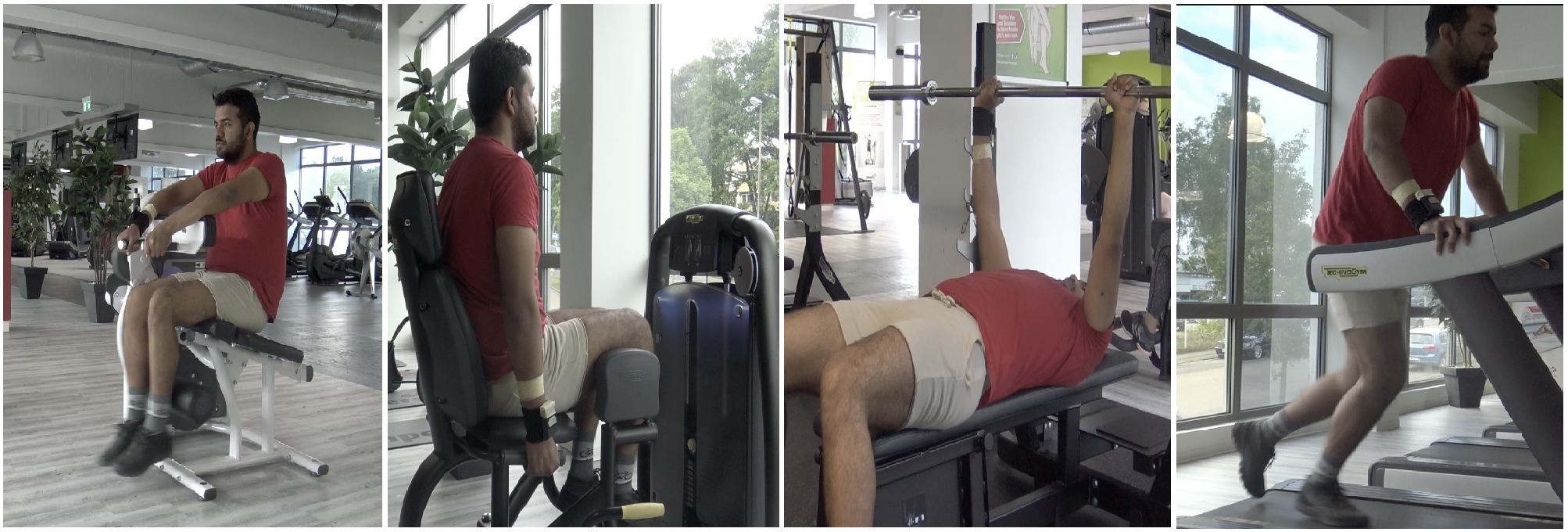}
\caption{Four examples of the 11 widely trained gym exercises(from left to right: Leg-Curl, Adductor, Bench-Press, Running), including both aerobic and anaerobic training. Especially the training where the arm is in a static state, meaning that the wrist-mounted $IMU$ losts its recognition ability in such training.}
\label{Calos_Gym}
\end{figure}

Our previous work\cite{bian2019passive} has demonstrated the feasibility of $HBC$ based fitness recognition. To expand the scale of recognized workouts with wearable devices, we combined the traditional motion sensor, IMU, and this ubiquitous biophysical sensing modality, $HBC$. We studied both aerobic and anaerobic workouts. As a popular topic, fitness recognition research appeared in a rich set of literary works, either mobile phones or stand-alone IMUs are used to detect the object's movements. Koskimaki et al.\cite{koskimaki2014recognizing} explored 30 exercises and got an outstanding recognition result. However, since the data was from only one subject, the validation accuracy was low. Morris et al.\cite{morris2014recofit} classified 13 exercises with up to 114 subjects, and got 86.8\% precision when leaving one subject out with one IMU on the arm. Wahjudi et al.\cite{wahjudi2019imu} deployed the IMU on shoes and analyzed the gait to recognize walking-related workouts. Chang et al.\cite{chang2007tracking} focused on free-weight exercises and got 90\% recognition accuracy over nine different exercises, with two accelerometers worn on the body. Depari et al.\cite{depari2019lightweight} also tried to recognize free-weight exercises and got 93\% accuracy with a single IMU. All those works' signal source is the IMU alone, which is normally only capable of sensing the motion of the body part where it is attached on. To verify the benefits of $HBC$ sensor, which is able to sense the motion of the body part where it is not attached on, we studied firstly 7 leg exercises with both IMU and $HBC$ sensors attached on the wrist from 5 participants in our laboratory and further to be more practical, we studied 11 widely trained exercises(as Figure \ref{Calos_Gym} depicts) in a gym studio from 10 participants with our custom prototype in three positions: in the pocket, on the calf and the wrist. We classified the exercises with both classical machine learning and deep neural network models and counted the exercises with the peak detection approach. 

\subsection{Related work of collaborative activity}

\begin{figure}[H]
\centering
\includegraphics[width=0.25\columnwidth,height=5cm]{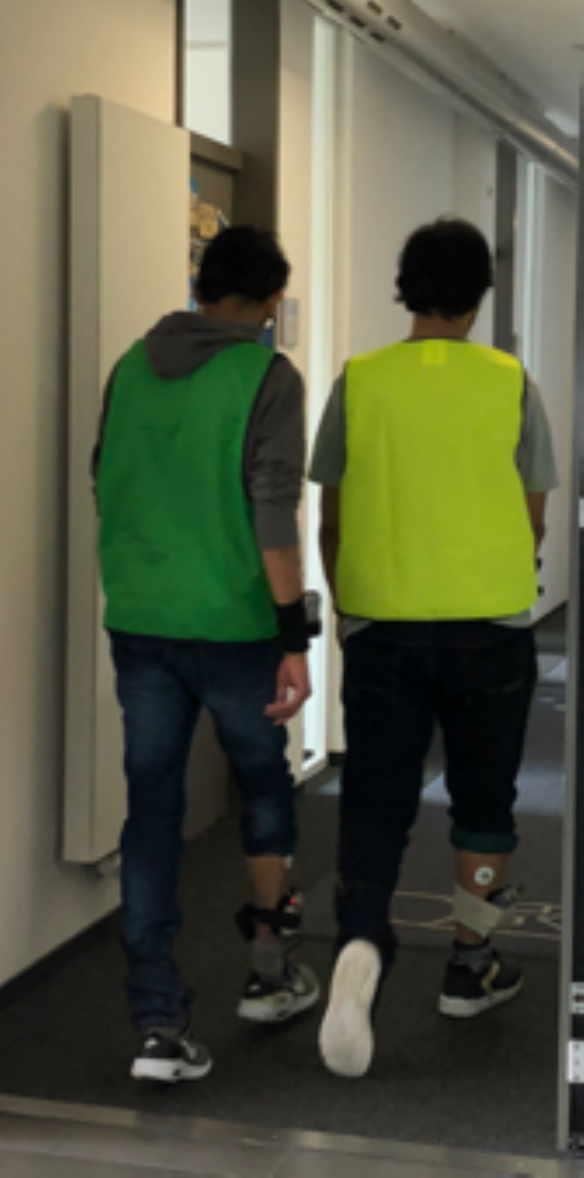}
\includegraphics[width=0.25\columnwidth,height=5cm]{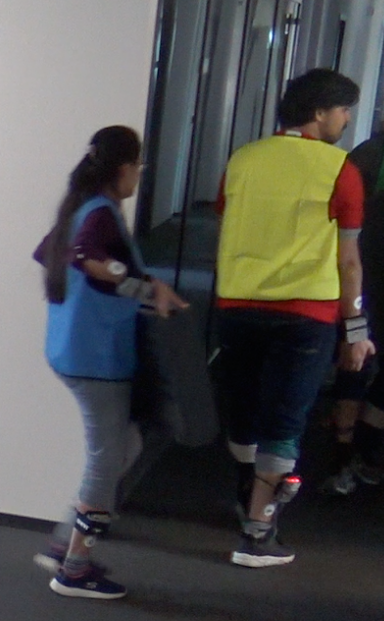}
\includegraphics[width=0.25\columnwidth,height=5cm]{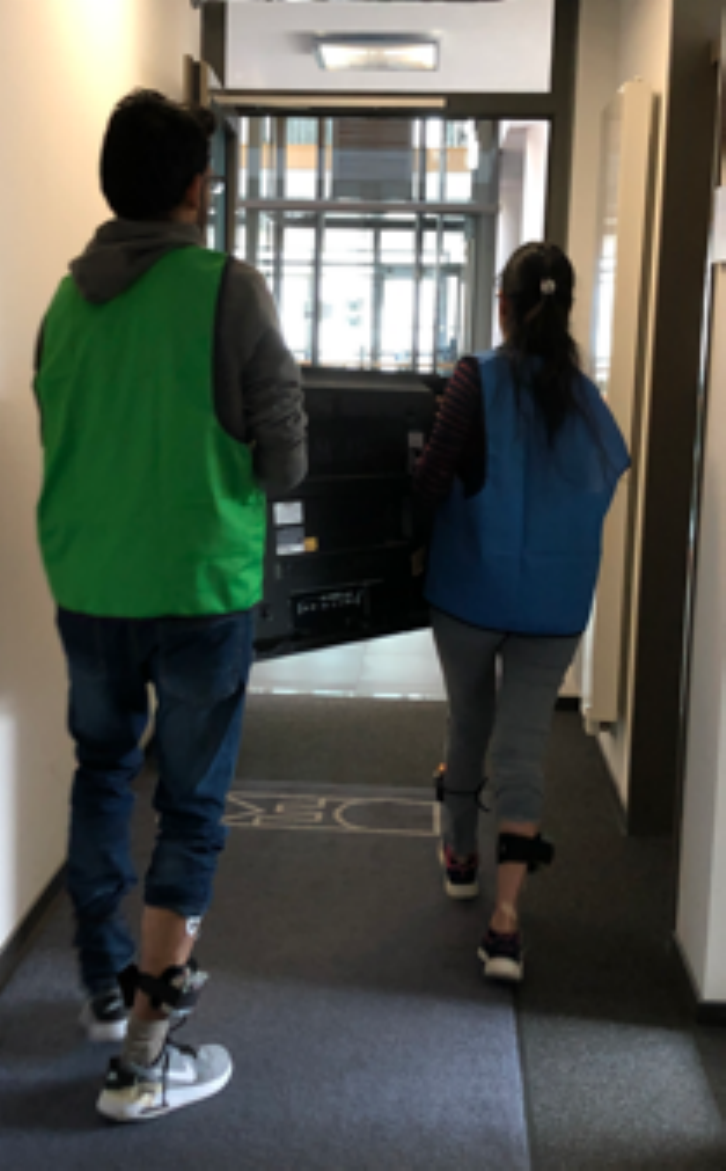}
\caption{Example of the type of recognition tasks that our work targets: distinguishing  two people among just walking next each other, walking next each other and carrying a heavy object alone and walking next to each other while jointly carrying a heavy object together. Co-location is not a good separator. The fact that people are actually holding the object between (rather then just occluding it) them is not easily recognized from the picture either.}
\label{capacitance_model}
\label{fig:problem}
\end{figure}

$HBC$ is a environmental-sensitive parameter, not only the motion from the body itself will change the body capacitance, but also the invasion of other bodies will influence the body capacitance\cite{bian2019wrist}. From this point, we moved our steps from the recognition of physical actions performed by an individual subject to physical actions collaboratively performed by several users \cite{gruenerbl2017detecting}, and tried to recognize physical activities that are a big challenge for the classical motion sensor (see e.g., \cite{bulling2014tutorial,lara2013survey} for an overview of IMU-based methods). 
Benefiting from our prototype's low power, small sensing hardware size, we used $HBC$ sensing modality to recognize the typical scenario of physical works related to the manipulation of objects and physical collaboration between users. The problem that we addressed is illustrated in Figure \ref{fig:problem}. In previous work, we have investigated using IMUs together with co-location detection of such recognition tasks\cite{ward2006activity, lukowicz2004recognizing,stiefmeier2008wearable}. The problem is that the motion and posture differences involved in for example handling an object together vs. each user handling an object alone are subtle and often overshadowed by differences resulting from the specific object being handled and inter-person variations. Furthermore, people just walking next to each other may be closer in terms of location than people jointly carrying a large object, so that co-location is also not a conclusive indication of collaboration.

Related work in group activity recognition was mainly based on surveillance videos from cameras. Most of this work focused on spatio-temporal relations among the people in the scene. Those works involved tracking multi-agents' spots, evaluating their appearance, aggregating independent and joint features, segmenting their movements, extracting their actions, and then perceiving their activities in a group\cite{gong2003recognition,ryoo2008recognition,zhang2008hierarchical,li2009learning,ni2009recognizing,chang2011probabilistic,yun2012two}. Video-based group activity recognition suffers from its heavy computational cost and other ethical issues. Our work focused on the sensor-based one-dimensional data source and aims to facilitate the detection of joint physical manipulation of objects which is more difficult to detect visually and try to supply a complementary piece of information.

Concerning the sensor-based analysis of multi-agents activities, Wilson et al. \cite{wilson2005simultaneous} offered a combination of motion detectors, break-beam sensors, pressure mats, and contact switches to detect agents' proximity and touch actions, thus correctly classified 84.6\% of the rudimentary activities(whether or not an occupant is moving). An infrared sensor-based scalable network \cite{wren2006toward} was also proposed for perceiving people's activities in an intelligent building and got better than 90\% recognition performance for low-level activities such as walking, loitering, and turning as well as meeting, visiting of multiple agents. Wang et al.\cite{wang2009sensor} developed a multi-modal wearable sensing platform and presented a theoretical framework to recognize both single-user and multi-user activities, in which temperature, humidity, light, audio, RFID, motion sensors were collected to contribute the recognition, they achieved an accuracy of 85.46\% for daily activities like washing face, eating meal and so on. Gordon et al. \cite{gordon2013towards}, explored a node with motion sensors, which can be attached to a mug, by deploying several nodes on mugs, the collaborative group activity was extracted at mobile devices' side, overall an F-score of 53.4\% for global group activity recognition by decision tree was achieved. Chen et al. \cite{chen2019framework} developed a framework for group activity detection and recognition by abstracting the similarity in motion, audio, and proximity from smartphone sensors and beacons and provided more than 89\% accuracy in group detection. Those sensor-based multi-agents activity recognition platforms were mostly too complex to use, and they usually focused on detecting whether the involved agents are gathering or dispersing as an indication of group activities. A high-level activity like the collaboration between the agents is a more challenging topic in this field and exists in plenty of practical scenarios like in manufactories. Ward et al. \cite{ward2017detecting} presented a sensing fusion with body-worn microphones and accelerometers to detect instances of physical collaborative activities between members in a group and achieved an F-score of 60.1\% with two classes "collaboration" and "no collaboration".

\subsection{Contribution}
Overall, we present the following contributions in this paper:
\begin{enumerate}
\item We explored the physiological signal $HBC$ and designed a wearable, low cost, low power-consumption, $IMU$-competitive motion sensing unit by monitoring the $HBC$ signal. We developed a motion tracking prototype with both $IMU$ and $HBC$  units integrated. With this prototype, we demonstrated the competitive/complementary role of $HBC$ to the traditional $IMU$ sensing for individual motion recognition and repetition counting, and collaborative group activity recognition.
\item The first activity recognition experiment was composed of 7 leg-only, machine-free exercises. Five volunteers participated in this experiment with our prototype mounted on the wrist. Ten sessions of $HBC$ and $IMU$ data were collected. With a random forest classifier we got 89\% F-score with capacitance-only data and 78\% F-score with $IMU$-only data while leaving one person out. The result shows that the $HBC$ signal outperforms in the recognition of leg-related exercises. Combining the two sensing sources didn't supply better result than $HBC$ alone. The $HBC$ signal also significantly outperforms in the repetition counting with an averaged accuracy of 98.2\%, especially in the leg exercises when the arm was in a static state so that the $IMU$ lost the leg monitoring ability.
\item We also studied 11 most popular gym workouts, including both aerobic and anaerobic activities. Compared with the leg-related exercises, those gym workouts involve more arm actions. With a random forest model(which shows better result than the CNN and LSTM based neural networks), we got the recognition F-score(leave-one-user-out) of 66\%, 83\%, 91\% with the prototype placed in the pocket, worn on the calf and wrist respectively, using the signal combination of $HBC$ and $IMU$. The recognition result with single signal source shows that $HBC$ doesn't help to improve the $IMU$-alone derived workouts classification result, unless the workout is the ones that only has leg movement to finish the workout, like adductor.  Although the $HBC$ didn't present competitive contribution in general gym workouts recognition compared with $IMU$, it supplies better result over the $IMU$ for workouts counting(0.800 vs. 0.756 for example, when wearing the sensors on the wrist). 

We also classified the ten volunteers with the result of 93\% F-score by combing the data of $IMU$ and $HBC$.
\item $HBC$ is a somewhat more elusive concept instead of a concrete physiological feature. By analyzing the potential influence factors of $HBC$, we demonstrated the robustness of $HBC$ in the gym tracking tasks.
\item The study of a collaborative experiment, TV-Wall assembling and dissembling, indicated that $HBC$ can contribute to recognizing joint activities when the subjects are well coupled or connected physically. By fusing capacitive sensing on wrist and accelerometer on wrist and calf, we achieved the classification accuracy of 71\%, 64\%, 72\%, 88\% for walking alone, carrying alone, carrying together with another subject and the left null state respectively when receiving test data from single-user, and 82\%, 91\% for carrying stuff jointly and the other performed activities when receiving data from both users pairwise with a logistic regression model. 16\% accuracy increase was observed by adding the $HBC$ sensing to a single wrist accelerometer for the collaborative activity recognition.

\item We summarized the work described in this paper as a toolkit in a public repository\cite{bian2021github}, where the hardware, firmware, datasets, machine learning models are accessible to promote other researchers for further exploration of $HBC$ in their specific applications.

\end{enumerate}

\subsection{Paper Structure}
In section \ref{Section_1}, we introduced the motivation of $HBC$ exploration in the field of sensor-based activity recognition, and described related work in individual and collaborative domains. The physical background, measurement principle of $HBC$ sensing, as well as our $IMU$ and $HBC$ integrated prototype was represented in section \ref{Section_2}. In Section \ref{Section_3}, we used the body-worn prototype to classify seven leg workouts performed in labor and ten workouts on a gym studio with different classification methods and count the workouts. We also researched the robustness of the prototype related its potential influence factors. Then we described a TV Wall assembling and disassembling collaborative activity in Section \ref{Section_4} and researched the contribution of $HBC$ sensing in collaborative activity recognition. Section \ref{Section_5} concluded our work and stated the future work.

\section{Physical Background and Sensing Prototype}\label{Section_2}
\begin{figure}
\begin{minipage}[t]{0.5\linewidth}
\centering
\includegraphics[width=0.6\textwidth,height=3.0cm]{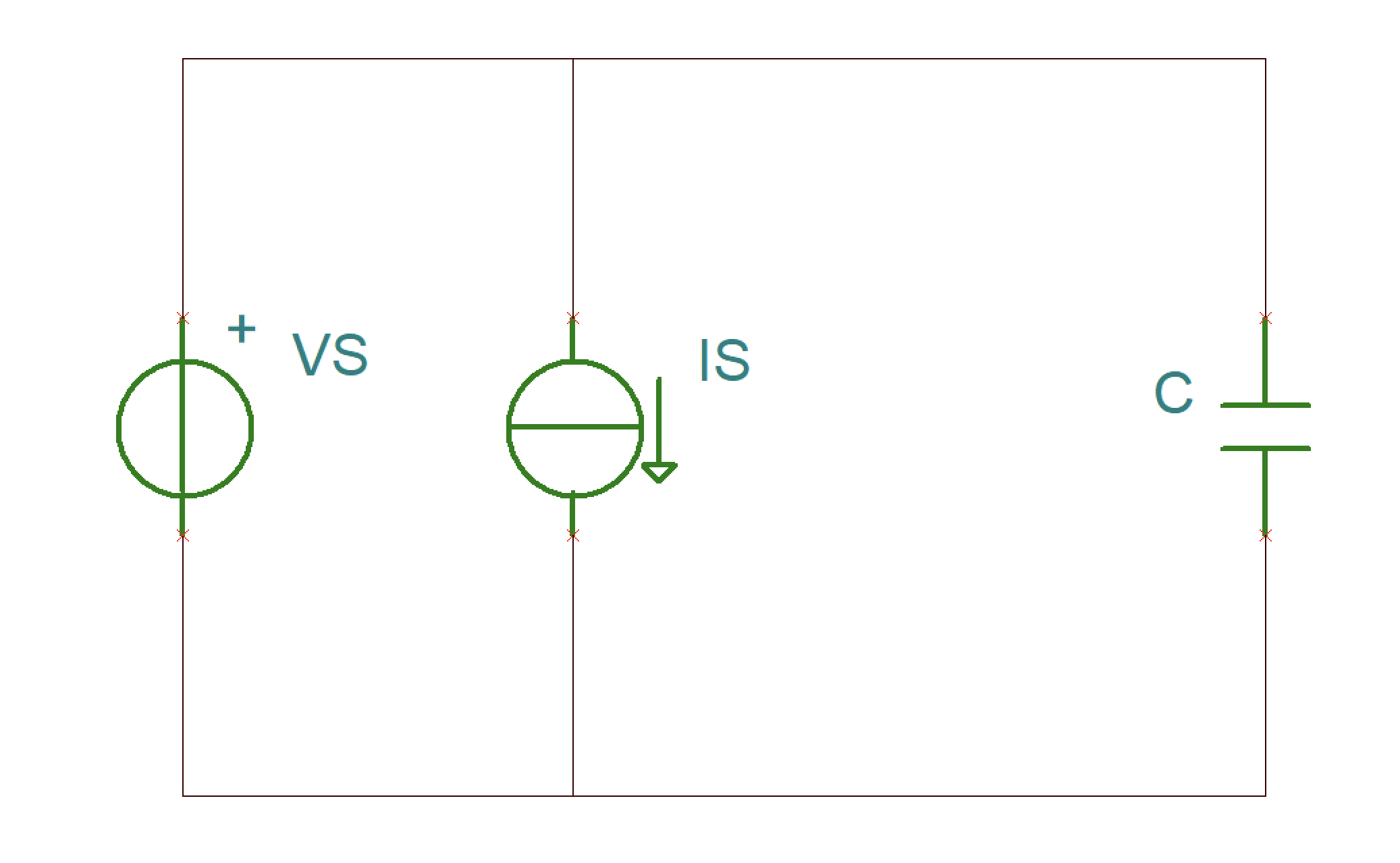}
\caption{Basic structure of a body capacitance sensing method, including a voltage source and current source}
\label{Sensor}
\end{minipage}
\quad
\begin{minipage}[t]{0.5\linewidth}
\centering
\includegraphics[width=0.7\textwidth,height=3.0cm]{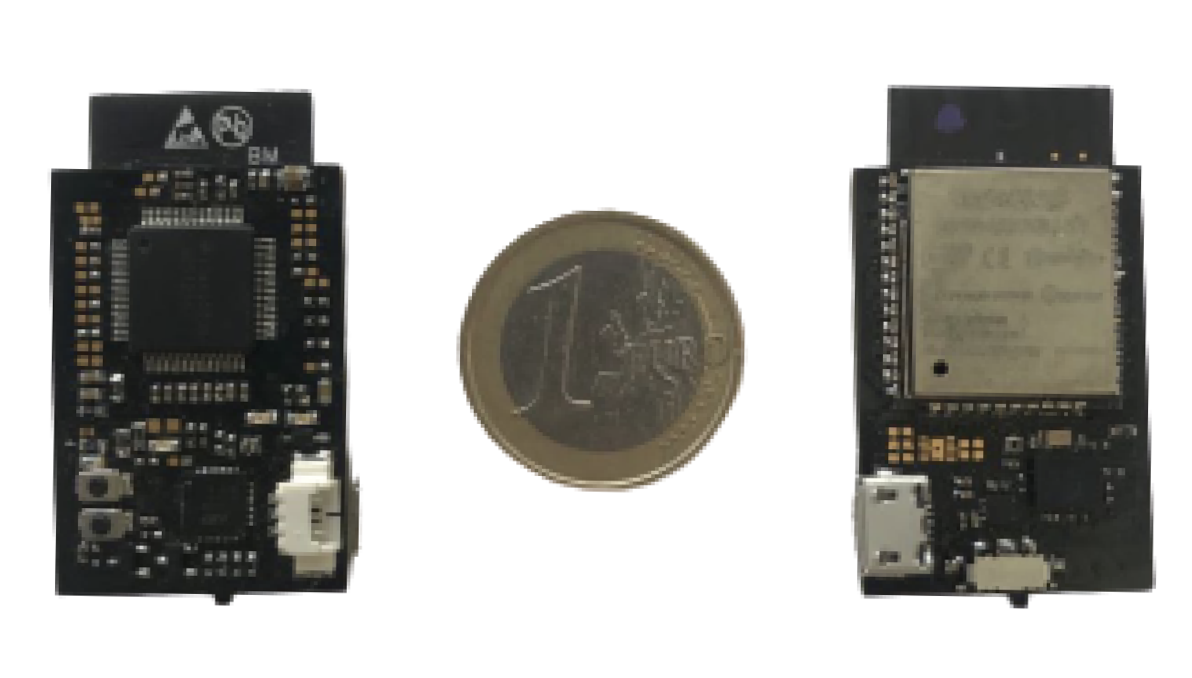}
\caption{A prototype for body motion sensing with IMU and HBC sensing integrated}
\label{Prototype}
\end{minipage}
\end{figure}

As we described above, the human body can store the electrons. Assuming that the charge on a human body is $Q_B$ and the instantaneous capacitance between body and ground is $C_B$, then the potential of the body $U_B$ could be described as:
\begin{equation}  
  U_B = \frac{Q_B}{C_B} 
\end{equation}
$C_B$ is depicted as $C2$ in Figure \ref{Capacitive Model}, which describes the capacitance between body and environment. $C1$ and $C3$ represent the coupling between the human body and sensing devices' local ground, the coupling between ground and devices' local ground. Among them, $C2$ plays the leading role in this capacitive system, namely the human body capacitance, which will dominate the sensor signal since $C1$ is relative constant when the devices are worn on wrist and $C3$ is insignificant because of its long-distance of the corresponding two conductive plates. Therefore mainly the change of C2 delivers the motion information of the body or invasion information of the surrounding. Instead of monitoring the body capacitance directly, we measured the potential of the body continuously. Variation in $C_B$ will cause a potential variation on the human body. Figure \ref{Sensor} depicts the principle of our sensing front end. $C$ is the sum of $C1, C2, C3$ from Figure \ref{Capacitive Model}. The voltage source maintains the potential of the body; the current source is the electrons supplier to $C$. Once $C$ varies, a potential change will occur. The potential returns to the level of $VS$ with the complement of the electrons from $IS$ later. The whole mechanics is a series of charging and discharging processes. A more detailed explanation is in our previous work\cite{bian2019passive}. Figure \ref{Prototype} is the prototype for the motion tracking, which is composed of an ESP32 processing unit, a 24 bits high-resolution ADC unit, an IMU, and the $HBC$ sensing front end. The standard 43mm EKG electrode is used as a connection medium between the electrode of the prototype and the human body. 

Table \ref{Comparision} lists the differences between IMU and $HBC$ sensing modality. The $HBC$ sensing modality enjoys the similar advantages of IMU in size, cost, and power consumption. It outperforms with the properties of non-attachment and the surrounding-sensitivity. The disadvantage of $HBC$ sensing is its non-linearity, subtle analog output, thus a high resolution analog to digital chip is needed for sampling.

\begin{table}[H]
\centering
\caption{IMU and $HBC$ Sensing Difference}
\label{Comparision}
\begin{tabular}{ p{0.6cm} p{0.6cm} p{0.6cm} p{0.6cm} p{1.2cm} p{1.2 cm} p{2.2cm} p{2.2cm} p{2.2cm}}
\toprule
& size & cost & power cost & linearity & output & circuit form & attach position on body & surrounding sensitive \\ 
\midrule
IMU & small & Euros & $uW$ & yes & digital & integrated circuit chip & motion part & no \\

$HBC$ & small & cents & $uW$ & no & analog & discrete components & motion/static part & yes \\ 
\bottomrule
\end{tabular}
\end{table}

\begin{figure}[!b]
\centering
\includegraphics[width=0.65\linewidth,height=10cm]{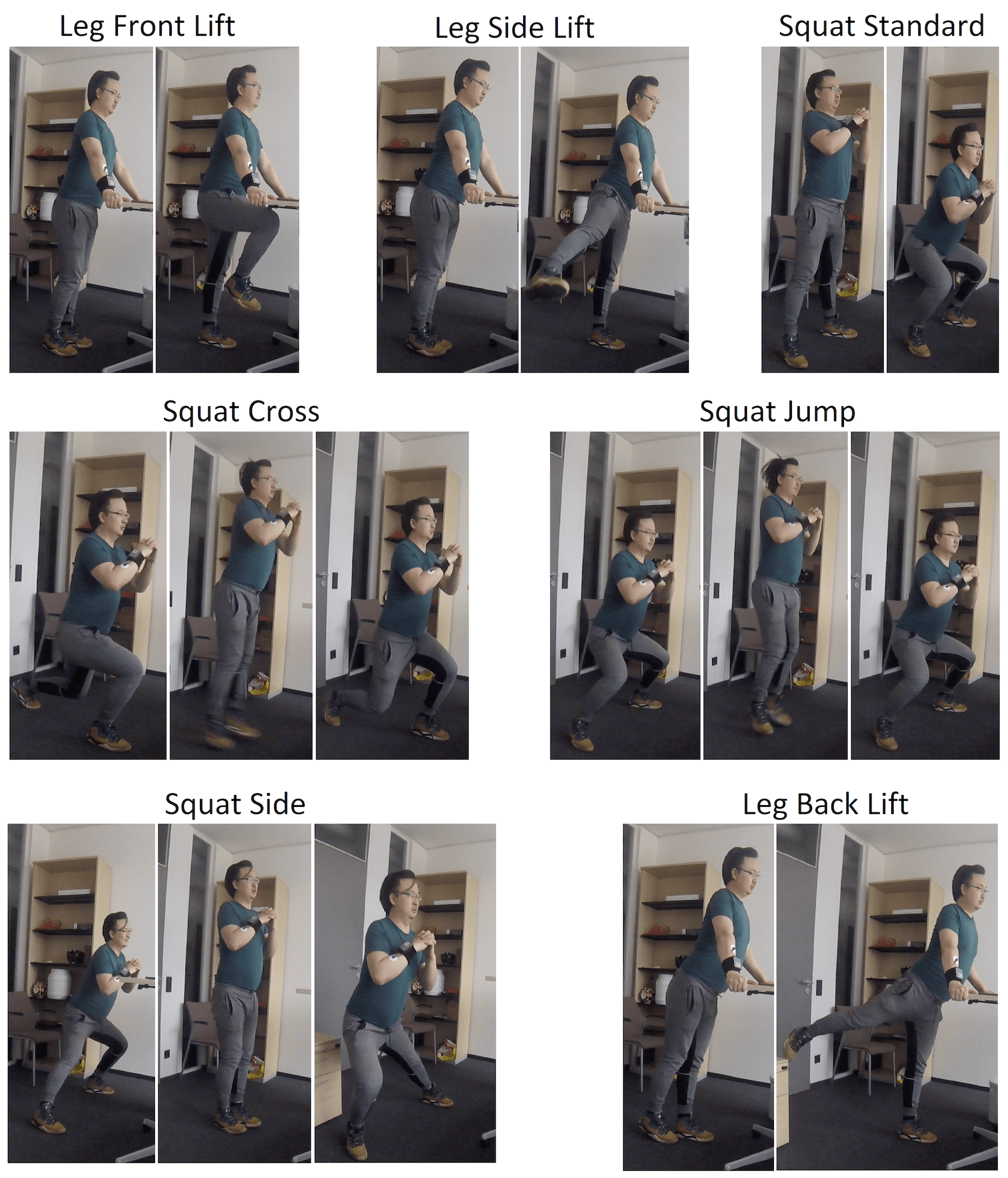}
\caption{Seven leg exercises( leg-front/side/back-lift with hands grasping the table and squat-standard/cross/jump/side with crossed hands in front of chest)}
\label{Leg_Exercise}
\end{figure}

As we explained above, since the $HBC$ sensing modality is sensitive to both action from body itself and from surrounding, we explored two practical activity recognition tasks. The first one is to recognize and count workouts performed by a single user in labor and gym studio respectively, the second one is to recognize a group activity including the physical collaboration: assembling and dissembling a TV-Wall.

\section{Individual Activity Exploration}\label{Section_3}

We explored two kinds of individual activity to research the contribution of $HBC$ to body motion sensing tasks: a preliminary experiment of machine-free leg-exercises in labor and another more practical experiment of general gym workouts in a fitness studio. The two experiments were explored by both traditional machine learning and deep neural network models for classification. The accuracy of activity recognition and repetition counting with signal sources from $HBC$ and $IMU$ in both experiments was described.

\subsection{Experiment in Labor: seven machine-free leg exercises}

This experiment was described in our previous regular paper\cite{bian2022using}, here we briefly summarized the experiment setup and the classification and counting result.

\subsubsection{Experiment Setup}

This preliminary experiment includes seven machine-free leg-exercises(Figure \ref{Leg_Exercise}): leg-front-lift, leg-side-lift, leg-back-lift, standard-squat, cross-squat, jump-squat, and side-squat. For the three leg-lift exercises, the volunteer's wrist was in a static state, while for the other four squat exercises, the wrist moved alongside the body movement. Five fitness enthusiasts aged 25 to 32 participated in the experiments with the prototype worn on the wrist. The exercises were performed in an office room and in an non-instructed way.


\begin{figure}[!b]
\centering
\includegraphics[width=0.95\linewidth,height=18cm]{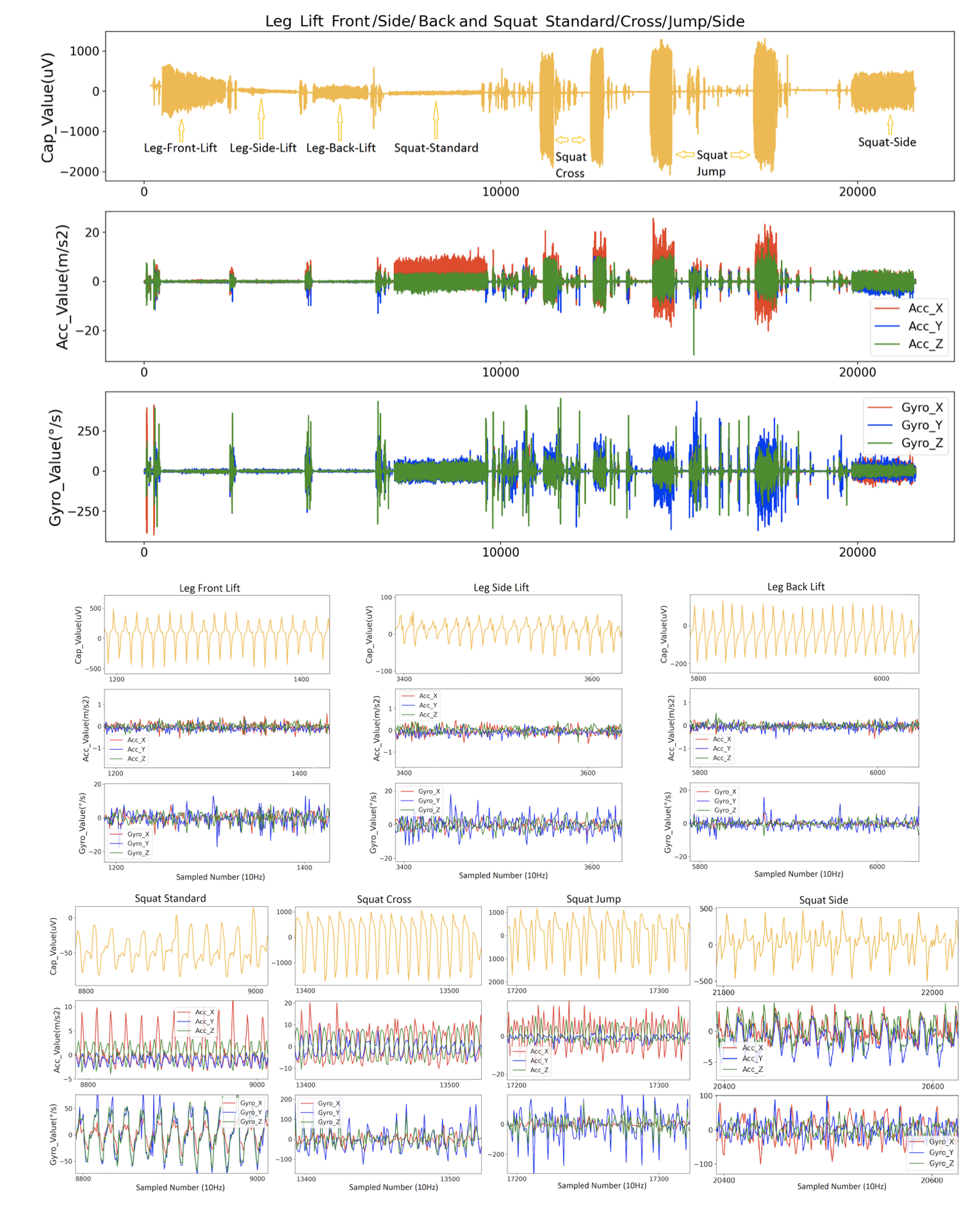}
\caption{A Full session and the close look of $HBC$ and $IMU$ signals of the seven leg-dominated exercises, $HBC$ sensing unit captures a more clear motion signal than the $IMU$ while doing the leg-lift exercises with hands in the relatively static state}
\label{Leg_Session}
\end{figure}



Figure \ref{Leg_Session} depicts a session of the experiment with both $HBC$ signal and $IMU$ signals and the close look of each exercises. It is obvious that the $HBC$ gives a more regular motion signal for all the seven exercises than the $IMU$ signals. However, since $HBC$ is an subtle physiological property, factors like the wearings could have an iinfluence on the sensitivity. Thus the sessions could have a different drawing of the $HBC$ signals. Since the volunteers performed the exercises with their preferred speed and scale and worn their daily sports clothes in the experiment, we explored the contribution of this physiological signal without considering the influence factors like the volunteer's wearing, body conditioning, and the environment. Another interesting point of the $HBC$ signal is that, as the Leg-Front-Lift signal in the figure shows, the amplitude varies even when the volunteer was doing the same repetition action. This variation can be observed in other sessions and different exercises. Whether it was caused by the environmental variation(like other bodies' proximity) or other body parameters (like the skin humidity) is still unknown, which will be one of our future topics focused on the potential influence factors of the $HBC$. From the close-look of each exercise, the sensed capacitance signal could capture the leg's repetitions, especially for the three leg-lift exercises where it is hard to get a sufficient signal from the $IMU$(which mostly gave irregular noisy signals). Overall, ten sessions of data from the five volunteers were collected, including 1500 leg-front-lift, 1500 leg-side-lift, 1500 leg-back-lift, 1500 standard-squat, 1000 cross-squat, 1000 jump-squat and 1000 side-squat.

\subsubsection{Classification Exploration}

\begin{enumerate}[label=(\Alph*)]
\item Random Forest
\end{enumerate}

We firstly utilized classical machine learning approach for exercise classification as the classical approached were proved in a mount of literary HAR explorations\cite{casale2011human, feng2015random, bayat2014study, nurwulan2020random}. A diverse of models were evaluated and the Random Forest provided the best result. We firstly split each session data with four seconds sliding window(two seconds overlap), then handcraft the features from each instance, and finally, feed the feature instances into the random forest model with grid-searched hyper-parameters(tree numbers with 20 and tree depth with 15). 

The features we used are:
\begin{itemize}
\item mean, standard deviation, max, min, difference between max and min
\item mad: Median absolute deviation 
\item energy: Sum of the squares divided by the number of values. 
\item IQR: Interquartile range 
\item Minimum distance of neighbor Peaks
\end{itemize}
We deduced 14 variations from the original seven signals:
\begin{itemize}
\item Cap, Acc$/$Gyro\_XYZ
\item Cap\_Jerk, Acc/Gyro\_Jerk\_XYZ
\end{itemize}

\begin{figure}[!t]
\centering
\includegraphics[width=0.99\linewidth,]{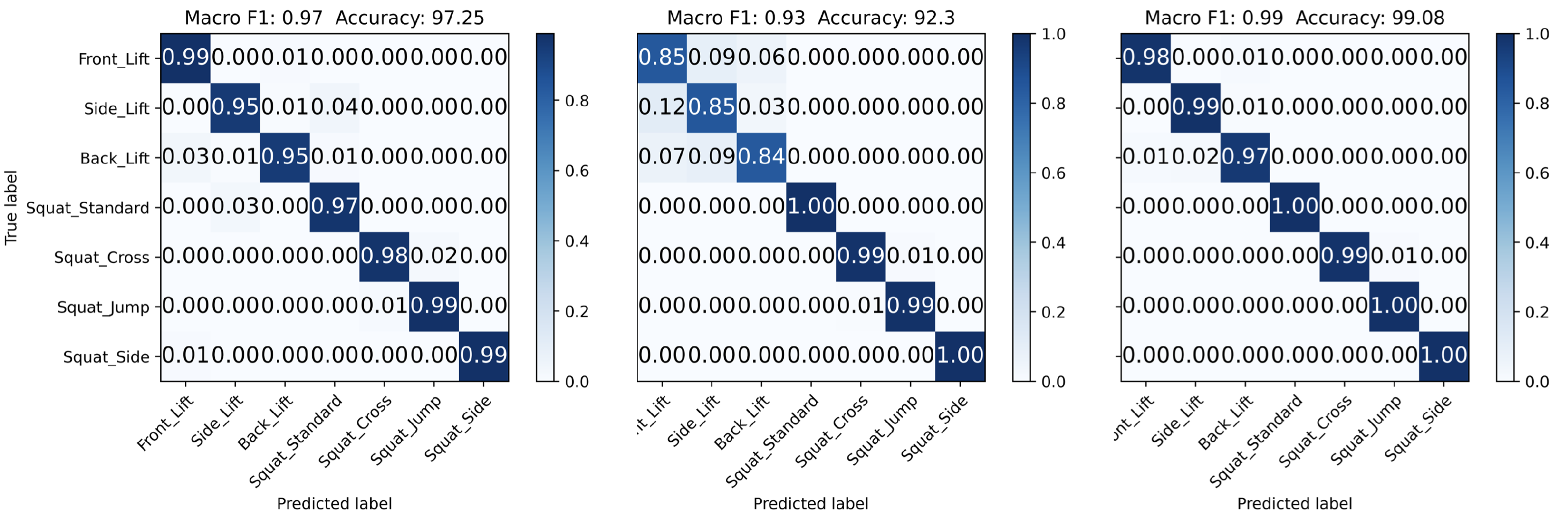}
\caption{Classification result with three-folds cross-validation with features from $HBC$, $IMU$, and both(from left to right), respectively. (Testing data was randomly chosen with a ratio of 0.3)}
\label{3_Fold}
\end{figure}

\begin{figure}[!t]
\centering
\includegraphics[width=0.80\linewidth,]{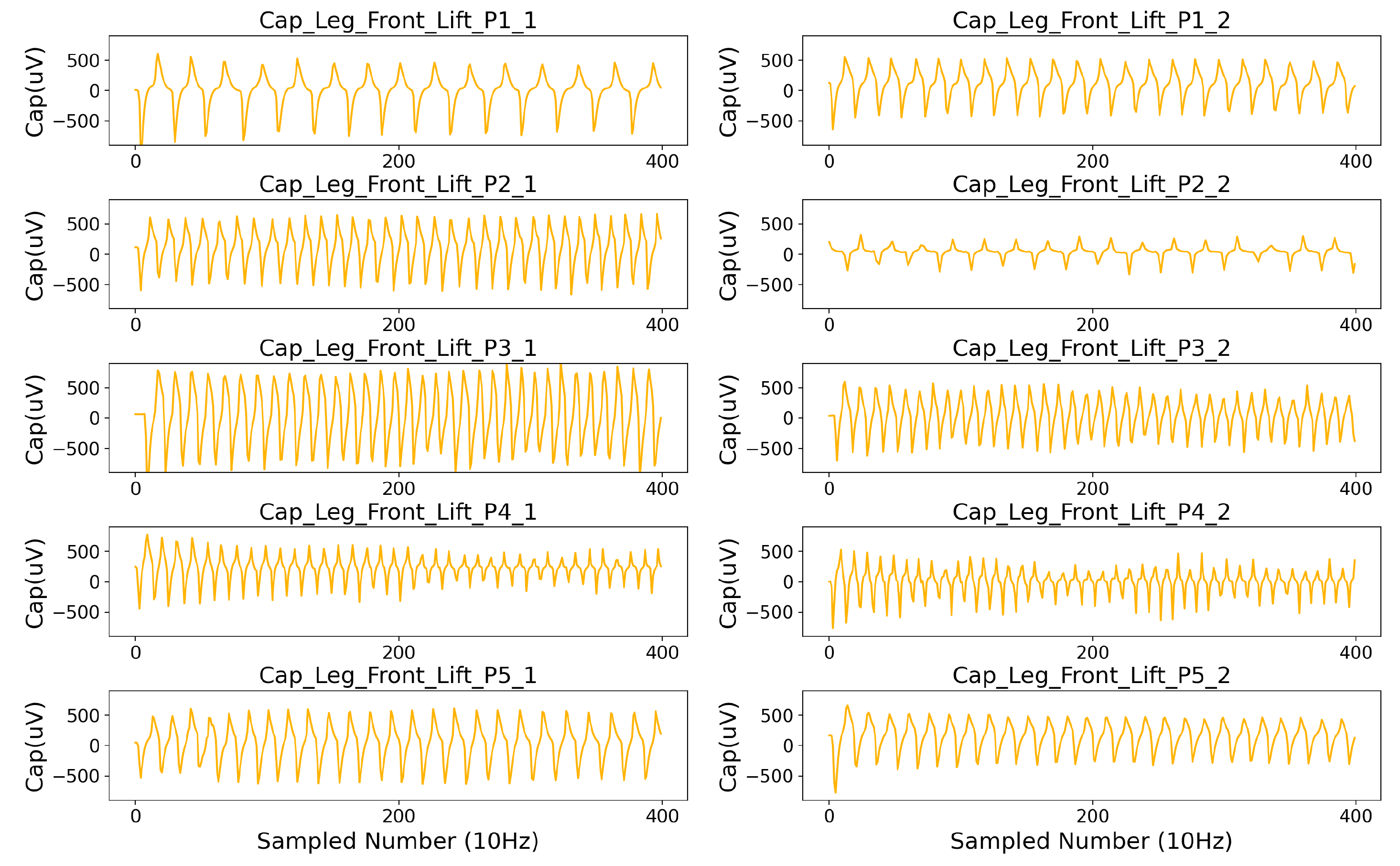}
\caption{A close look of the leg-front-lift's $HBC$ signal from each session, indicating the elusiveness of $HBC$ signal. An identical signal pattern doesn't exist because of volunteers' different moving scales, speeds, and wearings}
\label{Leg_Front_Lift_All}
\end{figure}

Where Jerk means the change rate of the signal\cite{hamalainen2011jerk}. In total, we utilized 18 features per window for $HBC$-based exercises classification, and 108 features per window for the $IMU$-based.
The minimum distance of neighbor peaks was calculated by firstly detecting the peaks in each window by a peak detection\cite{find_peaks} approach, then picking out the minimum distance of the neighbor peaks. The training data was balanced before feeding it into the model with the method of SMOTE\cite{chawla2002smote}. 
We firstly split all instances into four portions randomly(with the ratio of 0.3, 0.3, 0.3, and 0.1), then performed three-fold cross-validation(data with 0.3 rate as the testing data). 
The classification result with signal sources of $HBC$, $IMU$ and the combination is depicted as Figure \ref{3_Fold}. The $HBC$-sourced result shows a better F1 score and recognition accuracy than $IMU$-sources one, benefiting from its significant full-body motion sensing ability of the three leg-lift exercises. The $IMU$, however, also shows the considerable result for the first three exercises even the wrist was in a relatively static state, which indicates the high sensitivity of $IMU$ for motion sensing(like heart rate detection with wrist-worn accelerometer\cite{haescher2015study} from the muscular micro-vibrations), although the sensed signals are irregular and noise-like, as Figure \ref{Leg_Session} depicts. Combining both $HBC$ signal and $IMU$ signals, we got the classification accuracy over 99\%. However, this result is over-optimistic. Firstly, the data from the same repetition set have more considerable similarity compared with other sets. Secondly, the $HBC$ is an elusive signal compared with $IMU$, because of its many potential influence factors, like wearings(especially type and height of the sole) and postures, thus it is a challenge to grasp the pattern of $HBC$-variation caused-motion signal in an unregistered session. As Figure \ref{Leg_Front_Lift_All} shows, in each session, the $HBC$ signal of leg-front-lift gives varied signal regarding the amplitude and wave shape, even during the same set.

\begin{figure}[!t]
\centering
\includegraphics[width=0.99\linewidth,]{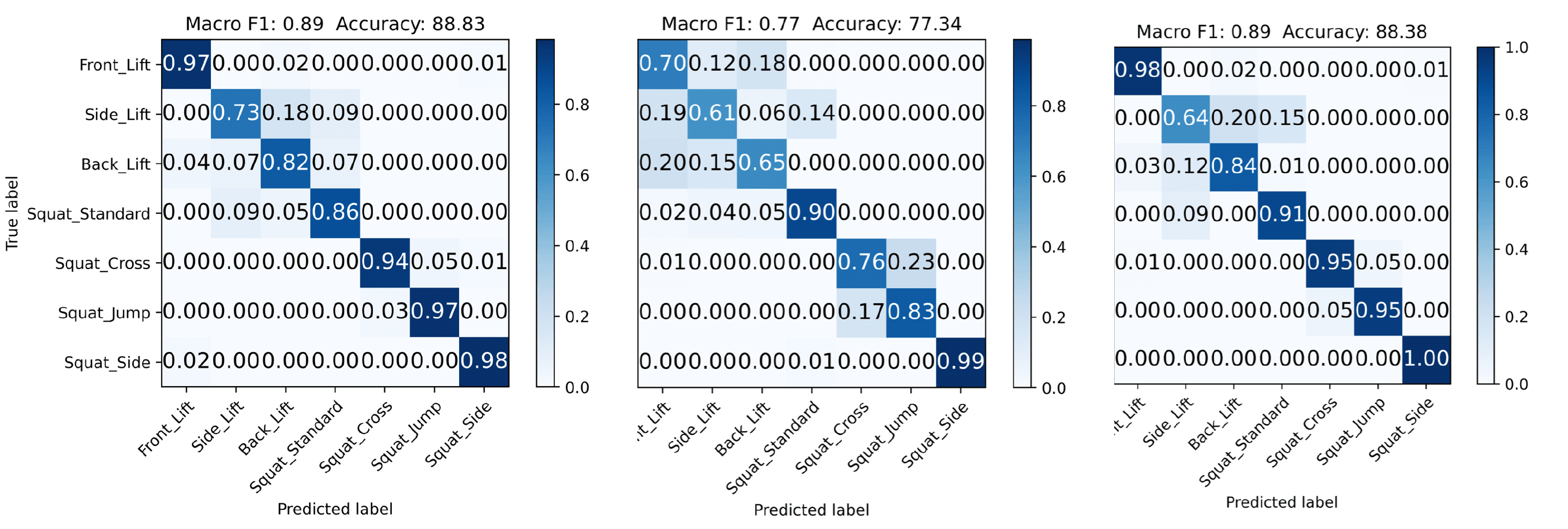}
\caption{Classification result with five-folds cross-validation with features from $HBC$, $IMU$, and both(from left to right), respectively. (Leave one person out)}
\label{5_Fold}
\end{figure}

\begin{figure}[!t]
\centering
\includegraphics[width=0.99\linewidth,]{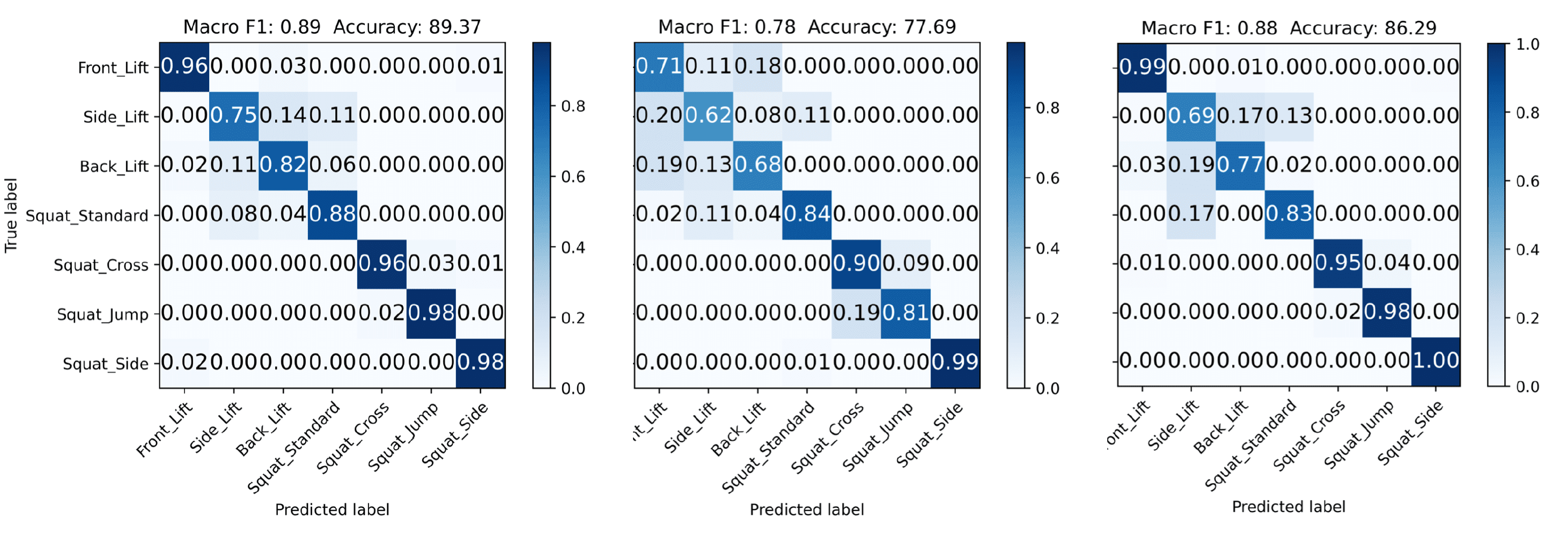}
\caption{Classification result with ten-folds cross-validation with features from $HBC$, $IMU$, and both(from left to right), respectively. (Leave one session out)}
\label{10_Fold}
\end{figure}

Exploration of how $HBC$ varies regarding the potential influence factors is covered in our previous work\cite{sizhen2021systematic}. To examine the robustness of the $HBC$ signal(as a source of leg exercise recognition) against encountering unregistered users as well as other possible influence factors, we performed cross-validation with a five-folds leave-one-user-out and a ten-folds leave-one-session-out. Before the cross-validation, we normalized the $HBC$ signal of each session so that the leg-front-lift has the same minimal and maximal signal(\-500 $uV$ and 500 $uV$). Figure \ref{5_Fold} and \ref{10_Fold} depict the results after normalization.
After adopting the unregistered sets as testing data, the classification performance gives more practical and reliable results with the two sensing modalities. As the source signal, $HBC$ gives better classification results for almost every class than $IMU$, particularly for the three leg-lift exercises, where it shows an increase of 0.12 to 0.27 in F-score.
Combining both signal sources doesn't supply a better classification.  As a result, we could conclude that the $HBC$ outperforms in the machine-free leg-exercises classification(0.89 vs. 0.78 in F-score).

\begin{enumerate}[label=(\Alph*)]\setcounter{enumi}{1}
\item Deep Neural Network
\end{enumerate}

Deep neural network models were developed for image classification, of which the model accepts a two-dimensional input representing an image’s pixels and color channels, in a process called feature learning. Over the last decade, the neural network-based computer vision shows significant advances from single object recognition to streaming video content analysis\cite{toutiaee2020video, amosov2020using}. The same advances were also presented in natural language processing\cite{torfi2020natural, nassif2019speech} so that machines are empowered to get a better understanding of the human language. There is no doubt that this usage of the deep neural network will be more comprehensive with the developments in computational power and the advent of large amounts of data. With the great success of deep neural network in computer version and natural language processing, researchers have deployed a scale of deep neural network-based models in sensor data based recognition tasks by automatically learning features from the raw sensor data\cite{wang2019deep}. In the field of human activity recognition, plenty of deep methods have been utilized to classify activities from sensor data and achieved a better understanding of people's behaviors, as summarised by Chen et al.\cite{chen2020deep}. The well-known Sussex-Huawei Transportation-Locomotion (SHL) Recognition Challenge\cite{wang2020summary} intends to recognize eight locomotion and transportation activities from the inertial sensor data of a smartphone. 
Different state-of-the-art result was presented from different models each year as the data volume increased, which indicates that the best way to get the best classification is to try different models.

\begin{figure}[!htb]
\centering
\includegraphics[width=0.6\linewidth,]{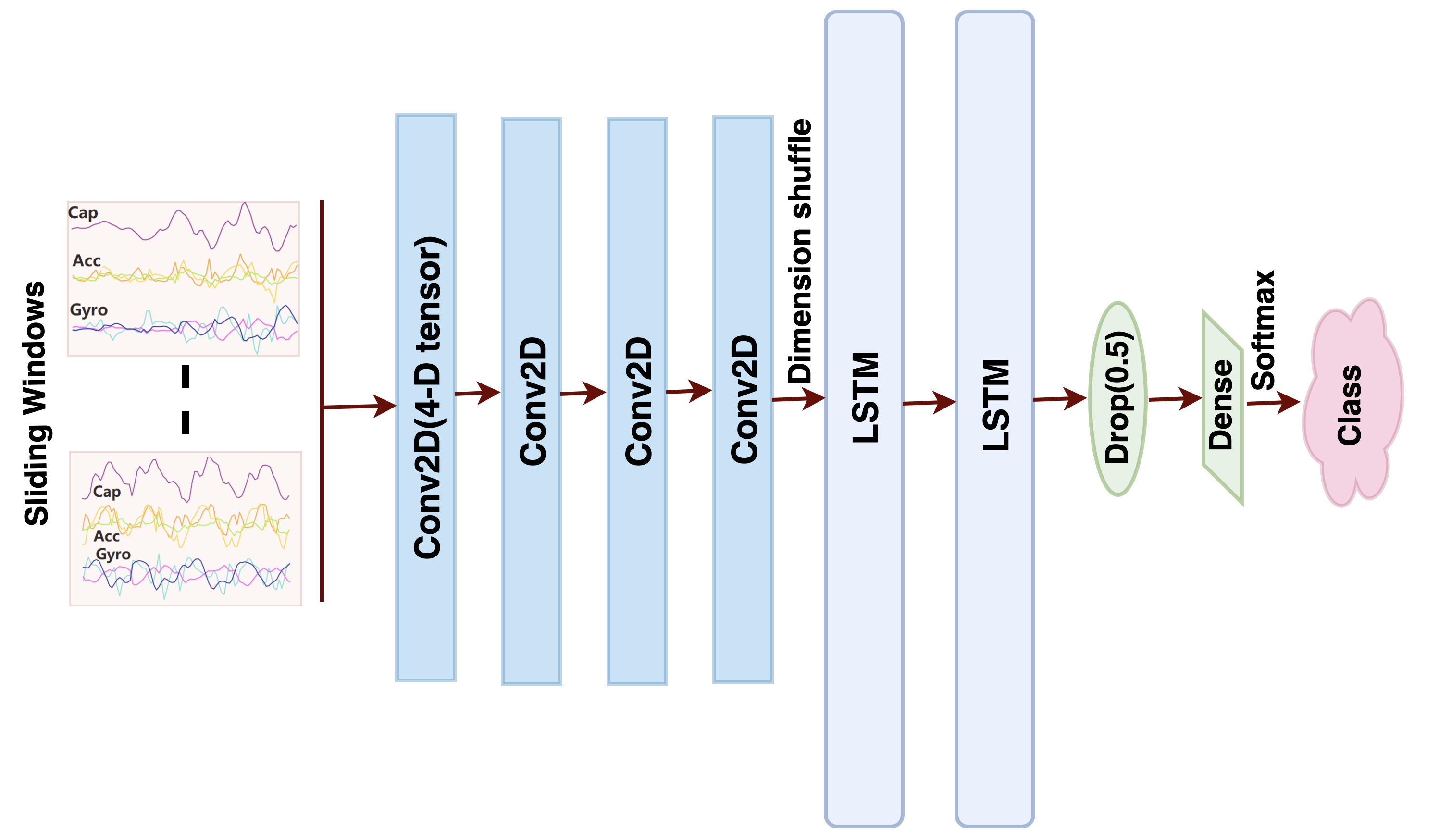}
\caption{Architecture of the DeepConvLSTM model implemented in Theano using Lasagne\cite{lasagne}}
\label{DeepConvLSTM_RAW}
\end{figure}

\begin{figure}[!htb]
\centering
\includegraphics[width=0.4\linewidth,]{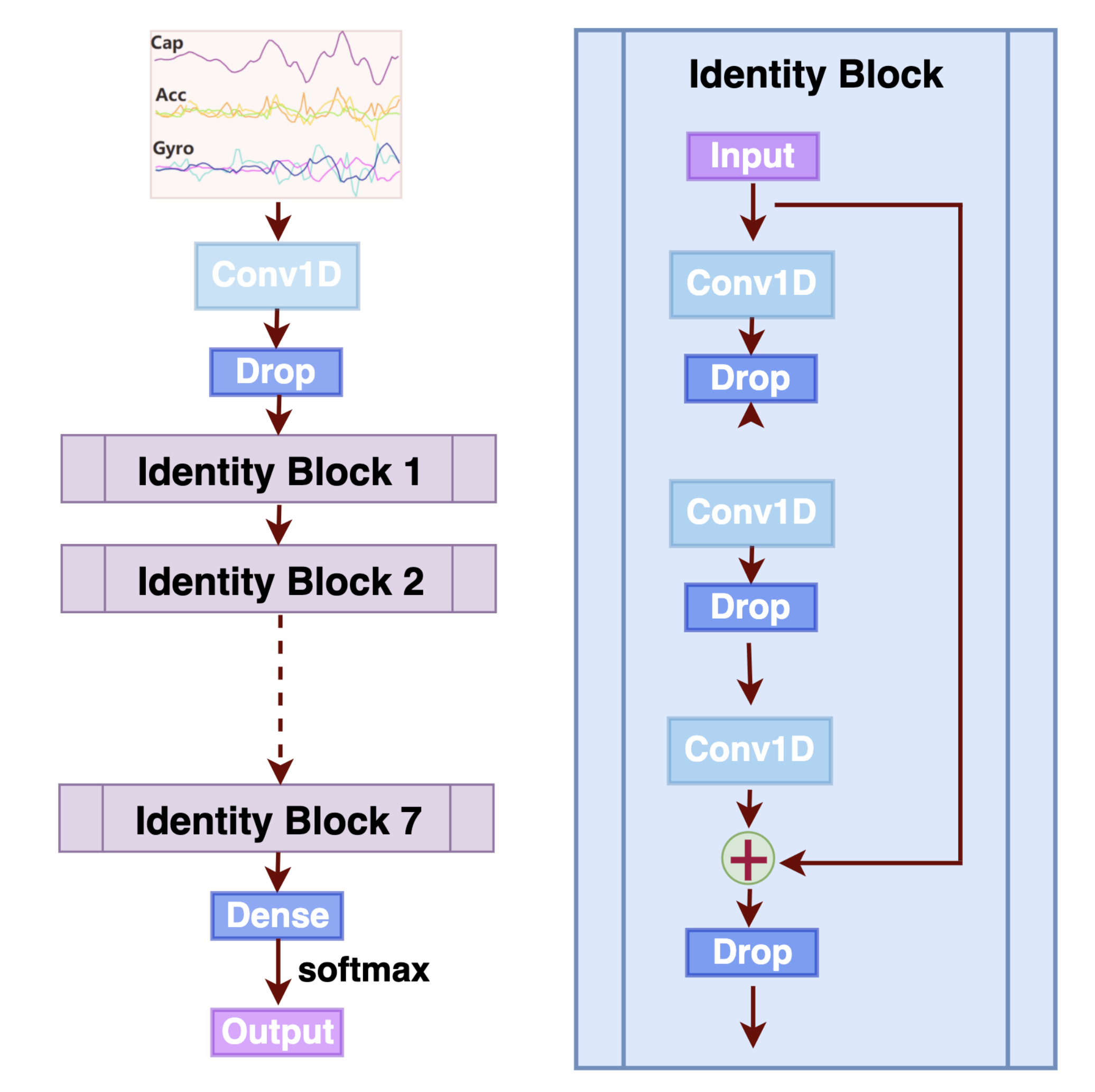}
\caption{Architecture of the residual network with 21 layers for the classification of the leg dominated exercises}
\label{Resnet21}
\end{figure}

In the last subsection, we use a random forest model to classify the seven leg exercises and get the leave-one-session-out classification F-score of 0.89 with $HBC$ signal, 0.78 with $IMU$ signal. In this section, we explored the performance of different deep neural network models(CNN, LSTM, etc.) for the classification task. Here we present two of the applied deep models, which were designed for similar sensor-based human activity recognition tasks and achieved state-of-art result in different public data set. The first model is named with "DeepConvLSTM" by Francisco J. et al. in their work\cite{ordonez2016deep}. 
The architecture of the DeepConvLSTM model is shown in Figure \ref{DeepConvLSTM_RAW}. The convolutional layers extract features and the recurrent layers model the temporal dynamics of the feature maps. 
The second model is a deep residual network inspired from Qin et al.\cite{qin2020imaging}, where the authors encoded the time series of sensor data as images and leveraged these transformed images to retain the necessary features. In our residual networks, we used the 1D convolutional neural networks supplied by Keras\cite{chollet2015keras} directly to extract features from sequences data and map the internal features of the sequence, which has been proved effective in related works\cite{cho2018divide, cruciani2020feature}. 
Figure \ref{Resnet21} shows the architecture of the residual networks used in our leg exercise classification. The network model was trained with the categorical cross-entropy loss function and the adam optimizer\cite{KingmaB14} with 0.0001 learning rate and $0.9$ and $0.999$ for $\beta_1$ and $\beta_2$, respectively. Each fold is trained for 4000 epochs with early stopping using a patience of 200 to avoid the over-fitting. The validation set used for the early stopping procedure consists of 20\% random samples of the training set. The batch size is set to 128. In each one-dimensional convolution layer, we use 64 filters with a size of 3. The activation function is $ReLu$, padding is $same$. All dropout rate was set to 10\%. 

The classification results of both deep models are listed in Table \ref{Resnet_Table}. Compared to the previously presented random forest performance, the deep neural network results are less competitive. This is reasonable since, firstly, neural network-based inference models are more dominant in domains where a massive data set is available, like the Image-Net, IMDB-reviews, etc. In which case, the generalized features can be precisely abstracted by shallow layers. Whereas the random forest with statistical features mostly gives better classification results on the limited data set and it is less prone to over-fitting because of the ensemble methods. Compared to computer vision applications, sensor-based human activity recognition still faces the problem of data insufficiency. The same result, where the random forest outperforms in human activity recognition, can be found in a mass of works\cite{wang2018summary, wang2019summary, gjoreski2020classical}. Secondly, sensor-based human activity recognition is a task that is highly user- and sensor-dependent. Variables, including the dynamic activity complexity, sensor orientation and placement, data quality, etc., will cause an unsatisfactory performance of models, especially the neural network-based ones, where the models struggle more for generalization with the end-to-end structure. From the literature, with the current public human activity recognition data set, there is still not a common ground of deep neural network-based model that supplies a stable impressive result\cite{hoelzemann2020digging, morales2016deep}.

\begin{table*}[htbp]

\centering
\begin{threeparttable}
\caption{Classification result with deep models: F-score/Accuracy}
\label{Resnet_Table}
\begin{tabular}{p{3.0cm}  p{3.0cm} p{2.0cm} p{2.0cm} p{2.4cm}}
\toprule
Deep model & Test approach & $HBC$ & $IMU$ & $HBC$+$IMU$ \\ 
\midrule
DeepConvLSTM & Leave one user out & 0.76 / 0.73 & 0.75 / 0.73 & 0.77 / 0.73 \\

&Leave one session out & 0.74 / 0.70 & 0.76 / 0.73 & 0.75 / 0.72 \\ 

Resnet 21 & Leave one user out & 0.75 / 0.75 & 0.65 / 0.65 & 0.71 / 0.74 \\

& Leave one session out & 0.76 / 0.77 & 0.60 / 0.61 & 0.69 / 0.72 \\ 

\bottomrule
\end{tabular}
\end{threeparttable}
\end{table*}

\subsubsection{Exercise Counting Exploration}

As previous figures show, with the the prototype worn on the wrist, the capacitance signal could give obvious peaks in the data stream. We thus used peak detection\cite{find_peaks} approach to count the exercise. Peaks were counted from the raw data of $HBC$ signal, noise-filtered Z-axis of accelerometer, and noise-filtered Y-axis of the gyroscope(both axes provide the best result compared to the other two). We use accuracy (Equation \ref{equation_accuracy}) to present the counting performance.


\begin{equation}
\label{equation_accuracy}
Accuracy =  1.0 - \frac{\abs{count_{detected} - count_{real}}}  { count_{real} }
\end{equation}

\begin{figure}[!t]
\centering
\includegraphics[width=0.8\linewidth,]{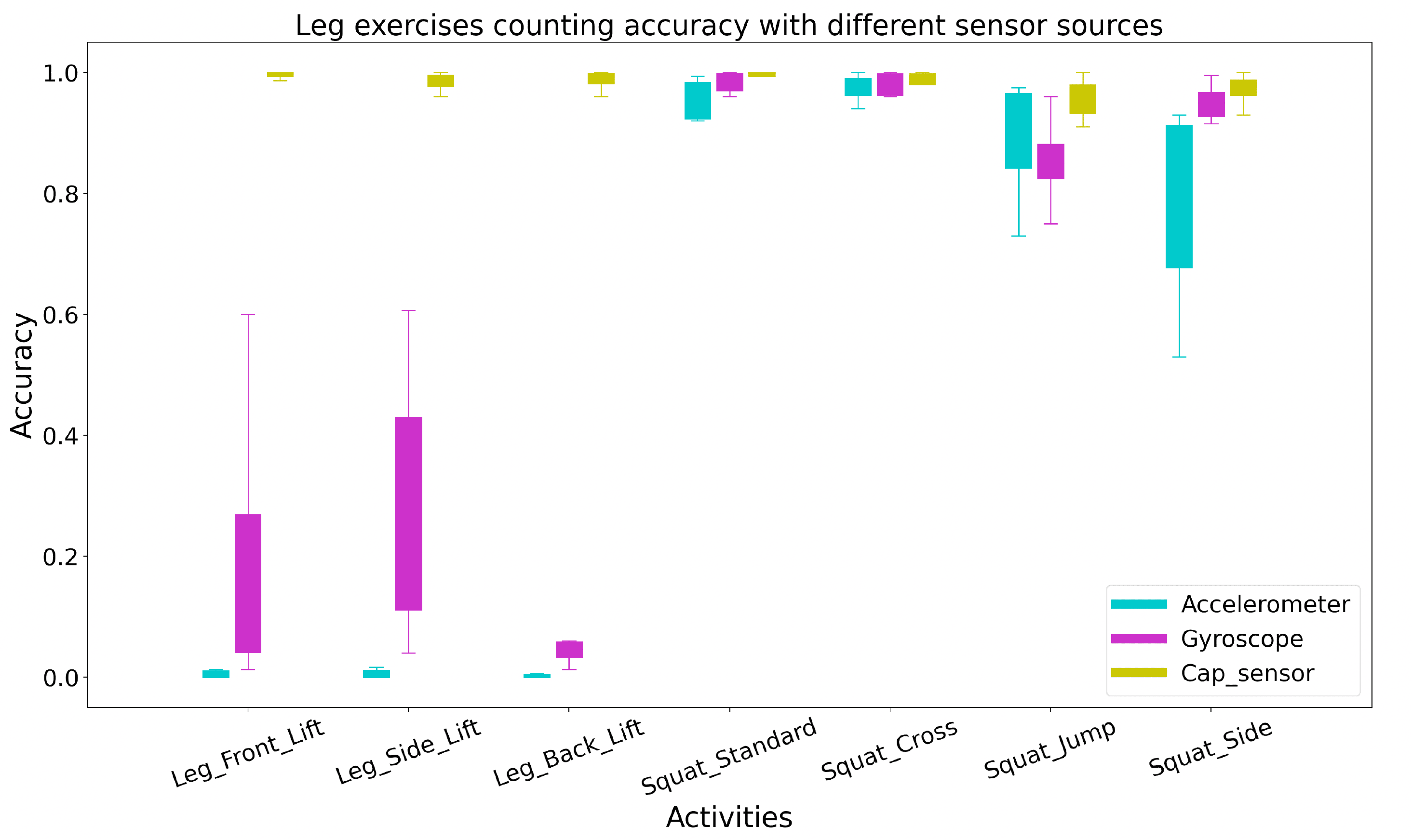}
\caption{Counting accuracy of the seven leg-exercises with signal source of $HBC$ and $IMU$}
\label{Leg_Count_Accuracy}
\end{figure}

\begin{figure}[!t]
\centering
\includegraphics[width=0.8\linewidth,]{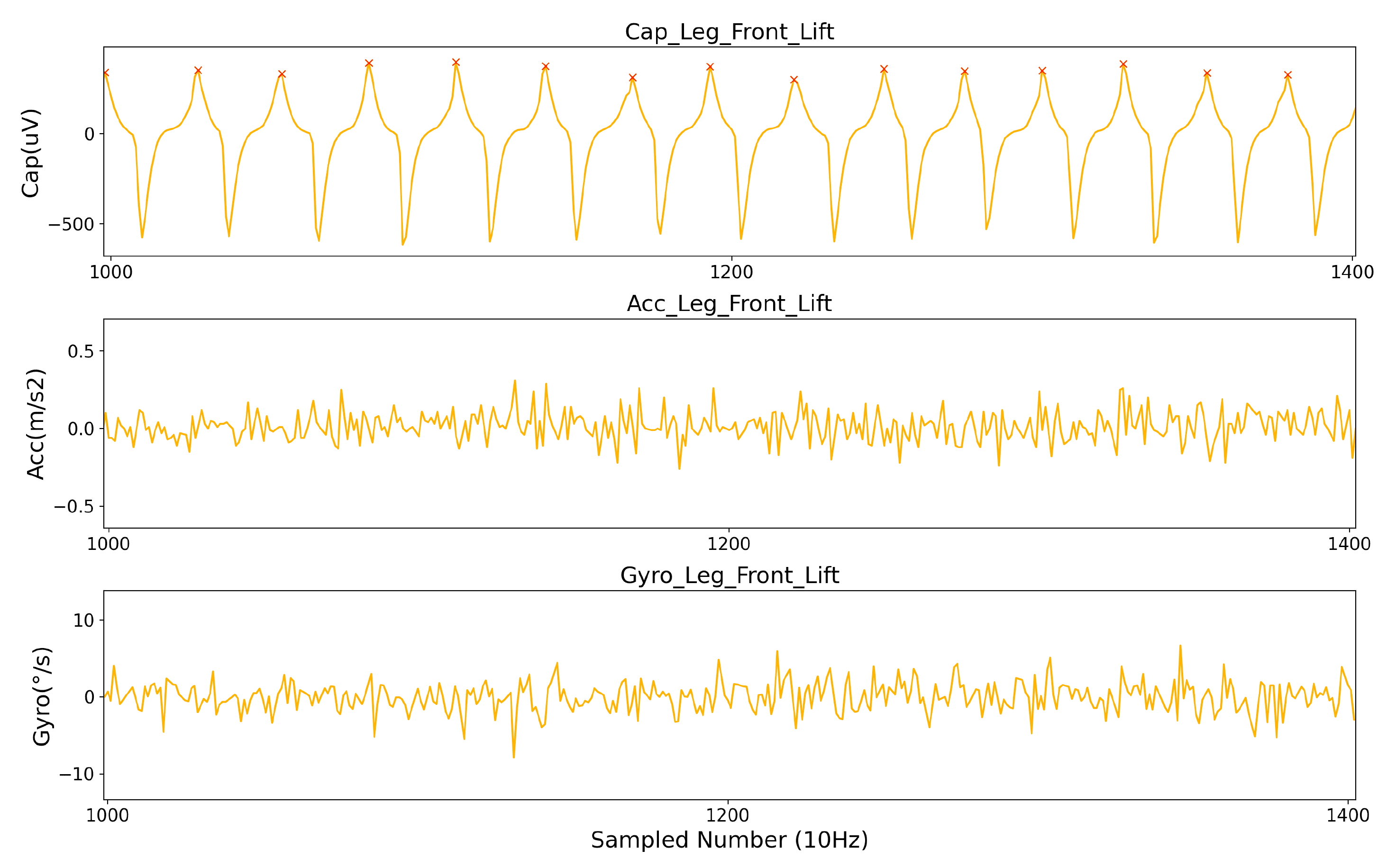}
\caption{Raw signal of $HBC$ and $IMU$ for leg-front-lift, indicating that $HBC$ signal can be used for counting in such exercises reliably, while $IMU$ only captures irregular micro-vibration}
\label{Leg_Count_Comparison}
\end{figure}

We use Boxplot to show the counting accuracy of each exercise with different signal sources(as Figure \ref{Leg_Count_Accuracy} shows). For the leg-lift exercises, the wrist-worn $IMU$ completely lost the repetition counting ability. In contrast, the $HBC$ could produce a reliable count number with over 95\% accuracy. For the other four squat-related exercises, $HBC$ also gave a better counting result than $IMU$. Table \ref{Counting_Leg_Exercise} lists the counting accuracy with the two sensing modalities, where the $HBC$ signals outperforms with an accuracy of 98.2\%.


\begin{table*}[htbp]
\centering
\begin{threeparttable}
\caption{Counting accuracy with signal of $HBC$(all seven exercises) and $IMU$(only four squat-related exercises)}
\label{Counting_Leg_Exercise}
\begin{tabular}{p{2.8cm} p{1.8cm} p{1.8cm} p{1.8cm} }
\toprule
Signal source & Acc & Gyro & $HBC$ \\ 
\midrule
Mean Accuracy(std) & 0.891 $\pm$0.119 & 0.938 $\pm$0.066 & \bfseries 0.982 $\pm$0.022 \\
\bottomrule
\end{tabular}
\end{threeparttable}
\end{table*}

\subsection{Experiment in Gym Studio: eleven popular gym workouts}
\subsubsection{Experiment Setup}
\begin{figure}[hbt]
\centering
\includegraphics[width=0.65\linewidth,height=11cm]{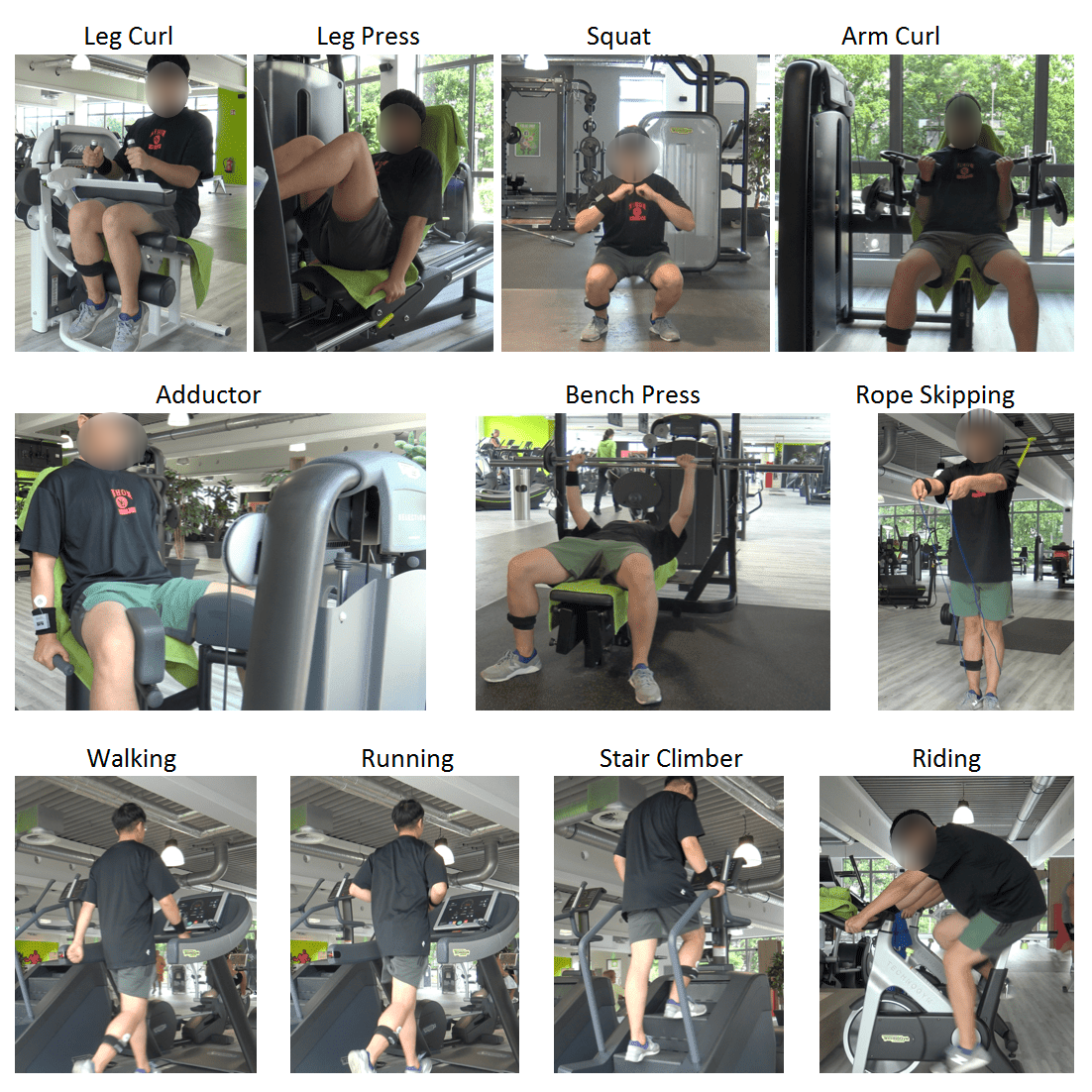}
\caption{Eleven gym workouts}
\label{Eleven}
\end{figure}

To explore our prototype's workouts recognition and counting ability in a more practical scenario, where the arm also plays a role in a complete workout action, we chose eleven most popular workouts to recognize and count individual activities performed in a gym studio. The exercises include both aerobic and anaerobic training: Adductor, Armcurl, Benchpress, Legcurl, Legpress, Riding, Ropeskipping, Running, Squat, Stairsclimber and Walking(as Figure \ref{Eleven} depicts). Both muscle strength and muscle endurance get trained. All the core muscle groups are considered within those 11 exercises, including pecs, quads, calves, biceps, triceps, gluteus, and hamstring. Running and Walking were performed on the treadmills with the speed of 5$\pm$0.2 km/h and 8$\pm$0.5 km/h for around 2 minutes in each session. Riding and Stairclimber were done at a self-determined pace and lasted about 2 minutes. The rest exercises were trained with gym instruments(except Squat) for 3$x$10 repetitions. Ten volunteers participated in this study, including five females and five males, with ages from 21 to 30, weight from 49kg to 85kg, and height from 158cm to 184cm. Eight of them go to the gym at least three times a week, two of them are novices. Each participant performed the above-listed exercises in 5 days. During the whole data collection phase, the temperature ranged from \SI{17.0}{\celsius} to \SI{27.5}{\celsius}, the relative humidity ranged from 45\% to 79\%(Data was from WetterKontor GmbH, measured by HMP45D). Since the triboelectric effect\cite{castle1997contact} will change the electron distribution on the body, we also considered the wearing of each subject: height of shoe sole, sole shoe material (PVC or rubber), clothes material (polyester or cotton). Table \ref{Subjects_Configuration} shows the configuration of the participants' wearing. This configuration aims to demonstrate the robustness of the $HBC$ based sensing modality. For all wearable devices, portability plays an important role, so we tested our prototype with three on-body deployments: on the calf, on the wrist, in the pocket.  

\begin{figure}[hbt]
\centering
\includegraphics[width=0.99\linewidth,height=8.5cm]{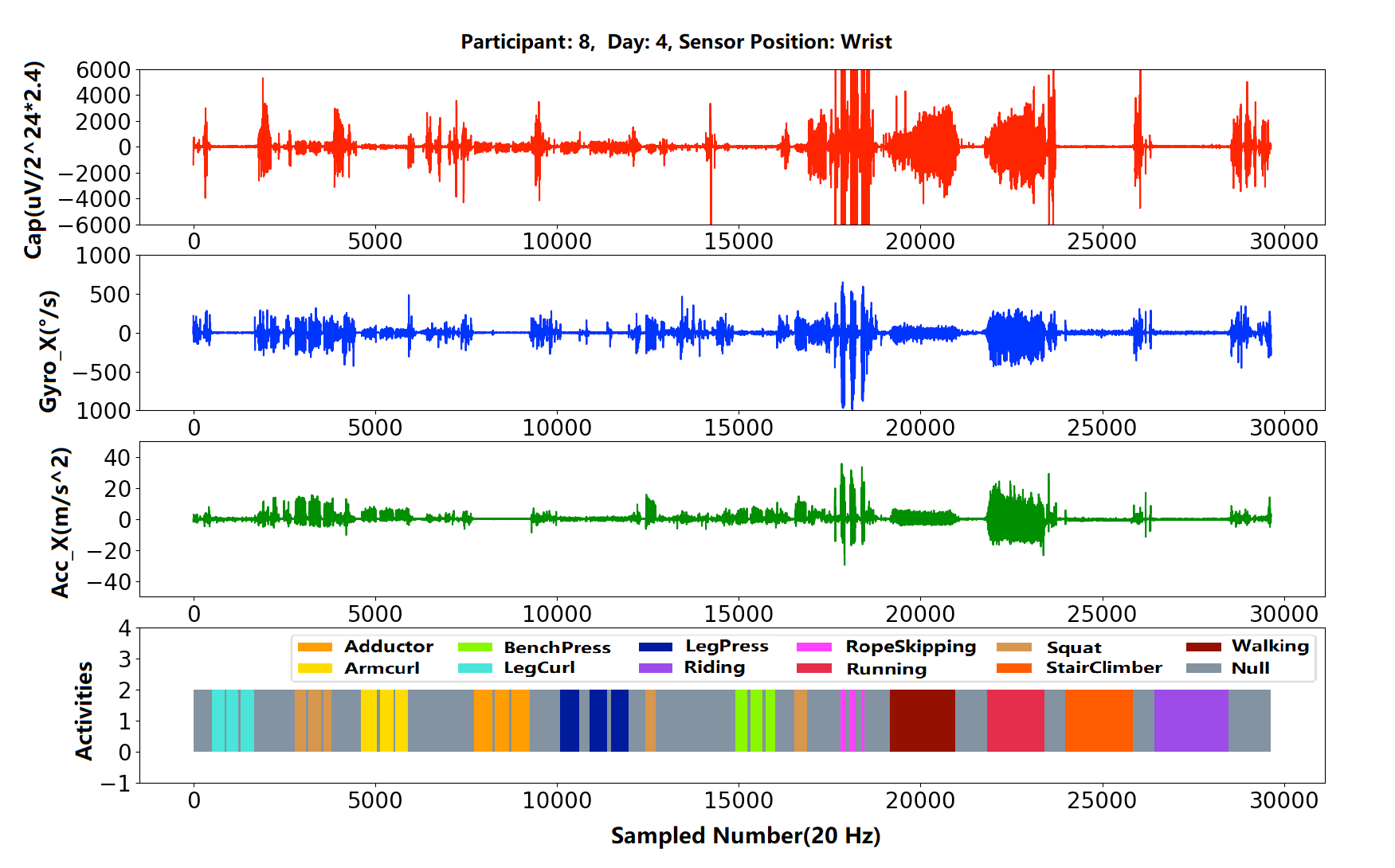}
\caption{Example of one session's initial measurement unit signal, capacitance caused potential variation signal and the exercise labels, including a null class.}
\label{Session}
\end{figure}

\begin{table*}[htbp]
\centering
\begin{threeparttable}
\caption{Participants' Configuration Across Days}
\label{Subjects_Configuration}
\begin{tabular}{ p{2.6cm} p{1.8cm} p{1.8cm} p{1.8cm} p{1.8cm} p{1.8cm}}
\toprule
& First Day & Second Day & Third Day & Fourth Day & Fifth Day \\ 
\midrule
Clothes Material & cotton & cotton & polyester & cotton & cotton\\

shoe sole height\tnote{a} & M & M & M & S & M  \\ 

shoe sole material & PVC & PVC & PVC & PVC & rubber  \\ 
\bottomrule
\end{tabular}
\begin{tablenotes}
    \item[a] For each user, M and S denote the height of shoe sole, with M meaning the height of the pair of shoes the user is used to wear, while S denotes the different height of another shoe belonging to the subject. Different users had different shoe heights of M and S. 
\end{tablenotes}
\end{threeparttable}
\end{table*}

We collected the data with a frequency of 20 Hz and developed a framework to get data from the Bluetooth of the prototype. The data then got stored and displayed locally on a computer. During the gym training, a second person labeled all the exercises with the help of the framework's user interface by simply choosing and clicking. Overall, we got five sessions of a whole day's training for each volunteer with each sensor position. Each session involved around one hours' IMU data and one hour's $HBC$ related body potential data. Within each session, there were three segments of each exercise(Adductor, Armcurl, Benchpress, Legcurl, Legpress, Ropeskipping, Squat), and one segment of each exercise(Riding, Running, Stairsclimber, Walking). Figure \ref{Session} depicts one whole session of IMU and $HBC$ related potential data from the eighth volunteer in the fourth training day with the prototype worn on the wrist. To be noticed, we did another two kinds of Squat in each session, as depicted with the color peru in Figure \ref{Session}. Each volunteer did the Squat on three ground types: concrete, wood, rubber. The purpose was to verify the robustness of $HBC$ based sensing modality related to different gym space.

\begin{figure}
\centering
\includegraphics[width=0.6\columnwidth,height=6.0cm]{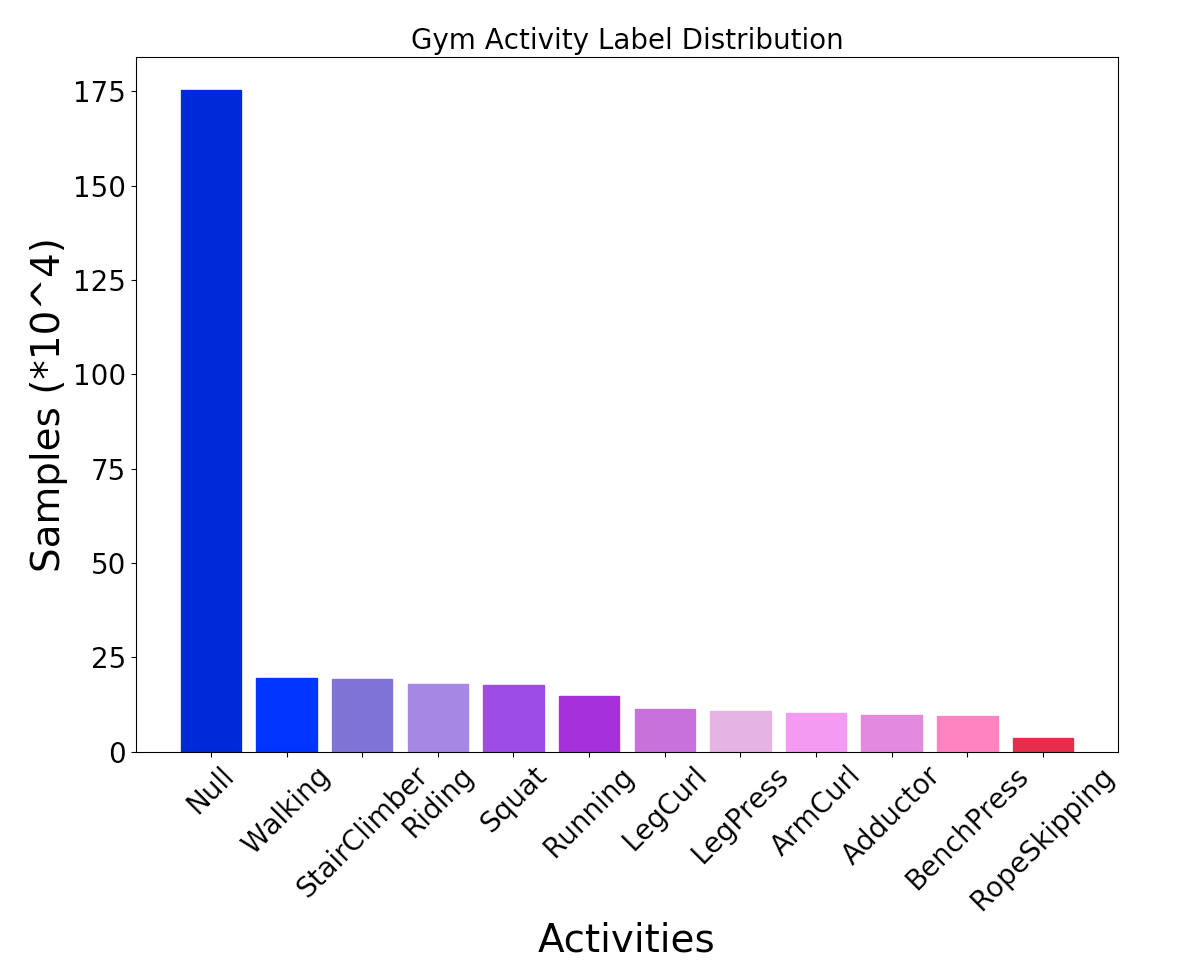}
\caption{Unbalanced class distribution in the whole sessions}
\label{DataUnbal}
\end{figure}

\subsubsection{Classification Exploration}
To classify the whole session's activity, we defined another class named "Null", indicating the process when the volunteer was not busy with the above-listed exercises. This class was depicted with color grey in Figure \ref{Session}. Figure \ref{DataUnbal} shows the sample distribution in one session after adding this class. Thus balancing the data was an indispensable step before classification.   
Researchers have presented a rich set of algorithms for activity recognition tasks\cite{zeng2014convolutional,jiang2015human,rana2015application,marinho2016new, ronao2016human,murad2017deep}. Both classic machine learning approaches, like SVM, KNN, and deep neural network approaches, like CNN, LSTM, were represented and showed outperforming recognition performance than others in their specific activity recognition tasks. In the following subsections, we use keras\cite{chollet2015keras} with the tensorflow\cite{tensorflow2015-whitepaper} backend to train our classification model. The results of algorithms from both approaches are presented.

\begin{figure*}[t!]
\centering
\includegraphics[width=0.6\columnwidth,height=17.0cm]{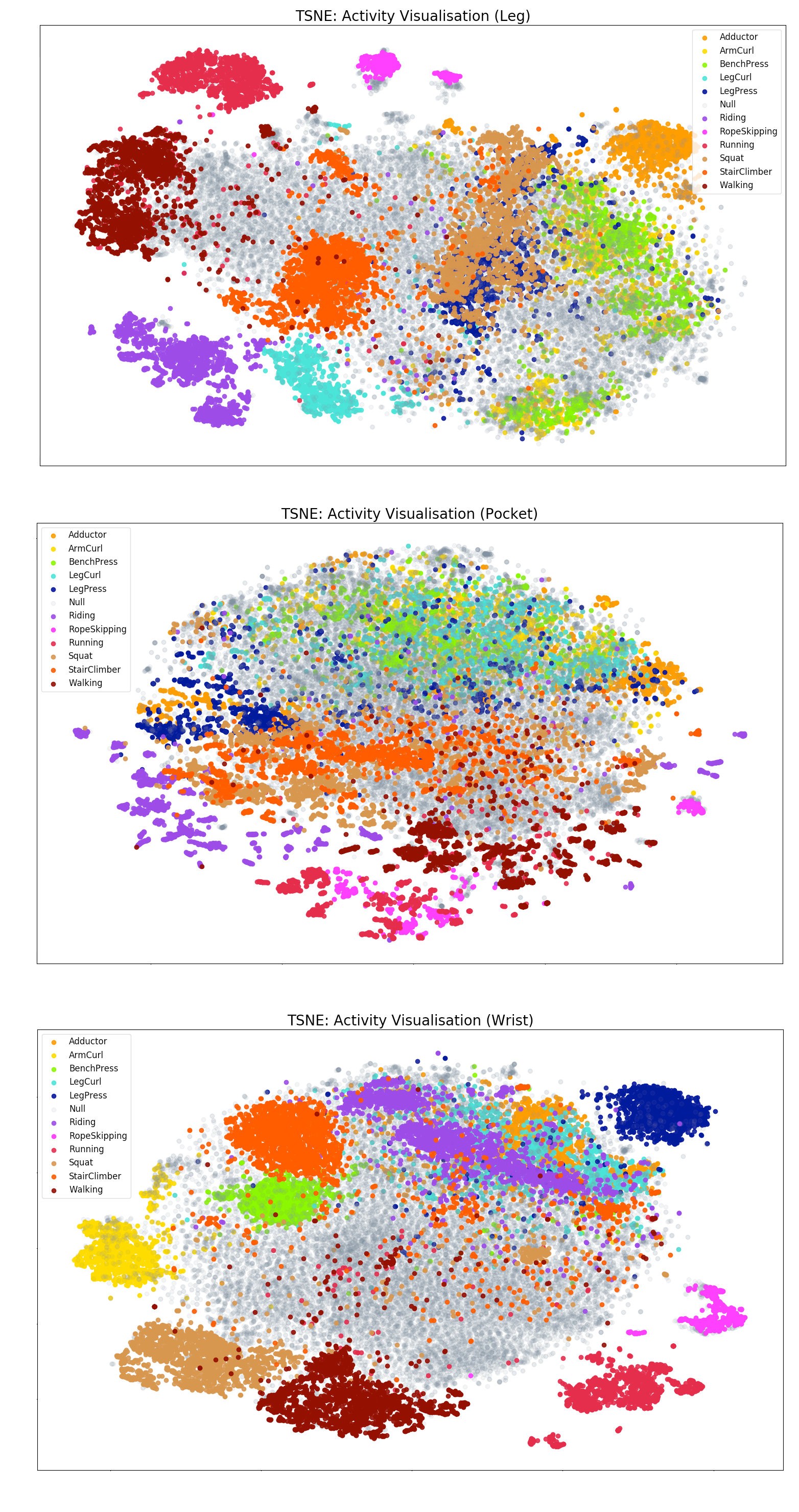}
\caption{t-SNE plots of the twelve exercises' feature distribution with sensing unit deployed on the leg, in the pocket and on the wrist, respectively}

\medskip
\begin{minipage}{0.95\columnwidth} 
{\footnotesize Note: as a nonlinear dimensionality reduction technique, t-SNE plots the visual clusters, thus the axes have no units \cite{maaten2008visualizing}\par}
\end{minipage}
\label{TSNE_Class}
\end{figure*}

For the beginning, we used the sliding window approach to get instances. The size of the window also plays a role for segmentation and classification\cite{laguna2011dynamic,banos2014window}, which is mainly a trade-off between the recognition speed and accuracy. We tried window size with 2, 4, and 6 seconds. The 4 seconds size performs best in our case, meaning each instance owes 80 readings. The overlapping size is 2 seconds.

Since we have ten volunteers, we run the models with ten-fold cross-validation with one volunteer out, aiming to guarantee an expected level of fit against unregistered users. The classification result will be represented by F-score and accuracy. The contribution of each sensing modality(IMU and $HBC$) is also exploited.


\begin{enumerate}[label=(\Alph*)]
\item Random Forest
\end{enumerate}

As we described above, classical approaches solving the problem of classifying sequences of sensor data involve two steps: handcraft the features from the time series data with the sliding window and feed the models with the features to train the models. In our work, we evaluated a diverse set of machine learning algorithms, including k-Nearest Neighbors, Support Vector Machine, Gradient Boosting Machine, Random Forest, etc. The Random Forest provided the best performance. 
At the very beginning of each tested model, we balanced the labels with the method of SMOTE\cite{chawla2002smote}.

\begin{figure}
\centering
\includegraphics[width=0.7\columnwidth,height=5.5cm]{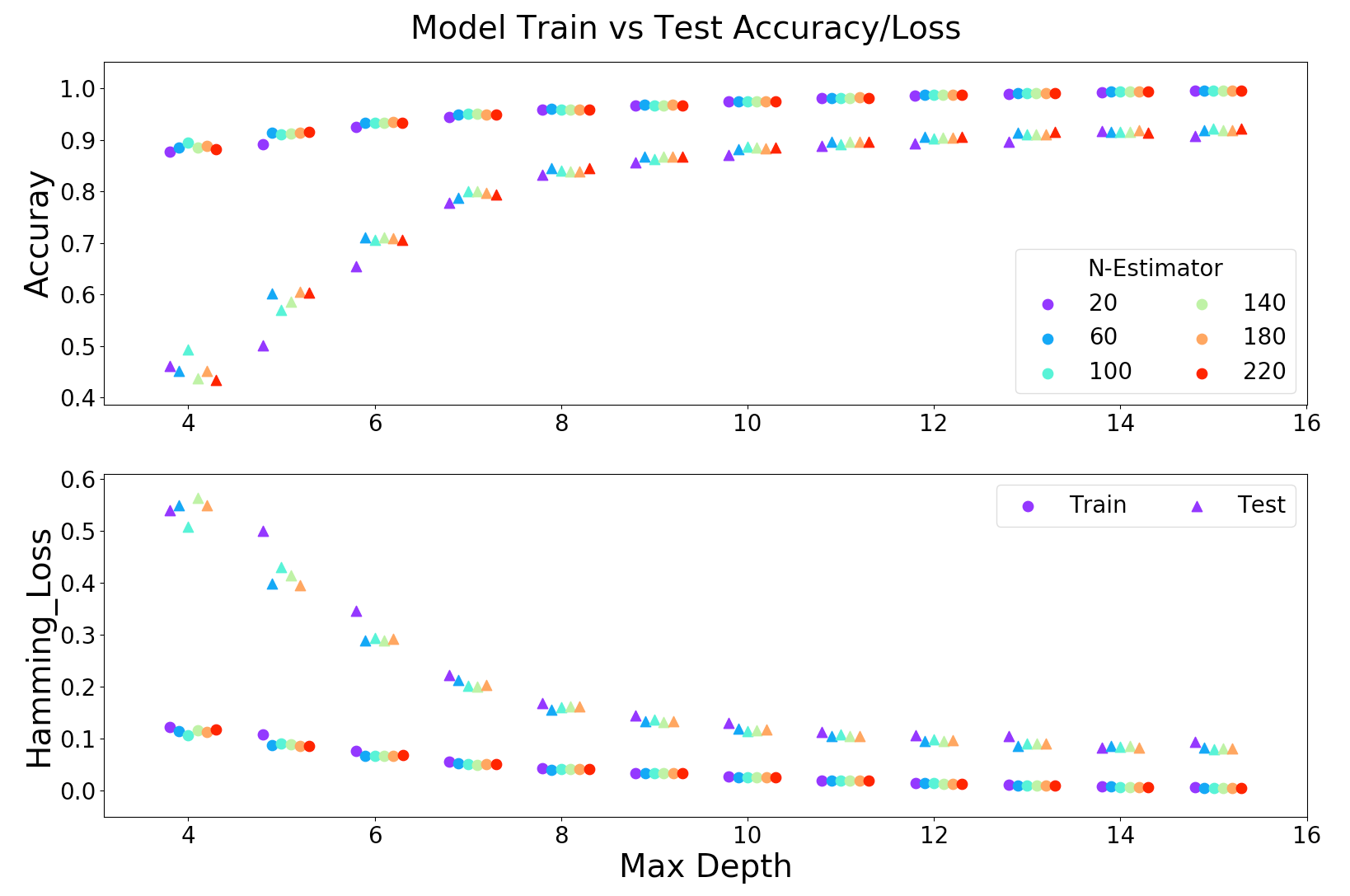}
\caption{Accuracy and loss of one fold with different hyper-parameters}
\label{RF_Acc_Loss}
\end{figure}

The performance of a machine learning model relies on the quality of the feature extraction result\cite{kuhn2019feature}. Within each window, we have 80$x$7 raw data, representing data from the three axes accelerometer, three axes gyroscope, and the body potential. We summarized the following mathematical features in both time and frequency domains:
\begin{itemize}
\item mean, standard deviation, max, min
\item mad: Median absolute deviation 
\item SMA: Signal magnitude area
\item energy: Sum of the squares divided by the number of values. 
\item IQR: Interquartile range 
\item entropy: Signal entropy
\item arCoeff: Autoregressive coefficients with Burg order equal to 4
\item correlation: correlation coefficient between two signals
\item maxInds: index of the frequency component with the largest magnitude
\item skewness: skewness of the frequency domain signal 
\item kurtosis: kurtosis of the frequency domain signal 
\item bands energy: Energy of a frequency interval within the 64 bins of the FFT of each window.
\end{itemize}
We deduced 36 variations from the original seven features:
\begin{itemize}
\item t$/$f\_Acc$/$Gyro\_XYZ
\item t$/$f\_Cap
\item t$/$f\_Acc\_Jerk$/$Gyro\_Jerk\_XYZ
\item t$/$f\_Cap\_Jerk
\item t$/$f\_Acc$/$Gyro\_Mag
\item t$/$f\_Acc\_Jerk$/$Gyro\_Jerk\_Mag
\end{itemize}
Where t\_f means the time and spectral domain, Mag means the magnitude of the $XYZ$ vector. In total, we utilized 615 features, so the input sample is now an array of 1$x$615 per window. All the features are then normalized to 0$-$1 with a threshold in both directions. The dataset is available in public\cite{bian2021github} for further studies by interest.

To check how the exercises' features distinct from each other, we visualized the features with the method of t-SNE\cite{maaten2008visualizing,van2014accelerating}, which is a commonly used dimensionality reduction technique. Thus for each window, we got an array of 1$x$2. Figure \ref{TSNE_Class} depicted the exercises distribution with two-dimensionality. It shows that after the feature engineering, a classification of the gym exercises was feasible.

\begin{figure}[!t]
\centering
\includegraphics[width=1.0\textwidth,height=12.5cm]{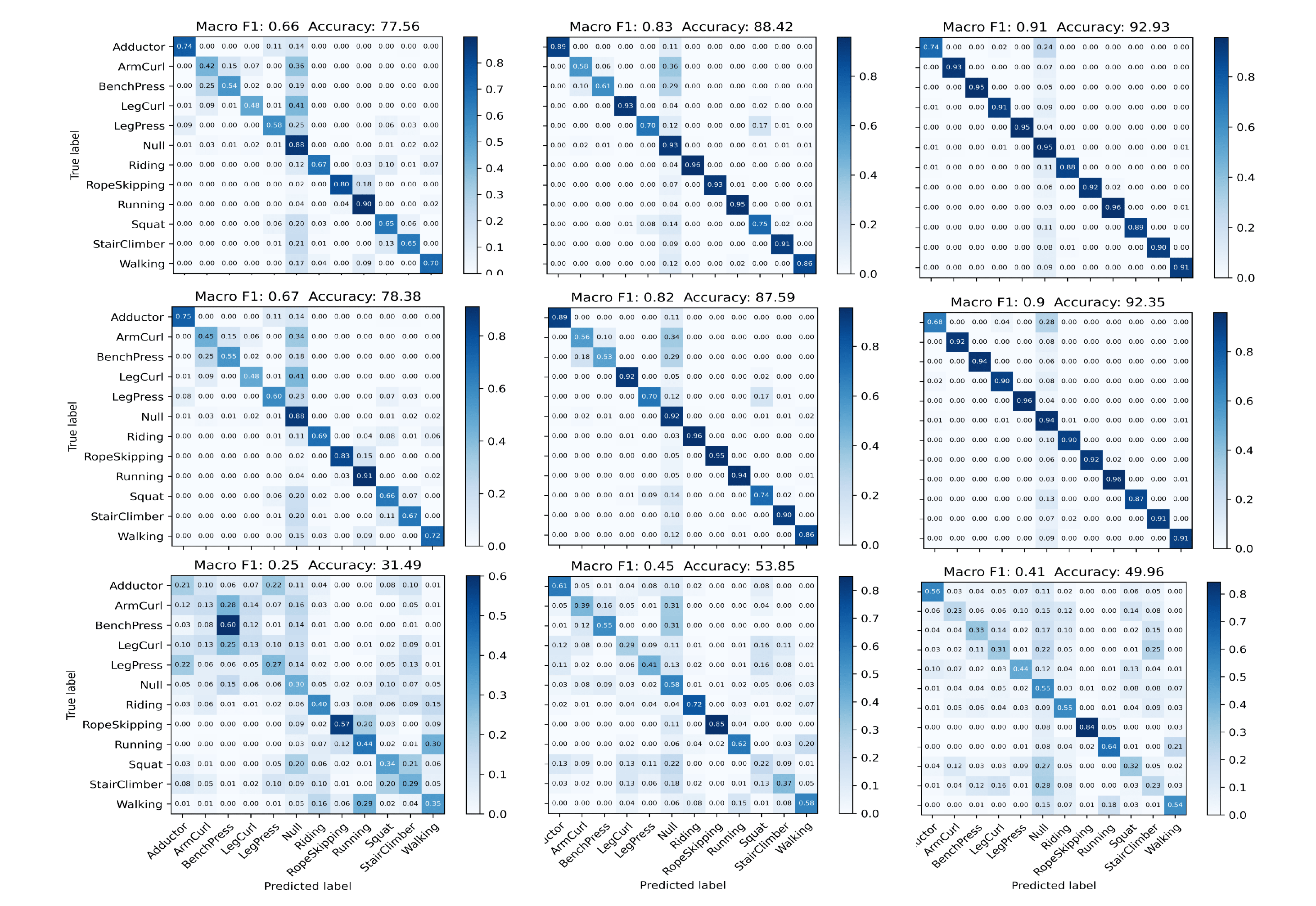}
\caption{Confusion matrix of classification result from the random forest model(From left column to right: prototype in the pocket, on the calf, on the wrist; From third row to first: signal source from $HBC$ sensing, IMU, and the combination)}
\label{RF_CM}
\end{figure}


Hyperparameter tuning relies more on experimental results than theory. Thus we determined the optimal settings by trying different combinations and evaluating the performance of each model. In our random forest model, we mainly tuned two hyper-parameters: n\_estimators, namely the number of trees, and max\_depth, indicating the tree's maximum depth. We used the grid searching method to find the best combination. Figure \ref{RF_Acc_Loss} shows the classification result of one fold with different settings. We used hamming loss, the fraction of labels that are incorrectly predicted, to check the over-fitting. As shown in this figure, the tree number does not have much influence on our model accuracy, unlike the maximum depth. We finally set the n\_estimators to 100 and max\_depth to 15 for the whole classification task.

\begin{figure}
\begin{minipage}[t]{0.5\linewidth}
\centering
\includegraphics[width=0.99\textwidth,height=5.5cm]{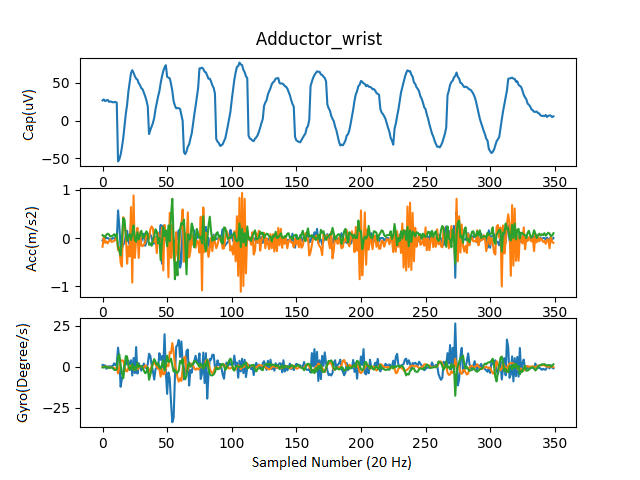}
\caption{An instance of Adductor's raw signal with prototype on the wrist}
\label{Cheng_Adductor_wrist}
\end{minipage}
\quad
\begin{minipage}[t]{0.5\linewidth}
\centering
\includegraphics[width=0.99\textwidth,height=5.5cm]{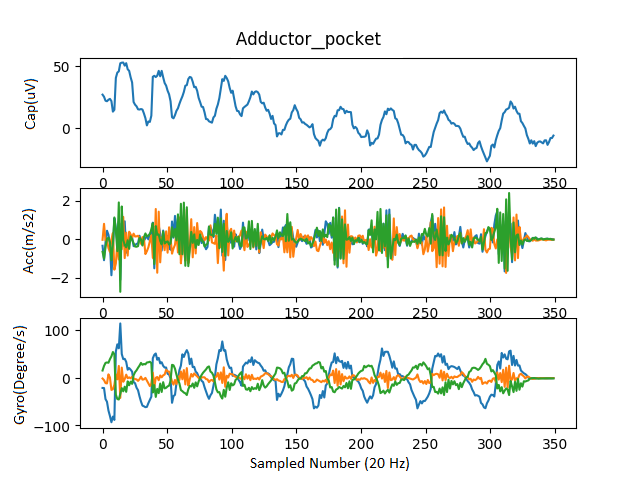}
\caption{An instance of Adductor's raw signal with prototype in the pocket}
\label{Cheng_Adductor_pocket}
\end{minipage}
\end{figure}

Figure \ref{RF_CM} depicts the classification result from our random forest model. Overall, we got the macro F-score of gym exercises recognition up to 0.66, 0.83, 0.91 with the combination sensing method when the prototype was in the pocket, on the calf, and wrist, respectively. The result from $IMU$ alone shows that $IMU$ alone is sufficient enough for the gym workouts recognition. The reason behind this is that most of the picked workouts have the arm movement involved in each repetition, which is much easier to be captured by $IMU$ sensors since $IMU$ supplies at least six signal sources(3-axes accelerometer, 3-axes gyroscope) also the combination of them(magnitude). $HBC$ could only supply help in the workouts that $IMU$ can not track, like the Adductor when the prototype was worn on the wrist(in a static state), and only the legs were moving apart and close. A raw signal instance is depicted in Figure \ref{Cheng_Adductor_wrist}. From the confusion matrix, the adductor classification accuracy is improved from 0.68 to 0.74 when combining the two signal sources. Similar improvements can also be observed in workouts like ArmCurl(0.56 to 0.58) and BenchPress(0.53 to 0.61) when the prototype was worn on the calf. The classification result of $HBC$ signal when the prototype was located in the pocket seems even worse, possible reason behind is that the sensing unit was not directly connected to the body with an electrode, the coupling between the body and sensing unit is week. The friction from the cloth also results in signal noises. A raw signal instance is depicted in Figure \ref{Cheng_Adductor_pocket} when the prototype was in the pocket. Overall, we could conclude that the $HBC$ doesn't supply significant help to improve the $IMU$-alone derived workouts classification result for the gym workouts. It helps only with certain workouts that the $IMU$ is incapable of monitoring the moving action, like adductor.

\begin{enumerate}[label=(\Alph*)]\setcounter{enumi}{1}
\item Deep Neural Network
\end{enumerate}

As described in our leg-alone exercise classification part, deep learning methods such as convolutional neural networks and recurrent neural networks have shown significant and even achieve state-of-the-art results by automatically learning features from the raw sensor data\cite{wang2019deep} in some works of human activity recognition\cite{zeng2014convolutional,jiang2015human, cho2018divide, murad2017deep}. With this approach, we feed the model with streams of raw data and predict the associated exercises with the possibilities. The deep neural network can extract high-level representation in deep layers while the conventional methods utilize only shallow features. 
Here we again use the same approaches as described in previous section to recognise the twelve gym activities including the null class. A ten-fold cross-validation(leave-one-user-out) was performed to address the overfitting or selection bias problem.

The classification result from this DeepConvLSTM and Resnet21 models are listed in Table \ref{Deep_Resualt}. The contribution from $HBC$ based sensing modality to the exercises classification from deep models is different with the its contribution from the random forest model. For example, the DeepConvLSTM shows that the $HBC$ improves the classification with a significant f-score change. However this is not the case in Resnet21 result when getting data from leg and wrist deployment. In any cases, the random forest model classifies much better in most source and sensor position cases. In essence, the random forest model and the deep neural network model are different types of feature learning approaches. One in common is that they have different use cases with the best performance\cite{liu2013comparison, were2015comparative, rodriguez2015machine}. The performance of classifiers is inherently data-dependent. To get the best one, the best way is to try them all. In another paper\cite{sizhen2023exploring}, we also explored the edge performance (power efficiency, throughput, latency, etc.) when running the deep model to recognize the workouts on different edge platforms.

\begin{table*}[htbp]
\centering
\begin{threeparttable}
\caption{Classification of deep models: F-score/Accuracy}
\label{Deep_Resualt}
\begin{tabular}{p{2.4cm} p{2.0cm} p{2.4cm} p{2.4cm} p{2.6cm}}
\toprule
Model & Position & $HBC$ & IMU & $HBC$+IMU \\ 
\midrule
&Pocket & 0.17 / 0.16 & 0.53 / 0.62 & 0.57 / 0.62 \\
DeepConvLSTM&Leg & 0.23 / 0.21 & 0.62 / 0.69 & 0.69 / 0.71 \\ 
&Wrist & 0.17 / 0.16 & 0.59 / 0.66 & 0.63 / 0.66\\

&Pocket & 0.15 / 0.17 & 0.55 / 0.60 & 0.60 / 0.63 \\
Resnet21 & Leg & 0.39 / 0.41 & 0.78 / 0.81 & 0.76 / 0.75 \\
&Wrist & 0.32 / 0.37 & 0.89 / 0.91 & 0.89 / 0.91\\ 
\bottomrule
\end{tabular}
\end{threeparttable}
\end{table*}


\subsubsection{Exercise Counting Exploration}
Since there is not too much contribution of $HBC$ signal to the general gym workout recognition, we wonder how the counting performed with different signal sources. The counting exploration was carried out directly on the original framework-splitted data, without first segmenting it using the classification pipeline. We did this in order to provide an upper bound to the possible counting performance. Once the classification result of different workouts is over 90\%, like the result from the wrist-worn prototype(showed in Figure \ref{RF_CM}, top-right), out counting approach can be reliably applied.

We have three sources of data being utilized for counting, accelerometer, gyroscope, and $HBC$. First, since during the gym workouts(mostly with machines), the attitude of our prototype does not have a single pattern(like the leg-exercises in labor where Z-axis of the accelerometer and Y-axis of gyroscope always provide best signal track for counting), we combined the three axes to one magnitude as the new track for counting.
Then, we used Fourier Transform and Inverse Fourier Transform to smooth the data, removing the undesired high frequencies. To be noticed, different parameters were given to the transforming process, since Running, Walking, Ropeskipping, and Riding have a higher frequency than the other exercises and also are easily recognized by the classification model.

Finally, we detected the peaks of the smoothed signal using the PeakUtils \cite{PeakUtils} python package. Two parameters were set to identify the peaks correctly. One is the threshold with relation to the highest value, another is the minimum distance between two peaks.

\begin{figure}[!t]
\centering
\includegraphics[width=0.6\columnwidth,height = 12cm]{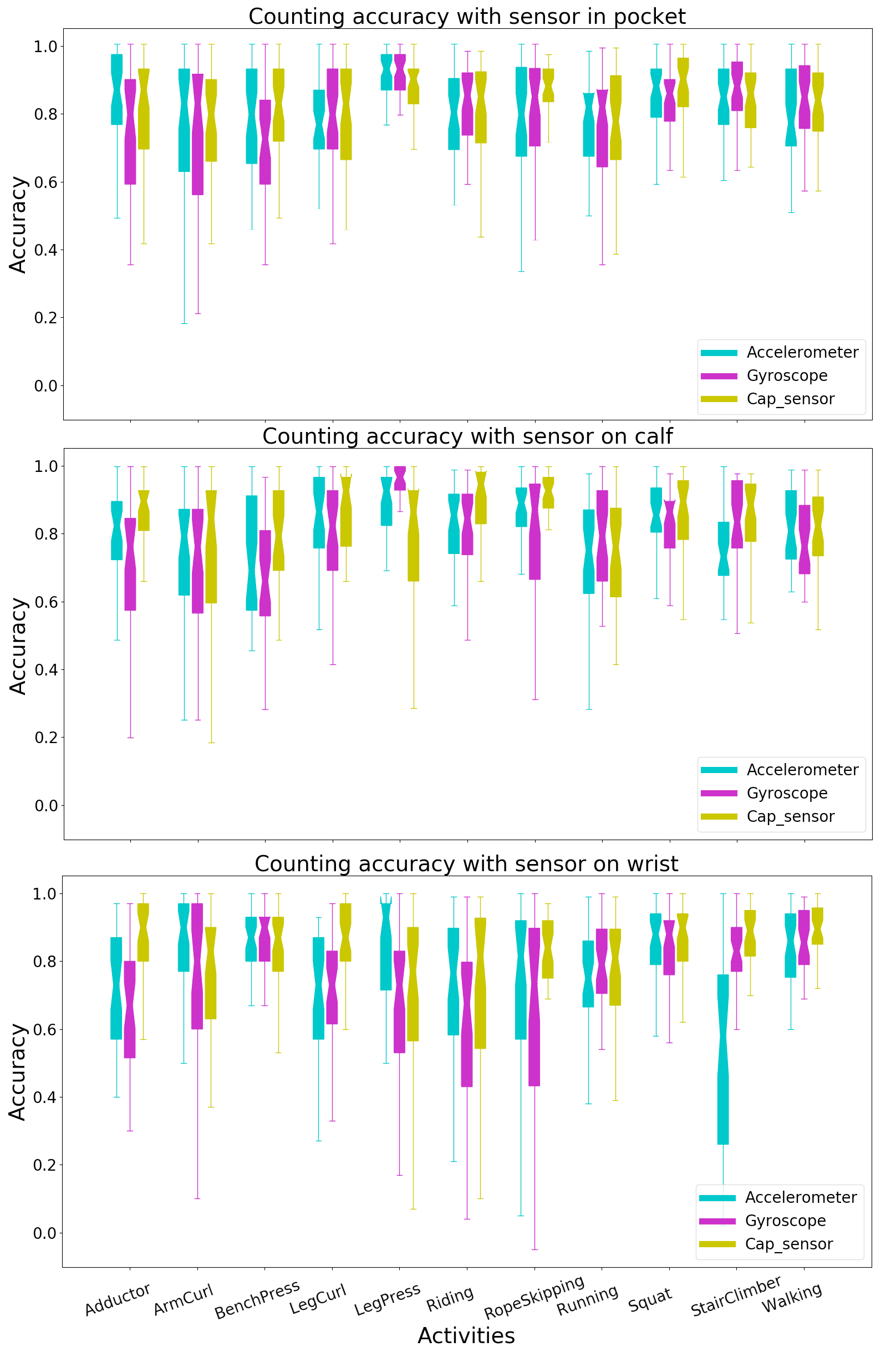}
\caption{Counting accuracy of each exercise from the three signal sources($HBC$, gyroscope, accelerometer) with the three prototype deployments(pocket, calf, wrist)}
\label{Counting_three}
\end{figure}

\begin{figure}[!t]
\centering
\includegraphics[width=0.6\columnwidth, height = 5cm]{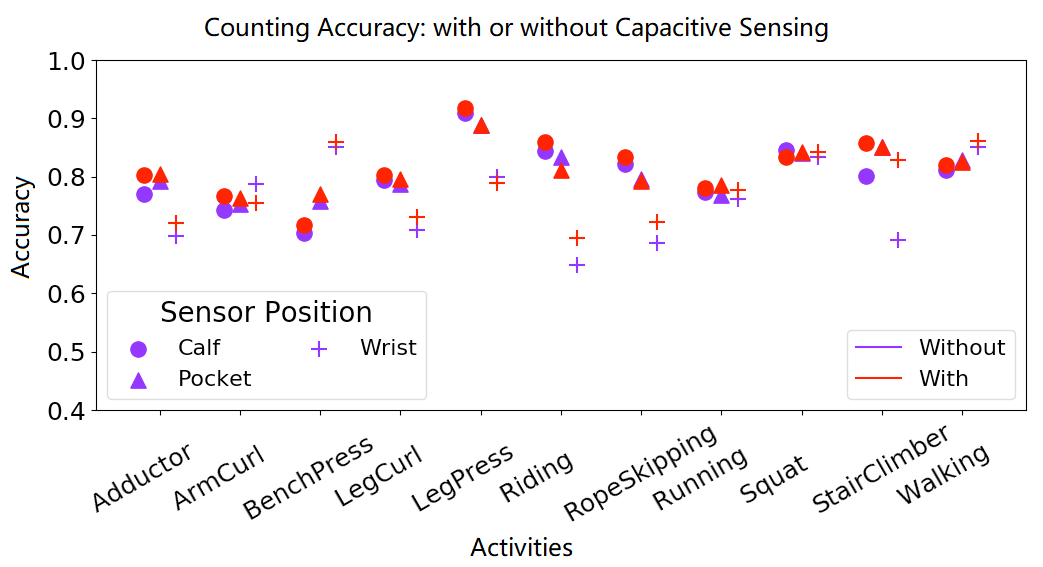}
\caption{Counting accuracy of each exercise with/without $HBC$ sensing modality}
\label{Counting_Cap}
\end{figure}

\begin{table*}[htbp]

\centering
\begin{threeparttable}
\caption{Counting with/without $HBC$ sensing modality}
\label{Counting_Cap_2}
\begin{tabular}{p{1.8cm} p{1.8cm} p{1.8cm} p{1.8cm} p{1.8cm} p{1.8cm}}
\toprule
Position & Acc & Gyro & $HBC$ & IMU & $HBC$ + IMU \\ 
\midrule
Pocket & 0.806 $\pm$0.159 & 0.797 $\pm$0.168 & \bfseries 0.811 $\pm$0.168 & 0.809 $\pm$0.152 & \bfseries 0.824 $\pm$0.142\\
Leg & 0.801 $\pm$0.158 & 0.788 $\pm$0.181 & \bfseries 0.820 $\pm$0.190 & 0.802 $\pm$0.160 & \bfseries 0.824 $\pm$0.144\\ 
Wrist & 0.752$\pm$0.223 & 0.739 $\pm$0.219 & \bfseries 0.800 $\pm$0.217 & 0.756 $\pm$0.190 & \bfseries 0.788 $\pm$0.171\\ 
\bottomrule
\end{tabular}
\end{threeparttable}
\end{table*}

Figure \ref{Counting_three} uses Boxplot and shows the counting result with signals from the accelerometer, gyroscope, and $HBC$ separately for each gym exercises. For leg exercises, the calf and pocket deployment ways gave higher counting accuracy, like LegPress, Riding; For arm exercises, the wrist deployment position outperformed in counting, like ArmCurl, BenchPress. In most cases, the average count accuracy could reach up to 80\% regardless of the prototype position, and the capacitance sensing performed mostly better than accelerometer or gyroscope for counting regarding the stability and accuracy, which was also represented in Table \ref{Counting_Cap_2} with the columns $Acc$, $Gyro$, $HBC$, where we summarized the counting accuracy regardless of exercise type. Among the three signal sources, the $HBC$ again provide the best counting accuracy. We then group the accelerometer and gyroscope together by averaging the calculated count as the counting result from $IMU$, and group the three by choosing the closest two and average it as the result of the combination($IMU$ and $HBC$), which is better than averaging the three directly, since it improves the reliability of the counting sources. The combination of accelerometer and gyroscope supplies higher counting accuracy. The combination of the three tracks shows better and more stable counting accuracy with sensing unit in the pocket and on the calf. While wearing the sensing unit on wrist, the $HBC$ always supplies best counting result regardless of the signal track combination. 
Figure \ref{Counting_Cap} shows the change of counting accuracy for each exercise when complementing the $HBC$ sensing modality to the traditional $IMU$ approach. Overall, the $HBC$ sensing has a positive effect on counting, which is also represented in Table \ref{Counting_Cap_2} with the columns $IMU$ and $IMU+HBC$ when grouping all the exercises.

\begin{figure}
\centering
\includegraphics[width=0.5\columnwidth,height=6.0cm]{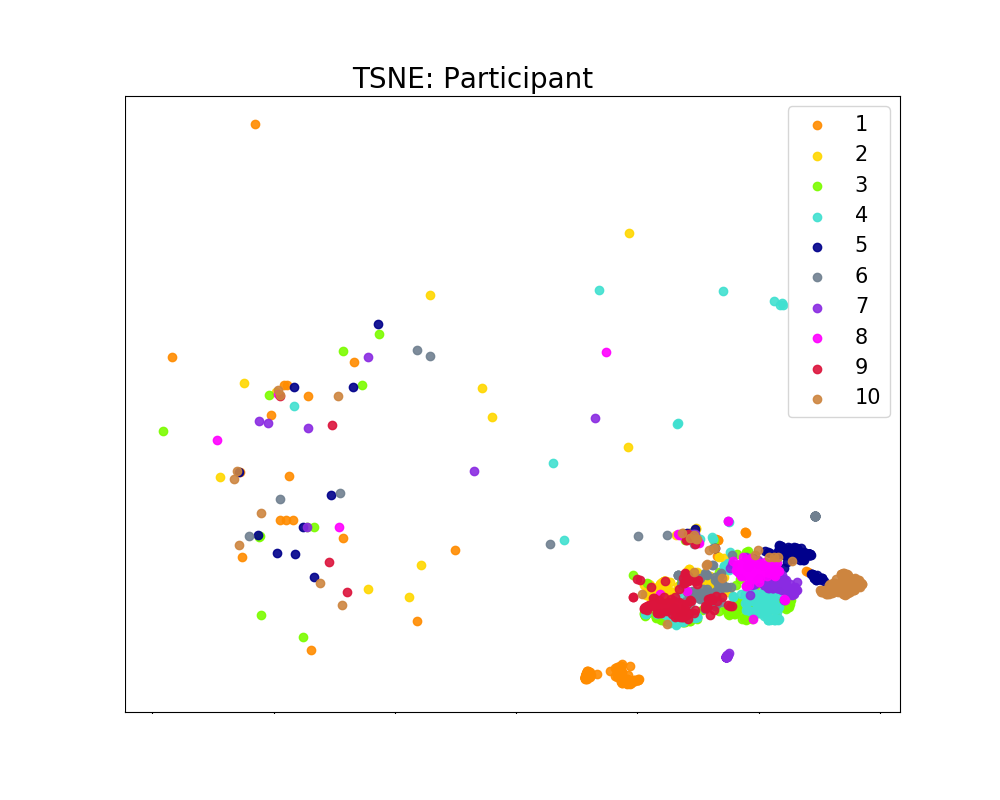}
\caption{t-SNE plot of the feature distribution from the ten participants(exercise of running with prototype on wrist)}
\label{Object}
\end{figure}

\subsubsection{Volunteers Recognition Exploration}
Besides the exploration of gym workouts classification and counting with our sensing prototype, we also exploited other potential recognition capabilities with the prototype. For example, we used the combination of both sensing modalities to recognize the ten volunteers. Figure \ref{Object} depicts the distribution of Running's features after the feature dimension reduction, which shows the possibility of volunteer classification. With the same random forest model described above, we got an F-score of 93\% for volunteers recognition from the workout of running with the prototype mounted on the wrist. Although the number of the classified volunteers are limited, this result is meaningful in the practical scenario, especially for a family shared gym tracking device. 

\begin{figure}[!t]
\centering
\includegraphics[width=0.7\columnwidth,height=12.0cm]{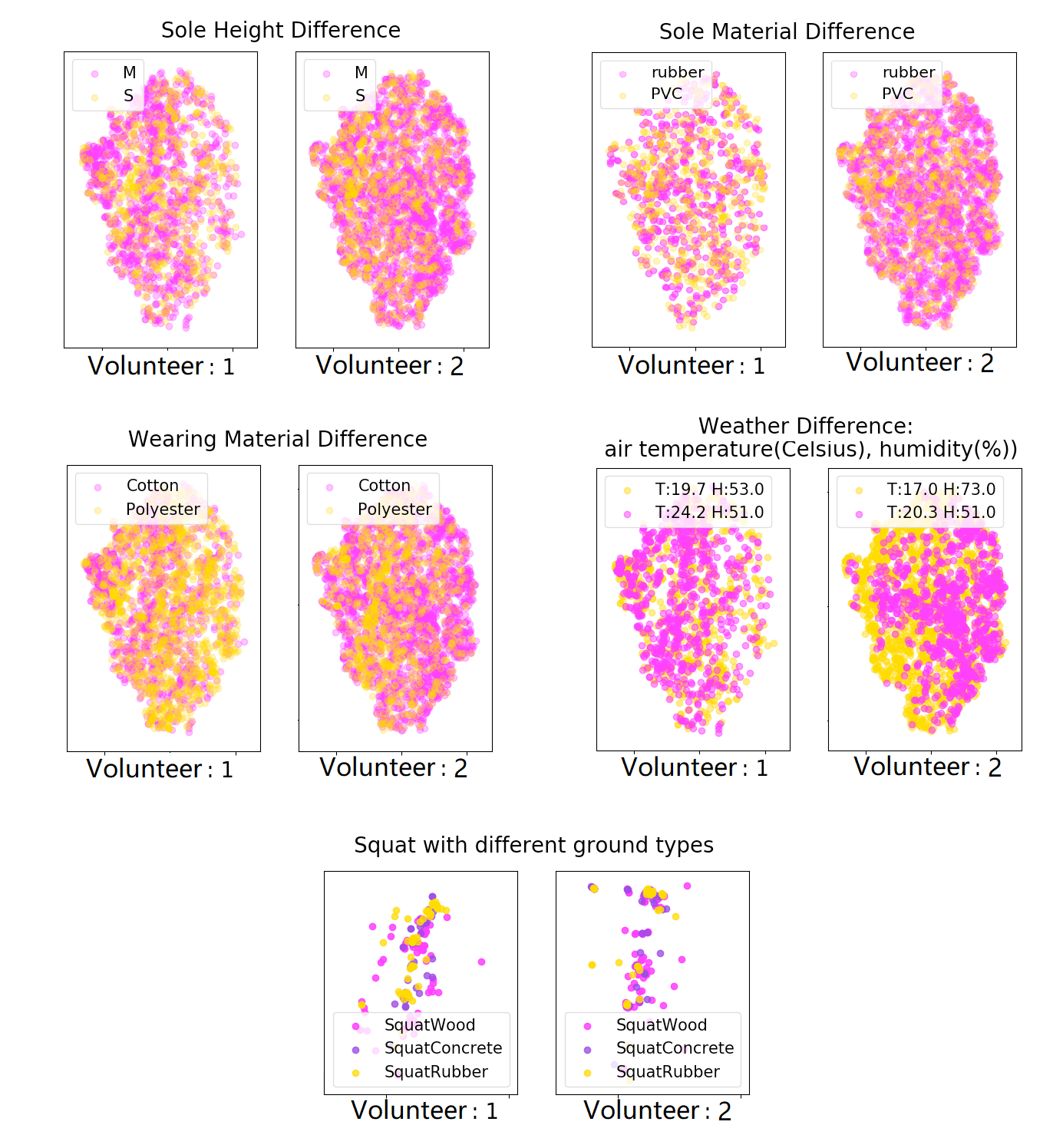}
\caption{t-SNE: Factors that may influence the robustness of $HBC$ based sensing modality}
\label{robustness}
\end{figure}

\subsubsection{Factors that impact the variation of $HBC$}
Finally, we analyzed the potential influence factors to the $HBC$ based sensing modality. It is clear that the wearing, especially the type and height of the sole, will change our body capacitance\cite{jonassen1998human}. Since our prototype does not measure the body capacitance directly, instead, it continuously measures the body potential's variation caused by the variation of body capacitance, we wonder how those factors influence the sensor reading with our prototype.
We inspected five possible factors, Sole Height(M or S), Sole Material(PVC or rubber), Wearing(cotton or polyester), Weather Condition, Ground Type(wood, concrete, or rubber). The configuration of wearing and the shoe is listed in Table \ref{Subjects_Configuration}. The weather condition was measured by HMP45D, offering the air temperature and the relative humidity. The ground type was verified by doing Squat on three ground types. 
To avoid differences from objects, we studied the five factors within a single volunteer. Since data from a single volunteer is deficiency for an adoption of feature classification method, we only use t-SNE here to roughly describe the feature distribution considering the above-listed five categories, as Figure \ref{robustness}(from two objects) depicted. Compared to Figure \ref{TSNE_Class} and Figure \ref{Object}, the t-SNE result of the factors does not show a separable distinction. Although the weather condition shows light separation on object 2's features, it does not give the same result to the other subjects. This unnoticeable 
feature distinction applies to all volunteers, thus we can say that the proposed $HBC$ based sensing modality is robust to the above listed potential factors, meaning that the sensing modality can be used for gym applications(repetition counting at least) regardless of wearing, weather condition, as well as the different gym studios.

\section{Collaborative Activity Exploration}\label{Section_4}

In Section \ref{Section_3}, we explored the contribution of $HBC$ motion-sensing modality to the recognition and repetition counting of leg-exercises and gym-workouts. The results show that this novel sensing approach could improve the $IMU$-based classification of leg-only exercises significantly, and of general gym workouts with a slight rate of 1\% to 3\%(with the sensing unit attached to the body), and produces the best counting accuracy with a significant advantage over the $IMU$-based repetition counting. As we described in Section \ref{Section_2}, $HBC$ is a parameter that is sensitive to both body motion, but also the environmental variation\cite{bian2019wrist}, like the intrusion of other bodies. This section utilized $HBC$'s characteristic of environmental sensitivity to explore the collaborative activity recognition.

\subsection{Experiment Setup}

\begin{figure*}[t!]
    \centering
    \begin{subfigure}[t]{0.5\textwidth}
        \centering
        \includegraphics[width=0.8\textwidth,height=5.0cm]{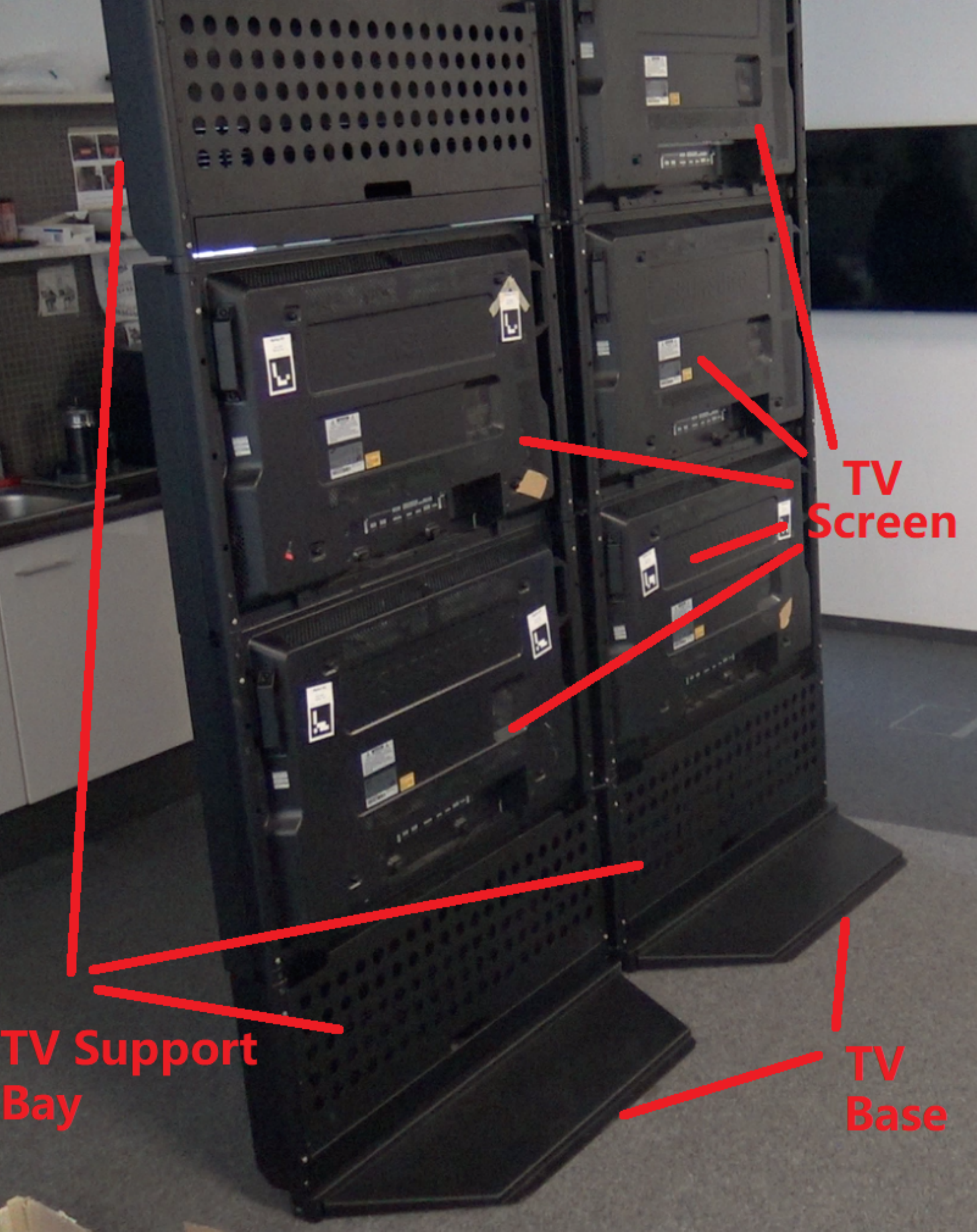}
        \caption{2.44m High TV-Wall}
        \label{TV_Wall}
    \end{subfigure}%
    ~ 
    \begin{subfigure}[t]{0.5\textwidth}
        \centering
        \includegraphics[width=0.8\textwidth,height=5.0cm]{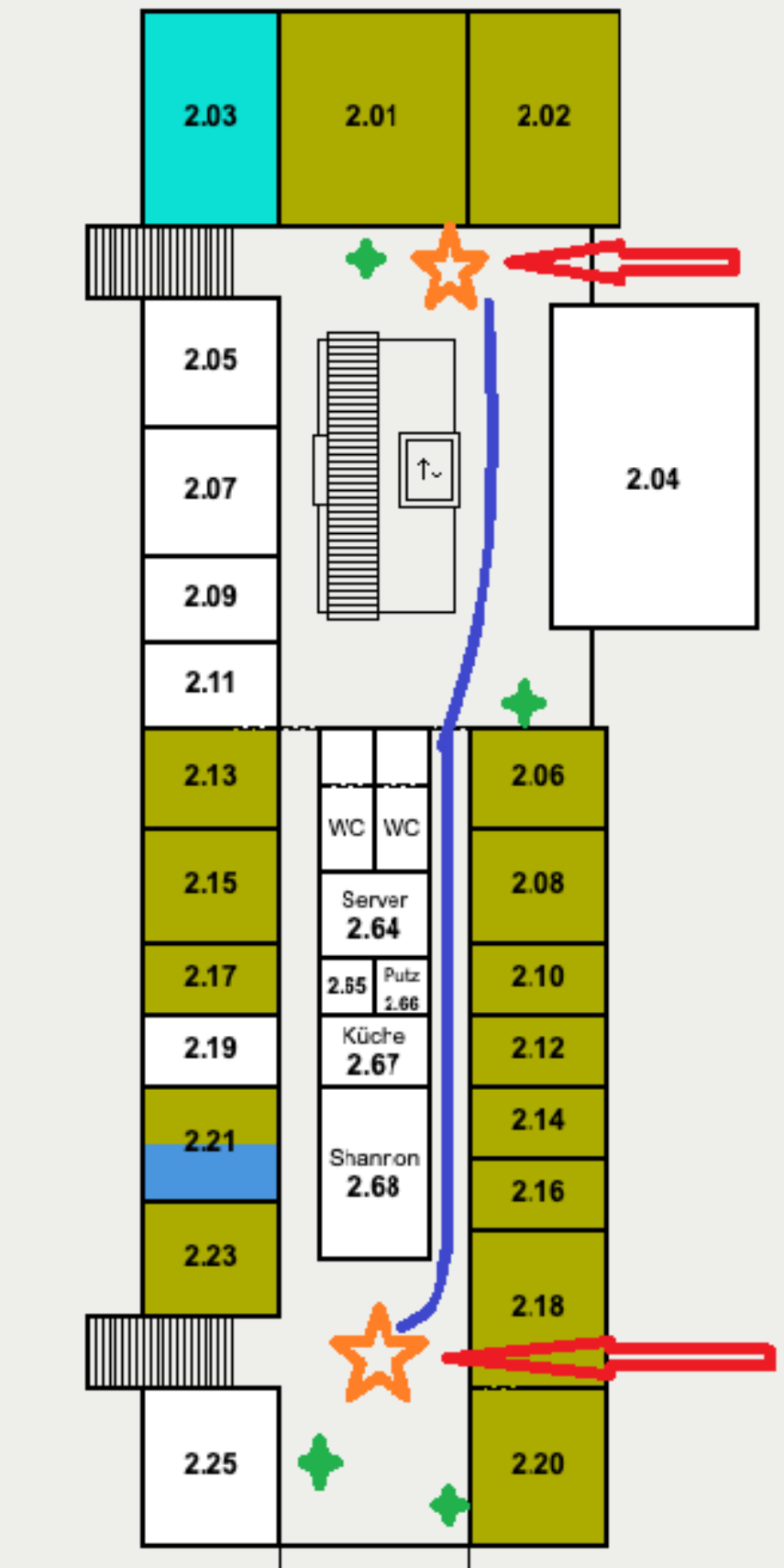}
        \caption{Collaborative activity's map}
        \label{Building}
    \end{subfigure}
    \\
    \quad
    \\
    \quad
    \\
    \begin{subfigure}[t]{0.5\textwidth}
        \centering
        \includegraphics[width=0.8\textwidth,height=5.0cm]{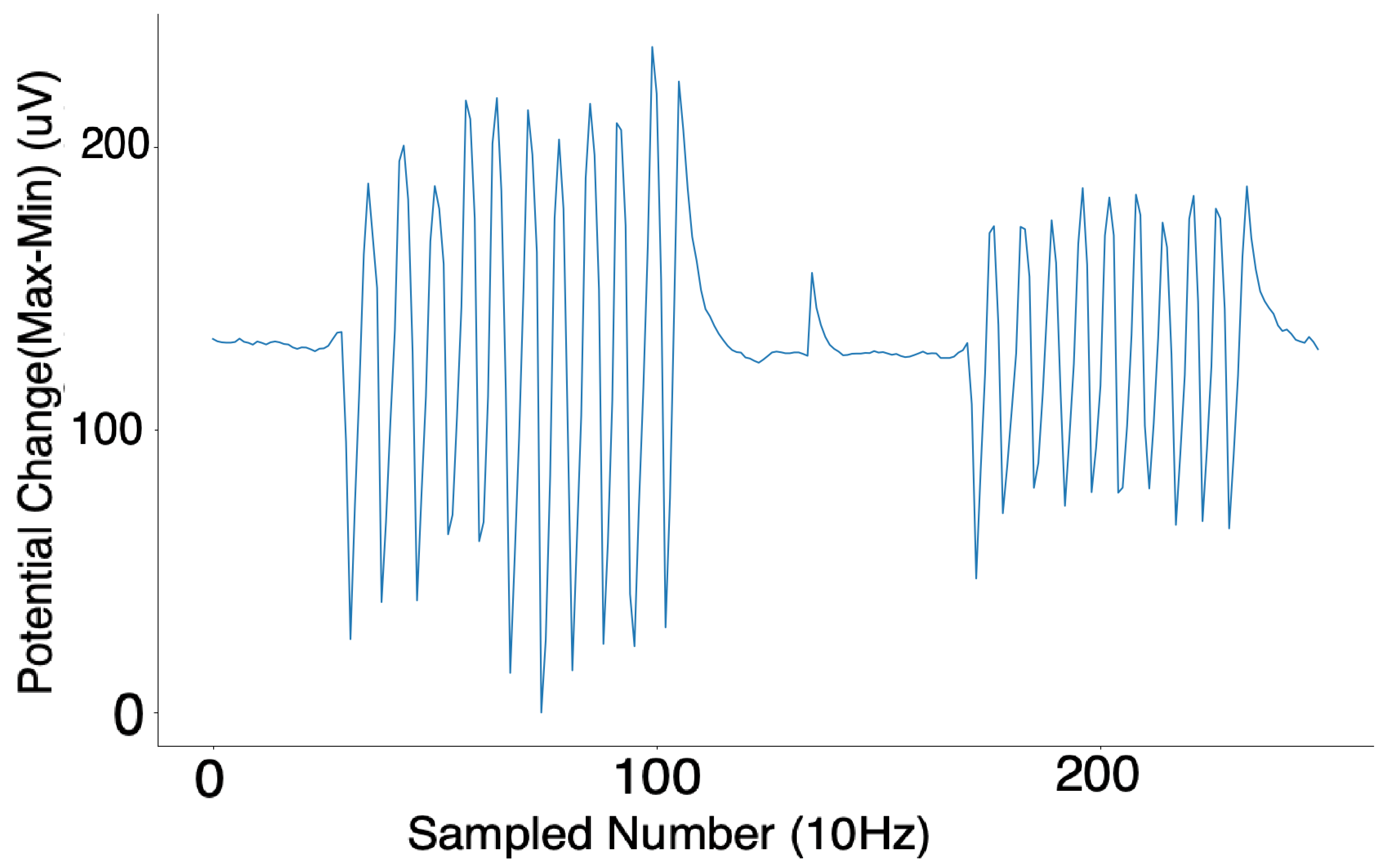}
        \caption{$HBC$ related signal of walking-alone and walking while touching another worker's hand}
        \label{Walk_Defference}
    \end{subfigure}%
    ~ 
    \begin{subfigure}[t]{0.5\textwidth}
        \centering
        \includegraphics[width=0.8\textwidth,height=5.0cm]{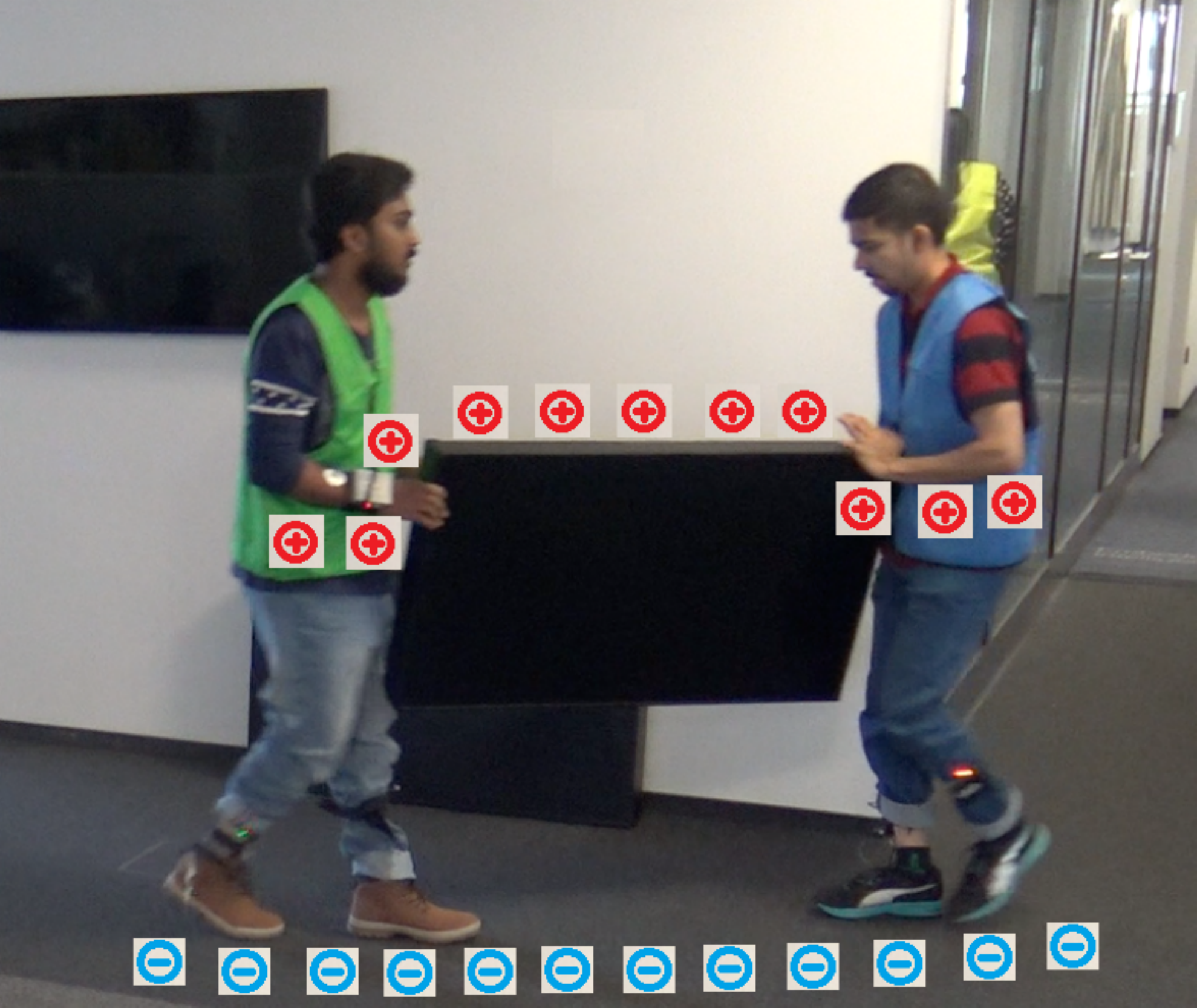}
        \caption{Electrically connection between multi-workers in a collaborative work}
        \label{Taking_TV}
    \end{subfigure}
    ~ 
    \\
    \quad
    \\
    \caption{Collaborative Work and the Sensing Modality}
    \label{HBC_Sensing}
\end{figure*}

To evaluate body capacitance-based sensing in group activity recognition, we planed a collaborative physical work including both independent and joint activities of each worker, building a TV-Wall. Twelve participants(ten male, two female) were divided into four groups, each of them carried some large TV screens from the storeroom to a task operating spot, assemble and disassemble a high TV-Wall, and carried them back. Each group performed this physical task 4 times in 4 days, where each time the task lasted around one hour. As Figure \ref{TV_Wall} presents, the 2.44 m TV-Wall is composed of 3 screen support bays weighting 10.3 kg for each, 2 TV bases weighing 22.1 kg for each and 5 TV screens weighing 23.2 kg for each. For the lighter ones, the participants could carry them alone. For the others, two participants took and carried them jointly. Figure \ref{Building} depicts the map where the activities were performed. The orange signs and red arrows indicate the original and operational location of those heavy metal. Participants carried the heavy from the top orange sign spot to the lower orange sign spot following the blue line, which was mainly a corridor. The route was around 36 meters. The green spots in the figure were the locations where cameras were placed. We used four cameras to record the whole working process to provide the ground truth, and every participant knew and agreed to the presence of cameras.   

In the experiment, each participant wore one prototype on one of their wrists and did the task naturally without any instruction. Finally, we got 39 sessions of valid data altogether; each of it contains around 1 hour's motion signals from both body capacitance and accelerometer. Since the gyroscope data was not enabled in the firmware in the first several sessions by misoperation, to keep the consistency of all sessions, we only consider accelerometer data in this study, which does not affect the demonstration of contribution from the capacitance sensor. We also placed an accelerometer on the calf of each participant, and it showed that by knowing the motion pattern of the leg, our recognition accuracy was significantly increased.

Figure \ref{Walk_Defference} briefly depicts the benefits of body capacitance for sensing collaborative work. It shows the sensed body capacitance caused potential signal from two types walking. The left signal was perceived when person A was walking in a typical office, then he shook hands with person B (the moment that the middle impulse occurs), while touching the other's hand, person A made some other walk around person B. The potential variation from the skin while walking $"$unconnected$"$ is different when two bodies are strongly $"$connected$"$ by way of shaking hands. By coupling or connection of two human bodies, the body capacitance will be enlarged, resulting in the decrease of the skin potential change. When two people are electrically good coupled or connected, charges on both bodies can flow until a balanced level is reached(Figure \ref{Taking_TV}). This feature contributes to recognizing if two workers are working collaboratively or separately. 

\subsection{Activity classes}

\begin{table*}[htb]
\caption{Type of Activities}\label{ActivityType}
\centering
  \begin{tabular}{c c c c}
    \toprule
    \small\textit{ID}&\small\textit{Activities}&\small\textit{Comments}\\
    \midrule
    A1&Start and Stop & 10 on site steps \\
    A2&Doing nothing & stand still without any movement\\
    A3&Walk alone & normal walk without carring anything\\
    A4&Carry alone & walk and carrying the 10.3kg metal pieces\\
    A5&Carry together & walk and carrying the over 20kg metal pieces with another person\\
    A6&Lift & \makecell{ touch and lift the metal pieces from the box,ground and TV Wall}\\
    A7&Drop & \makecell{drop the metal pieces into the box, on ground and to the TV Wall} \\
    A8&Turn screw & turn the screws with an electric screw driver\\
    A9&No definition & \makecell{activities belongs to none of above listed like drinking, tying shoe during the work}\\
    A10&Out of camera & \makecell{where the participant walked out of camera's  field of view} \\
    \bottomrule
\end{tabular}
\end{table*}

The work process was divided into nine primitives, as Table  \ref{ActivityType} lists. For simple, we use A1 $-$ A10 to indicate all involved activity classes. Start and Stop activity is the ten on-site steps performed at the very beginning and end as well as in the middle of the whole task, aiming to synchronize sensor data and the videos. A2 occurred mostly when the participants took a rest. A3, A4, A5 are the most relative primitives to verify our sensing modality in collaborative group activity recognition, while carrying together means the participants were well coupled or connected together, which caused the charge redistribution on both bodies. The $HBC$ signal's response to motion, like walk, could give a different context when performed individually(A3). A4 is an independent activity, but with a heavy load at hand, which is different from A3 while the load enlarges the conductive plate at the body side. Also A3 is distinctive with A5 while the movement source was only from single person's walk instead of two. Lift and Drop are primitives that can happen independently or jointly since those two activities had only a small motion scale between arms and ground(without moving feet in most cases), so there was not too much difference at the $HBC$ signal whether they were performed individually or jointly. As explained before, the body capacitive signal only contributes when a body movement occurs, so while labeling, we did not differentiate whether Lift and Drop were performed individually or jointly. In the section of classification, we differentiate Lift/Drop individually or jointly by checking if the time slot of Lift/Drop was overlapped or not while receiving data from workers pairwise. A8 was performed with an electric screwdriver, which only caused a movement of the fingertip, so this activity did not generate a much useful information just as A2. A9 were activities that occasionally happened like drinking, tying shoes. Activities of A10 are not possible to label because the participant was out of the camera's field of view.

Four cameras were deployed to record the work process at a rate of 30 fps, after synchronizing the videos, we labeled data of each participant manually. The time slot of each activity was labeled with $1/30$ second precision.

\subsection{Classification Exploration}

We generated instances by applying a sliding window. Labels for each window were determined using majority voting inside it. The accelerometer data and the sensed potential data were detrended and normalized at first.

\begin{figure}
    \centering
    \includegraphics[width=0.6\columnwidth,height=7cm]{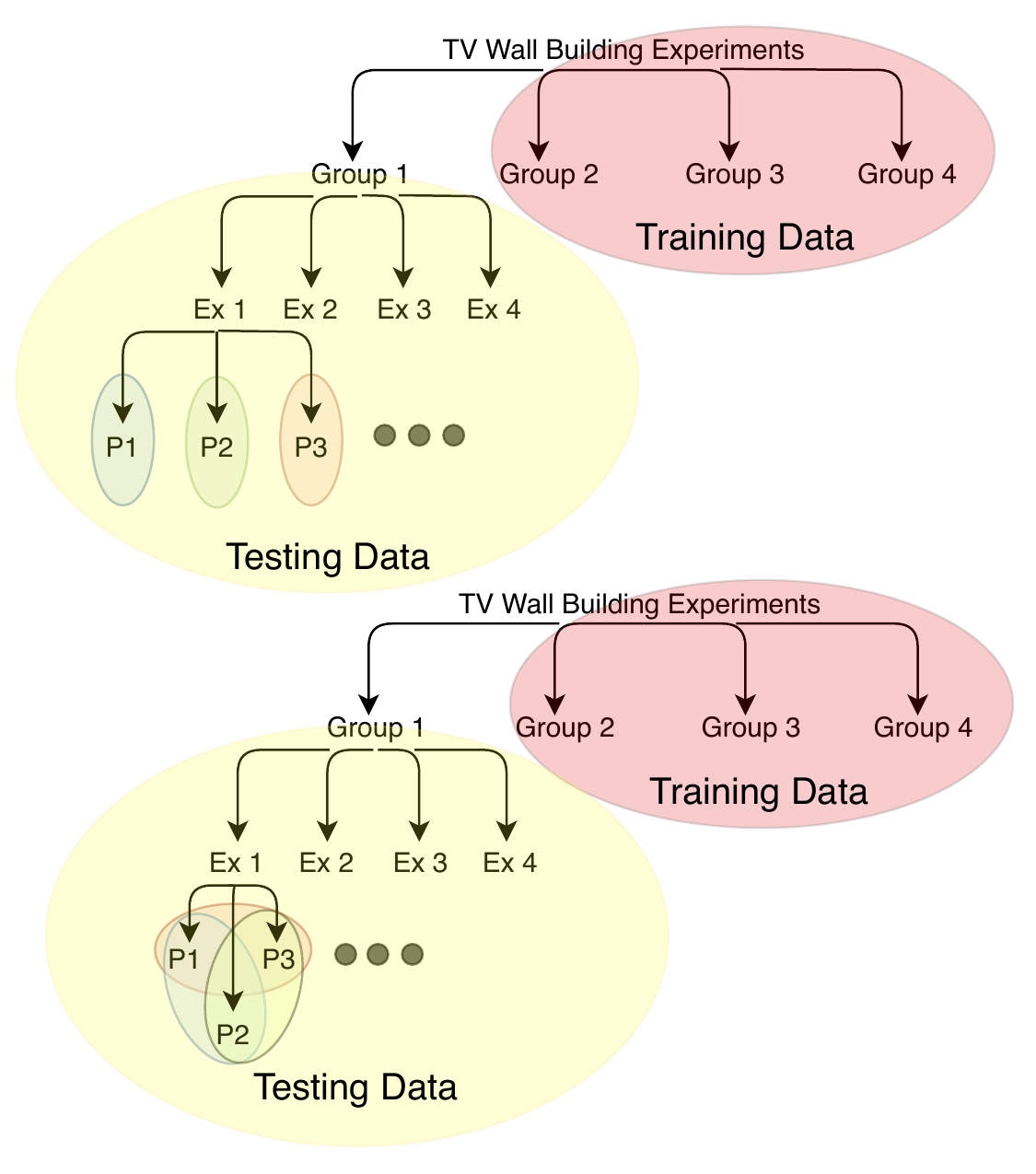}
    \caption{Modalities of activity recognition we used for testing: Receiving data from one user and Receiving data from a pair of users.}
    \label{fig:testType}
\end{figure}

We performed activity recognition in two modalities:
\begin{itemize}
    \item Receiving test data from a single user and predicting the activities of A3, A4, A5, A6, A7. Activities of A2, A8 are moved into a new class named null class. 
    \item Receive test data from both users and predicting the collaborative primitives of A5, A6(together), A7(together), with A2, A3, A4, A6(alone), A7(alone), A8 being considered into the null class.
\end{itemize}
The testing procedure can be better understood with the help of Figure \ref{fig:testType}.

\begin{figure*}
\centering
\includegraphics[width=1.0\textwidth,height= 7.0cm]{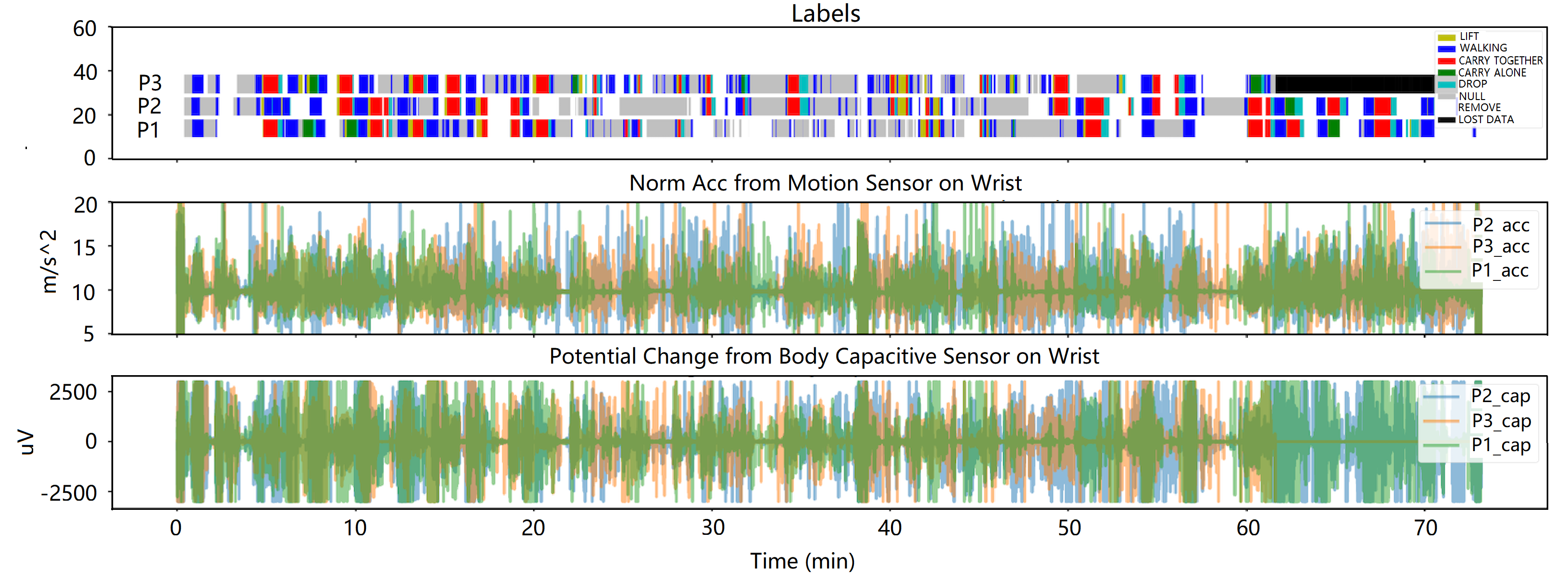}
\caption{Labeling and preprocessed Signal From Group 3's first Session}
\label{Labelling}
\end{figure*}

There are three cases that the generated sliding windows were discarded. The first case is the activity of A1, A9 and A10, where the activity is beyond our research interest(A9, A1) and the ground truth is not possible to annotate(A10). The second case is where labeling information is missing, no related activities were performed or where the participants were out of the camera, those intervals were marked as white. The last case is data loss caused by some occasional hardware problem, data was failed being written into SD card. This can be seen in the upper part of Figure \ref{Labelling} where missing data is marked as black. The middle and lower images of Figure \ref{Labelling} depict the pre-processed data from the accelerometer and body capacitive sensor on the wrist.

For each sliding window we generated for each sensor the statistical features referencing previous work. For classification, we trained different machine learning models including neural network ones, the logistic regression model using one versus all gave the best result. Since the data set is imbalanced, containing more null class instances than other activity types, every training instance is weighted based on the labels present inside the window. The total weight of a window is inversely proportional to the frequency of its labels in the dataset and is calculated based on the labels of a window $W$ as $$ \sum_{wi}^{W} \frac{ N }{ count(cl_{wi}) } $$ where $N$ represents the summed total number of timesteps in all training windows, $count(cl_{i})$ represents the number of those timesteps that belong to class $cl_i$, and $cl_{wi}$ is the class label at timestep $wi$.
Classifier predictions are smoothed by deciding the label for each window using soft voting where the current window, $3$ windows forward and backward vote, 
which helps to smooth predictions as well as to take into account of quick actions like touching metal, which can be sensed by the capacitive sensor.

\begin{figure*}
\centering
\includegraphics[width=1.0\textwidth,height=23cm]{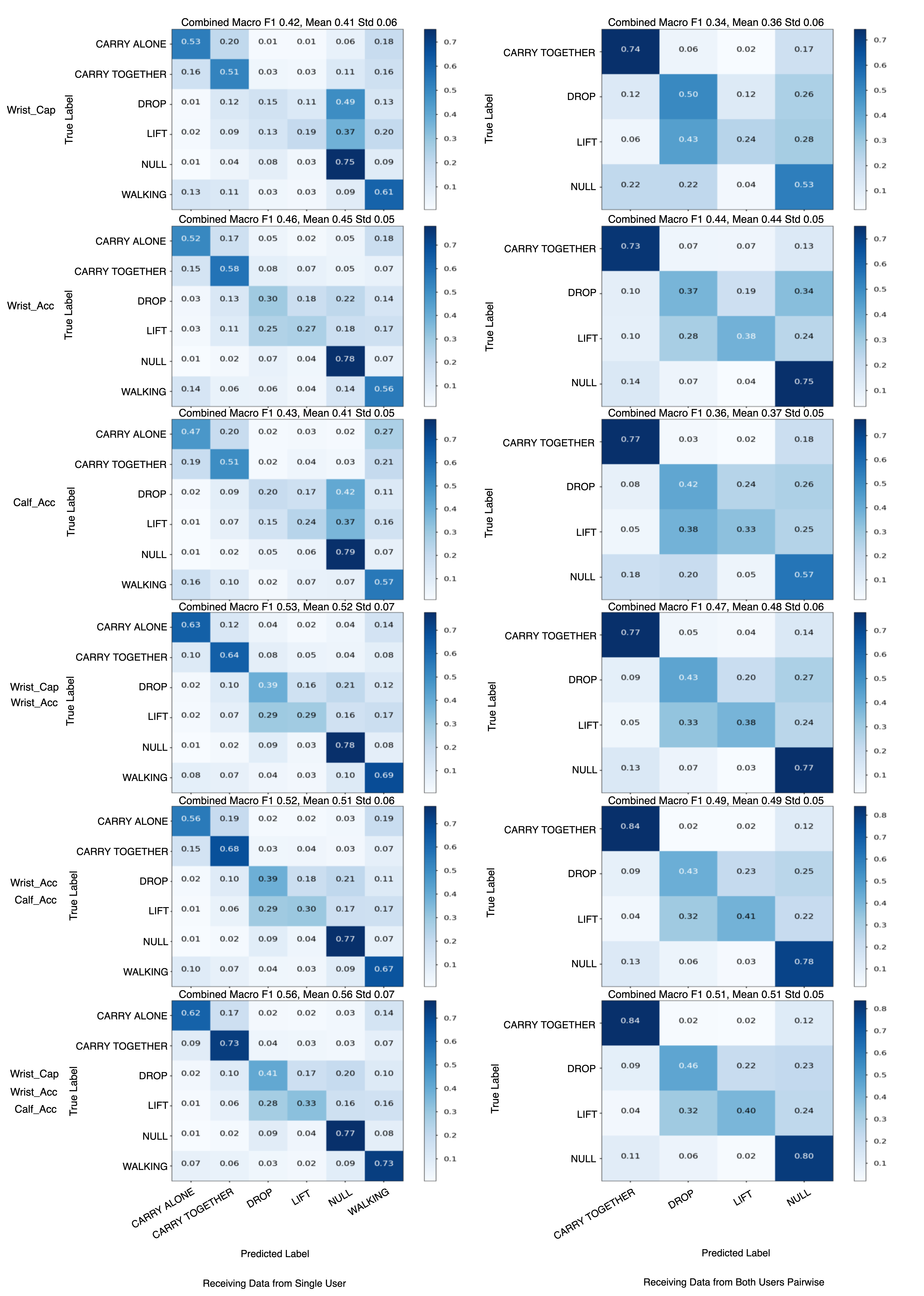}
\caption{Leave one Group out when receiving test data from single user/both users with different sensor sources}
\label{Matrix_BodyNet}
\end{figure*}

\begin{figure*}
\centering
\includegraphics[width=1.0\textwidth,height=23cm]{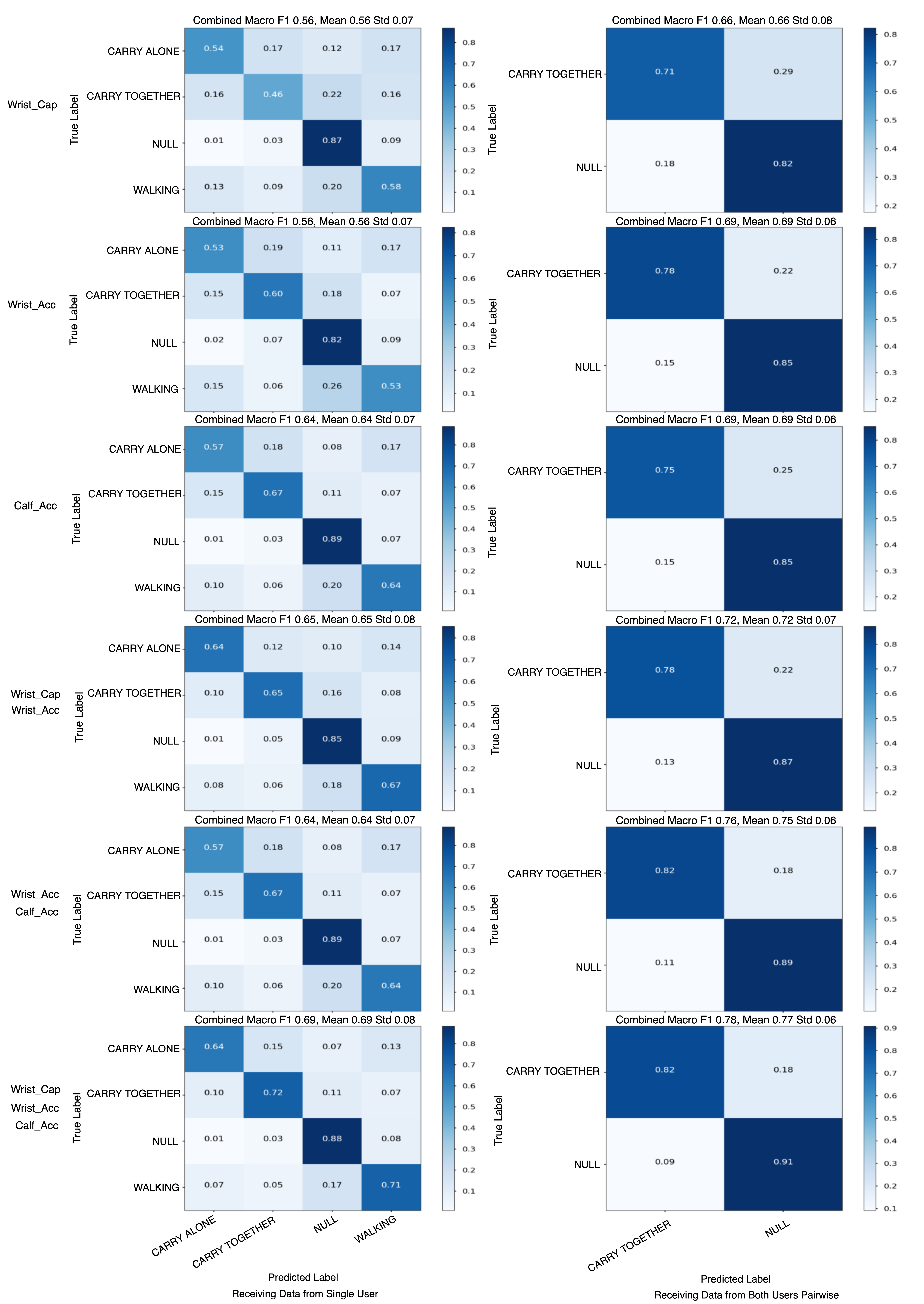}
\caption{Move Drop and Lift into Null Class}
\label{Matrix_BodyNet_Into_Null}
\end{figure*}



\subsection{Classification Results}
Figure \ref{Matrix_BodyNet} shows the classification result of different modalities with different sensor sources. As described before, in order to show that we can learn to recognize activities across groups, we employed a leave one group out procedure where, for each fold, the test set contains all days of one group, while the training set contains all days for the remaining groups. 
When the test data was from a single participant, we got a combined macro F-score with 0.42, 0.46, and 0.43 when the data source was a capacitive sensor on the wrist, accelerometer on the wrist and calf separately. For accelerometer, it is not able to recognise collaboration theoretically, since it has no ability of perceiving actions of other participants. In practical, the action of participant will variate slightly with the invasion of others. The classification ability of calf accelerometer located in the swing and stance phase, step and stride duration of the gait in the different action primitives. Accelerometer on wrist also supplied perception ability of variant wrist motion model in a certain level. As for the capacitive sensor on wrist, as we explained before, $"$Carrying together$"$ means more than doubling of the body capacitance, $"$Carrying alone$"$ means enlarging it by the metal load in their hands and $"$Walking alone$"$ keeps the human body capacitance as itself. All three sensor sources together contributed F-score of 0.56. The right column of Figure \ref{Matrix_BodyNet} shows the recognizing result when predicting only inter-group collaboration in non-scripted scenarios, that means we predicted here only primitives of $"$Carrying together$"$, $"$Drop together$"$, $"$Lift together$"$ and the others. The recognition was performed by receiving data from pairs of collaborators. Again here we predicted the recognition rate with different sensor sources and sensor fusion. When leaving $"$Drop$"$ and $"$Lift$"$ as independent classes, the capacitive sensor improved the combined F-score from 0.44 with a single wrist accelerometer to 0.47, and the calf accelerometer further assisted this value to 0.51.


We wonder how the recognition performs with the only wrist-worn sensors, which is more comfortable to use in practical scenarios. Combining the two wrist sensors, we got an increased combined F-score of 0.53 receiving data from single user and all the three primitives($"$Carrying together$"$, $"$Carrying together$"$ and $"$Walking alone$"$) got better recognition accuracy, meaning that the capacitive sensing modality granted a raise of 0.07(15\%) in combined macro F-score for the single wrist accelerometer. When receiving data from both users, the recognition increase from the capacitive sensing was 7\%.

As we described in the activity classes, the primitives of Lift and Drop were time-short and motion-lack actions(only the arms were stretching out and drawing back), the features were hard for our classifier to recognize, thus they both were frequently being recognized as $"$Null$"$ class or mixed with each other. Another reason for this inaccuracy came from the manually labeling process. As in practice, the participants' motions were mixed, sometimes they dropped or lifted stuff while walking, and walking caused capacitive signal easily overlapped the drop or lift caused capacitive signal. Our labeling result was time-in-series, the concurrency of the primitives was not considered. 

Since the most negative influence for recognizing was from the primitives of $"$Drop$"$ and $"$Lift$"$, and we were more interested in $"$Carrying together$"$, $"$Carrying alone$"$ or $"$Walking alone$"$, we moved $"$Drop$"$ and $"$Lift$"$ into $"$Null$"$ state, and got classification result of Figure \ref{Matrix_BodyNet_Into_Null}. The classification result was highly improved compared with original classes. This result is also impressive compared reseach work from Ward et al. \cite{ward2017detecting}, where the authors utilized body-worn microphones and accelerometers to detect instances of physical collaborative activities between members in a group and achieved an F-score of 60.1\% with two classes "collaboration" and "no collaboration". With all the three data sources we got combined F-score of 0.69 and 0.78 for each recognition modalities separately. Each of the data sources provided acceptable recognition of collaboration. When not considering the accelerometer signal from the calf, the wrist accelerometer gave 0.56 F-score when receiving data from single user, the wrist capacitive sensor improved the F-score to 0.65(16\% increase).

\begin{figure*}[t!]
    \centering
    \begin{subfigure}[t]{0.45\textwidth}
        \centering
        \includegraphics[width=1.0\textwidth,height=5.0cm]{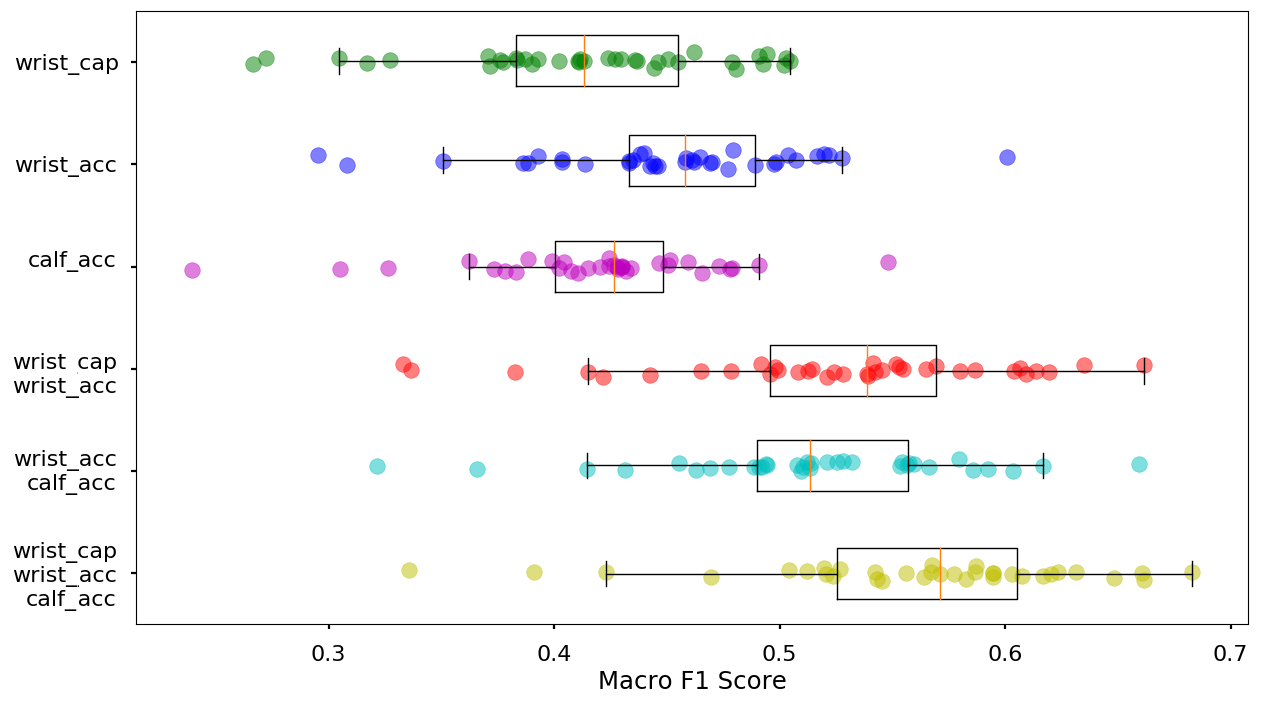}
        \caption{Receiving data from single user}
        \label{Data_Single_HARD}
    \end{subfigure}
    ~
    \begin{subfigure}[t]{0.45\textwidth}
        \centering
        \includegraphics[width=1.0\textwidth,height=5.0cm]{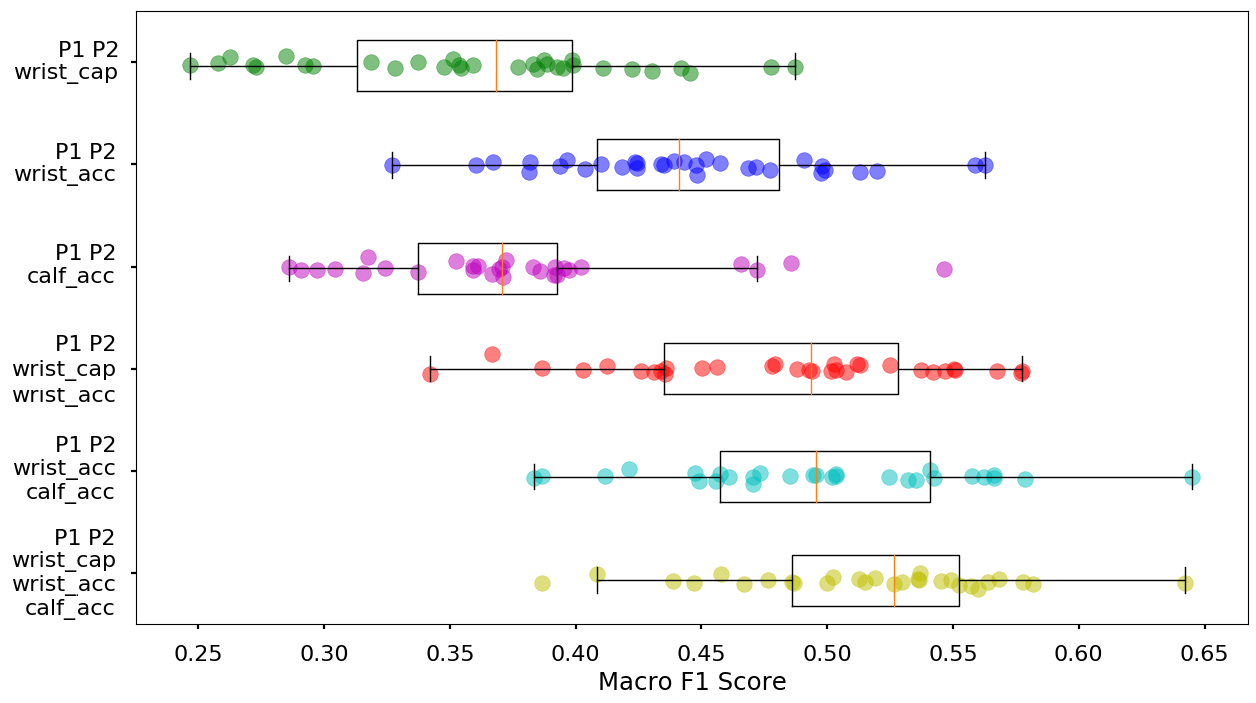}
        \caption{Receiving data from both users pairwise}
        \label{Data_Both_HARD}
    \end{subfigure}

    \caption{Macro F-Score with Lift/Drop as Separate Classes when using single Sensor and Sensor Fusion}
    \label{Sensorfusion_HARD}
\end{figure*}

\begin{figure*}[t!]
    \centering
    \begin{subfigure}[t]{0.45\textwidth}
        \centering
        \includegraphics[width=1.0\textwidth,height=5.0cm]{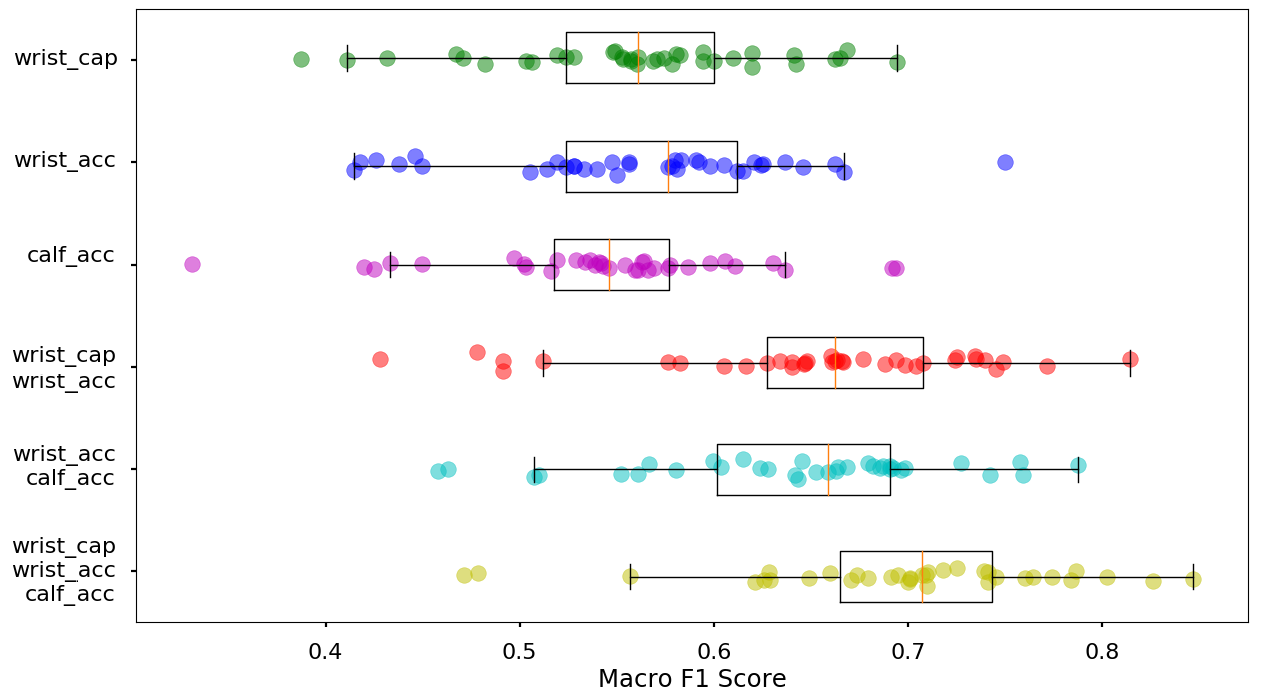}
        \caption{Receiving data from single user}
        \label{Data_Single}
    \end{subfigure}
    ~
    \begin{subfigure}[t]{0.45\textwidth}
        \centering
        \includegraphics[width=1.0\textwidth,height=5.0cm]{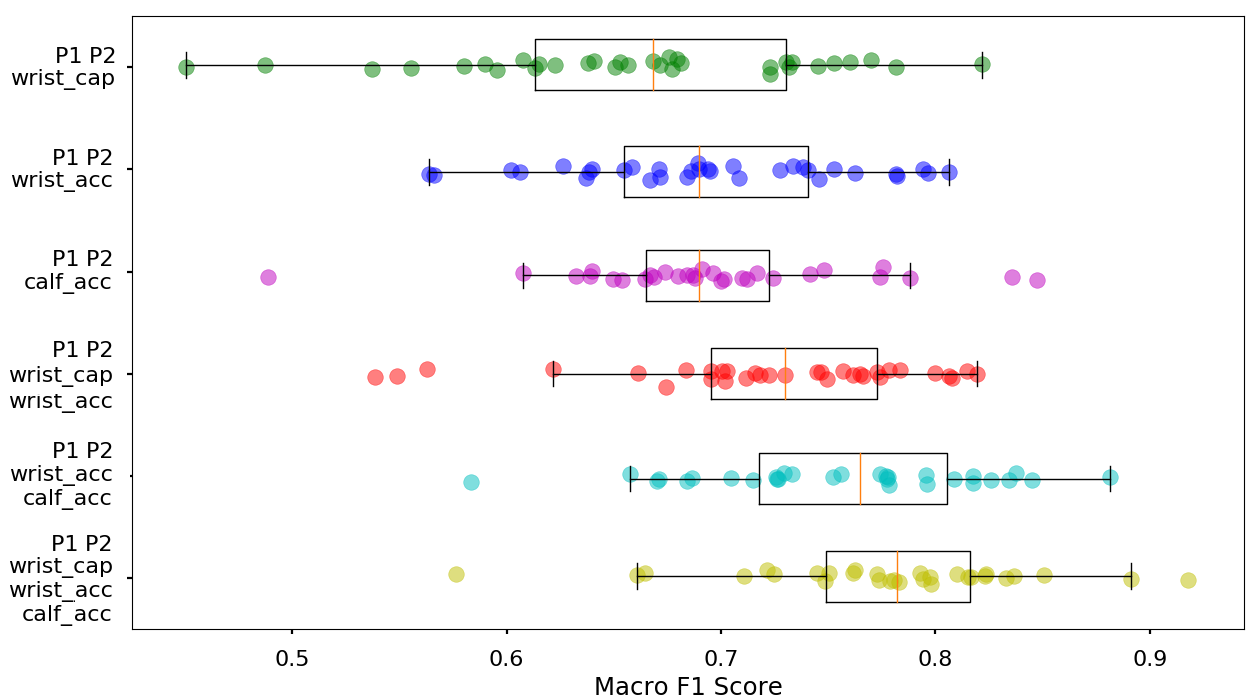}
        \caption{Receiving data from both users pairwise}
        \label{Data_Both}
    \end{subfigure}

    \caption{Macro F-Score with Lift/Drop into Null Class when using single Sensor and Sensor Fusion}
    \label{Sensorfusion}
\end{figure*}

To show the classification result with each sensor source and their fusion more evidently and seemingly, we provide Figure \ref{Sensorfusion_HARD} and Figure \ref{Sensorfusion}, which depict the result with F-score of all folds, with $"$Drop$"$ and $"$Lift$"$ as separate classes as well as $"$Null$"$ class. In each data receiving approach and class configuration, the wrist-worn system(fusion of wrist accelerometer and wrist capacitive sensor) benefits from the $HBC$ sensing approach, which increases the single wrist accelerometer with 15\%, 7\%, 16\%, 4\% separately in the four situations. In conclusion, the body capacitive sensing modality can supply impressive support to the traditional motion sensor regarding the recognition of collaborative activity.


\section{Conclusion}\label{Section_5}
In this work, we described a wearable human body motion tracking prototype composed of two sensing modalities, IMU, and $ HBC $, to track individual and collaborative activities. $HBC$ sensing modality is a motion-sensing approach with characters of low cost, low power consumption, and enjoys non-attachment and surrounding-sensitive advantages over the IMU. We explored the recognition and counting of seven machine-free leg-related exercises in labor and eleven popular gym workouts in a gym studio for individual activities tracking. We placed our prototype on the wrist while doing the labor experiment and in three positions while doing the gym experiment: in the pocket, on the calf, on the wrist, to track the activity, respectively. In the machine-free leg exercises experiment, the $HBC$ signal shows a significant better classification than $IMU$-based classification(0.89 vs. 0.77 in F-score when leave one user out), and a significant advantage for the exercise counting compared with $IMU$-based approach, especially for the exercises where the wrist is in a static state and $IMU$ completely lost the counting capability. In the gym workout experiment, we studied gym exercise classification and counting with both $IMU$ and $HBC$ signals and their combination, respectively. Results show that $HBC$ based sensing modality supplies very slight support to the traditional $IMU$-based individual workout classification tasks(around 1\% to 3\% increase while attaching the prototype to the body). The best workouts recognition F-score(leave one user out) is 91\%, which is achieved with the prototype attached on the wrist by using both $HBC$ and $IMU$ signals. 
Although the contribution of $HBC$ to workouts classification is very slight, $HBC$ signal supplies the best workout counting accuracy no matter where the prototype is worn. For example, the $HBC$ gives a counting accuracy of 80\% while wearing the prototype on the wrist, while the $IMU$ supplies 75.2\%. We also demonstrated the robustness of the $HBC$ sensing modality concerning the variation of wearing, weather conditions, and ground type. For collaborative activities tracking, we described a sensor-based collaborative group activity recognition approach related to manipulating objects and physical collaboration between users by utilizing $HBC$ sensing and accelerometers. The capacitive sensing can contribute to recognizing joint activities when the workers are well coupled or linked. By fusing capacitive sensing on wrist and accelerometer on wrist and calf, we achieved the classification accuracy of 71\%, 64\%, 72\%, 88\%(walking alone, carrying TV alone, carrying TV together with another worker, and the left null state) when receiving test data from single-user, and 82\%, 91\%(carrying TV jointly, and the other activities) when receiving data from workers pairwise with a logistic regression model. The $HBC$ signal contributes an increase of 16\% to the four primitive activities when receiving data from a single user. This work demonstrates the effectiveness of $HBC$ as an alternative/complementary approach for individual and collaborative activity recognition and repetitive action counting.

\bibliographystyle{unsrt}  
\bibliography{sample-base}  

\begin{thebibliography}{100}

\bibitem{park2003enhancing}
Sungmee Park and Sundaresan Jayaraman.
\newblock Enhancing the quality of life through wearable technology.
\newblock {\em IEEE Engineering in medicine and biology magazine},
  22(3):41--48, 2003.

\bibitem{caporusso2019comparative}
Nicholas Caporusso, Angela Walters, Meng Ding, Devon Patchin, Noah Vaughn,
  Daniel Jachetta, and Spencer Romeiser.
\newblock Comparative user experience analysis of pervasive wearable
  technology.
\newblock In {\em International Conference on Applied Human Factors and
  Ergonomics}, pages 3--13. Springer, 2019.

\bibitem{ward2017detecting}
Jamie~A Ward, Gerald Pirkl, Peter Hevesi, and Paul Lukowicz.
\newblock Detecting physical collaborations in a group task using body-worn
  microphones and accelerometers.
\newblock In {\em 2017 IEEE International Conference on Pervasive Computing and
  Communications Workshops (PerCom Workshops)}, pages 268--273. IEEE, 2017.

\bibitem{patel2015wearable}
Mitesh~S Patel, David~A Asch, and Kevin~G Volpp.
\newblock Wearable devices as facilitators, not drivers, of health behavior
  change.
\newblock {\em Jama}, 313(5):459--460, 2015.

\bibitem{naslund2016wearable}
John~A Naslund, Kelly~A Aschbrenner, Emily~A Scherer, Gregory~J McHugo, Lisa~A
  Marsch, and Stephen~J Bartels.
\newblock Wearable devices and mobile technologies for supporting behavioral
  weight loss among people with serious mental illness.
\newblock {\em Psychiatry research}, 244:139--144, 2016.

\bibitem{bian2022state}
Sizhen Bian, Mengxi Liu, Bo~Zhou, and Paul Lukowicz.
\newblock The state-of-the-art sensing techniques in human activity
  recognition: A survey.
\newblock {\em Sensors}, 22(12):4596, 2022.

\bibitem{holzemann2018using}
Alexander H{\"o}lzemann and Kristof Van~Laerhoven.
\newblock Using wrist-worn activity recognition for basketball game analysis.
\newblock In {\em Proceedings of the 5th international Workshop on Sensor-based
  Activity Recognition and Interaction}, page~13. ACM, 2018.

\bibitem{kaewkannate2016comparison}
Kanitthika Kaewkannate and Soochan Kim.
\newblock A comparison of wearable fitness devices.
\newblock {\em BMC public health}, 16(1):433, 2016.

\bibitem{alzahrani2016wearable}
Saeed Alzahrani, Abdulhakim Giadedi, Ross Lamberth, Tania Lilja, Obinna~Charles
  Mbagwu, and Edwin~Vilanova Velez.
\newblock Wearable technology: Diabetes monitoring in the healthcare industry.
\newblock 2016.

\bibitem{hansel2015challenges}
Katrin H{\"a}nsel, Natalie Wilde, Hamed Haddadi, and Akram Alomainy.
\newblock Challenges with current wearable technology in monitoring health data
  and providing positive behavioural support.
\newblock In {\em Proceedings of the 5th EAI International Conference on
  Wireless Mobile Communication and Healthcare}, pages 158--161. ICST
  (Institute for Computer Sciences, Social-Informatics and~…, 2015.

\bibitem{parker2018interplay}
Jayson~L Parker, Qasim Muhammad, John Kedzierski, and Sana Maqbool.
\newblock The interplay between regulation and design in medical wearable
  technology.
\newblock {\em Wearable Technology in Medicine and Health Care}, page 291,
  2018.

\bibitem{wang2018intelligent}
Ker-Jiun Wang, Quanbo Liu, Yifan Zhao, Caroline~Yan Zheng, Soumya Vhasure,
  Quanfeng Liu, Prakash Thakur, Mingui Sun, and Zhi-Hong Mao.
\newblock Intelligent wearable virtual reality (vr) gaming controller for
  people with motor disabilities.
\newblock In {\em 2018 IEEE International Conference on Artificial Intelligence
  and Virtual Reality (AIVR)}, pages 161--164. IEEE, 2018.

\bibitem{cha2018towards}
Seung~Hyun Cha, Joonoh Seo, Seung~Hyo Baek, and Choongwan Koo.
\newblock Towards a well-planned, activity-based work environment: Automated
  recognition of office activities using accelerometers.
\newblock {\em Building and Environment}, 144:86--93, 2018.

\bibitem{uddin2015wearable}
Mostafa Uddin, Ahmed Salem, Ilho Nam, and Tamer Nadeem.
\newblock Wearable sensing framework for human activity monitoring.
\newblock In {\em Proceedings of the 2015 workshop on Wearable Systems and
  Applications}, pages 21--26. ACM, 2015.

\bibitem{magalhaes2015wearable}
Fabricio Anicio~de Magalhaes, Giuseppe Vannozzi, Giorgio Gatta, and Silvia
  Fantozzi.
\newblock Wearable inertial sensors in swimming motion analysis: a systematic
  review.
\newblock {\em Journal of sports sciences}, 33(7):732--745, 2015.

\bibitem{kozlovszky2018imu}
Miklos Kozlovszky, P{\'a}l Bogdanov, K~Kar{\'o}ckai, G~Garaguly, and Gernot
  Kronreif.
\newblock Imu based human movement tracking.
\newblock In {\em 2018 41st International Convention on Information and
  Communication Technology, Electronics and Microelectronics (MIPRO)}, pages
  0240--0244. IEEE, 2018.

\bibitem{presta1983measurement}
Elio Presta, Jack Wang, Gail~G Harrison, P~Bj{\"o}rntorp, Wesley~H Harker, and
  Theodore~B Van~Itallie.
\newblock Measurement of total body electrical conductivity: a new method for
  estimation of body composition.
\newblock {\em The American journal of clinical nutrition}, 37(5):735--739,
  1983.

\bibitem{cochran1986total}
William~J Cochran, William~J Klish, William~W Wong, and Peter~D Klein.
\newblock Total body electrical conductivity used to determine body composition
  in infants.
\newblock {\em Pediatric research}, 20(6):561, 1986.

\bibitem{aliau2012fast}
Carles Aliau~Bonet and Ramon Pall{\`a}s~Areny.
\newblock A fast method to estimate body capacitance to ground.
\newblock In {\em Proceedings of XX IMEKO World Congress 2012, September 9-14,
  Busan South Korea}, pages 1--4, 2012.

\bibitem{aliau2013novel}
Carles Aliau-Bonet and Ramon Pallas-Areny.
\newblock A novel method to estimate body capacitance to ground at mid
  frequencies.
\newblock {\em IEEE Transactions on Instrumentation and Measurement},
  62(9):2519--2525, 2013.

\bibitem{buller2006measurement}
William Buller and Brian Wilson.
\newblock Measurement and modeling mutual capacitance of electrical wiring and
  humans.
\newblock {\em IEEE Transactions on Instrumentation and Measurement},
  55(5):1519--1522, 2006.

\bibitem{sizhen2021systematic}
Sizhen Bian and Paul Lukowicz.
\newblock A systematic study of the influence of various user specific and
  environmental factors on wearable human body capacitance sensing.
\newblock In {\em EAI International Conference on Body Area Networks}, pages
  247--274. Springer, 2021.

\bibitem{fujiwara2002numerical}
Osamu Fujiwara and Takanori Ikawa.
\newblock Numerical calculation of human-body capacitance by surface charge
  method.
\newblock {\em Electronics and Communications in Japan (Part I:
  Communications)}, 85(12):38--44, 2002.

\bibitem{jonassen1998human}
Niels Jonassen.
\newblock Human body capacitance: static or dynamic concept?[esd].
\newblock In {\em Electrical Overstress/Electrostatic Discharge Symposium
  Proceedings. 1998 (Cat. No. 98TH8347)}, pages 111--117. IEEE, 1998.

\bibitem{goad2016ambient}
N~Goad and DJ~Gawkrodger.
\newblock Ambient humidity and the skin: The impact of air humidity in healthy
  and diseased states.
\newblock {\em Journal of the European Academy of Dermatology and Venereology},
  30(8):1285--1294, 2016.

\bibitem{egawa2002effect}
Mariko Egawa, Motoki Oguri, Tomohiro Kuwahara, and Motoji Takahashi.
\newblock Effect of exposure of human skin to a dry environment.
\newblock {\em Skin Research and Technology}, 8(4):212--218, 2002.

\bibitem{cohn2012humantenna}
Gabe Cohn, Daniel Morris, Shwetak Patel, and Desney Tan.
\newblock Humantenna: using the body as an antenna for real-time whole-body
  interaction.
\newblock In {\em Proceedings of the SIGCHI Conference on Human Factors in
  Computing Systems}, pages 1901--1910. ACM, 2012.

\bibitem{bian2019wrist}
Sizhen Bian, Vitor~F Rey, Junaid Younas, and Paul Lukowicz.
\newblock Wrist-worn capacitive sensor for activity and physical collaboration
  recognition.
\newblock In {\em 2019 IEEE International Conference on Pervasive Computing and
  Communications Workshops (PerCom Workshops)}, pages 261--266. IEEE, 2019.

\bibitem{kan2015social}
Viirj Kan, Katsuya Fujii, Judith Amores, Chang~Long Zhu~Jin, Pattie Maes, and
  Hiroshi Ishii.
\newblock Social textiles: Social affordances and icebreaking interactions
  through wearable social messaging.
\newblock In {\em Proceedings of the Ninth International Conference on
  Tangible, Embedded, and Embodied Interaction}, pages 619--624. ACM, 2015.

\bibitem{cheng2010active}
Jingyuan Cheng, Oliver Amft, and Paul Lukowicz.
\newblock Active capacitive sensing: Exploring a new wearable sensing modality
  for activity recognition.
\newblock In {\em International conference on pervasive computing}, pages
  319--336. Springer, 2010.

\bibitem{sizhen2021capacitive}
Sizhen Bian and Paul Lukowicz.
\newblock Capacitive sensing based on-board hand gesture recognition with
  tinyml.
\newblock In {\em Adjunct Proceedings of the 2021 ACM International Joint
  Conference on Pervasive and Ubiquitous Computing and Proceedings of the 2021
  ACM International Symposium on Wearable Computers}, pages 4--5, 2021.

\bibitem{bian2019passive}
Sizhen Bian, Vitor~F Rey, Peter Hevesi, and Paul Lukowicz.
\newblock Passive capacitive based approach for full body gym workout
  recognition and counting.
\newblock In {\em 2019 IEEE International Conference on Pervasive Computing and
  Communications (PerCom}, pages 1--10. IEEE, 2019.

\bibitem{koskimaki2014recognizing}
Heli Koskim{\"a}ki and Pekka Siirtola.
\newblock Recognizing gym exercises using acceleration data from wearable
  sensors.
\newblock In {\em 2014 IEEE Symposium on Computational Intelligence and Data
  Mining (CIDM)}, pages 321--328. IEEE, 2014.

\bibitem{morris2014recofit}
Dan Morris, T~Scott Saponas, Andrew Guillory, and Ilya Kelner.
\newblock Recofit: using a wearable sensor to find, recognize, and count
  repetitive exercises.
\newblock In {\em Proceedings of the SIGCHI Conference on Human Factors in
  Computing Systems}, pages 3225--3234. ACM, 2014.

\bibitem{wahjudi2019imu}
Fanuel Wahjudi and Fuchun~Joseph Lin.
\newblock Imu-based walking workouts recognition.
\newblock In {\em 2019 IEEE 5th World Forum on Internet of Things (WF-IoT)},
  pages 251--256. IEEE, 2019.

\bibitem{chang2007tracking}
Keng-Hao Chang, Mike~Y Chen, and John Canny.
\newblock Tracking free-weight exercises.
\newblock In {\em International Conference on Ubiquitous Computing}, pages
  19--37. Springer, 2007.

\bibitem{depari2019lightweight}
Alessandro Depari, Paolo Ferrari, Alessandra Flammini, Stefano Rinaldi, and
  Emiliano Sisinni.
\newblock Lightweight machine learning-based approach for supervision of
  fitness workout.
\newblock In {\em 2019 IEEE Sensors Applications Symposium (SAS)}, pages 1--6.
  IEEE, 2019.

\bibitem{gruenerbl2017detecting}
Agnes Gruenerbl, Gernot Bahle, and Paul Lukowicz.
\newblock Detecting spontaneous collaboration in dynamic group activities from
  noisy individual activity data.
\newblock In {\em 2017 IEEE International Conference on Pervasive Computing and
  Communications Workshops (PerCom Workshops)}, pages 279--284. IEEE, 2017.

\bibitem{bulling2014tutorial}
Andreas Bulling, Ulf Blanke, and Bernt Schiele.
\newblock A tutorial on human activity recognition using body-worn inertial
  sensors.
\newblock {\em ACM Computing Surveys (CSUR)}, 46(3):33, 2014.

\bibitem{lara2013survey}
Oscar~D Lara and Miguel~A Labrador.
\newblock A survey on human activity recognition using wearable sensors.
\newblock {\em IEEE communications surveys \& tutorials}, 15(3):1192--1209,
  2013.

\bibitem{ward2006activity}
Jamie~A Ward, Paul Lukowicz, Gerhard Troster, and Thad~E Starner.
\newblock Activity recognition of assembly tasks using body-worn microphones
  and accelerometers.
\newblock {\em IEEE transactions on pattern analysis and machine intelligence},
  28(10):1553--1567, 2006.

\bibitem{lukowicz2004recognizing}
Paul Lukowicz, Jamie~A Ward, Holger Junker, Mathias St{\"a}ger, Gerhard
  Tr{\"o}ster, Amin Atrash, and Thad Starner.
\newblock Recognizing workshop activity using body worn microphones and
  accelerometers.
\newblock In {\em International conference on pervasive computing}, pages
  18--32. Springer, 2004.

\bibitem{stiefmeier2008wearable}
Thomas Stiefmeier, Daniel Roggen, Georg Ogris, Paul Lukowicz, and Gerhard
  Tr{\"o}ster.
\newblock Wearable activity tracking in car manufacturing.
\newblock {\em IEEE Pervasive Computing}, 7(2):42--50, 2008.

\bibitem{gong2003recognition}
Shaogang Gong and Tao Xiang.
\newblock Recognition of group activities using dynamic probabilistic networks.
\newblock In {\em ICCV}, volume~3, page 742, 2003.

\bibitem{ryoo2008recognition}
MS~Ryoo and JK~Aggarwal.
\newblock Recognition of high-level group activities based on activities of
  individual members.
\newblock In {\em 2008 IEEE Workshop on Motion and video Computing}, pages
  1--8. IEEE, 2008.

\bibitem{zhang2008hierarchical}
Weidong Zhang, Feng Chen, Wenli Xu, and Youtian Du.
\newblock Hierarchical group process representation in multi-agent activity
  recognition.
\newblock {\em Signal Processing: Image Communication}, 23(10):739--753, 2008.

\bibitem{li2009learning}
Ruonan Li, Rama Chellappa, and Shaohua~Kevin Zhou.
\newblock Learning multi-modal densities on discriminative temporal interaction
  manifold for group activity recognition.
\newblock In {\em 2009 IEEE Conference on Computer Vision and Pattern
  Recognition}, pages 2450--2457. IEEE, 2009.

\bibitem{ni2009recognizing}
Bingbing Ni, Shuicheng Yan, and Ashraf Kassim.
\newblock Recognizing human group activities with localized causalities.
\newblock In {\em 2009 IEEE Conference on Computer Vision and Pattern
  Recognition}, pages 1470--1477. IEEE, 2009.

\bibitem{chang2011probabilistic}
Ming-Ching Chang, Nils Krahnstoever, and Weina Ge.
\newblock Probabilistic group-level motion analysis and scenario recognition.
\newblock In {\em 2011 International Conference on Computer Vision}, pages
  747--754. IEEE, 2011.

\bibitem{yun2012two}
Kiwon Yun, Jean Honorio, Debaleena Chattopadhyay, Tamara~L Berg, and Dimitris
  Samaras.
\newblock Two-person interaction detection using body-pose features and
  multiple instance learning.
\newblock In {\em 2012 IEEE Computer Society Conference on Computer Vision and
  Pattern Recognition Workshops}, pages 28--35. IEEE, 2012.

\bibitem{wilson2005simultaneous}
Daniel~H Wilson and Chris Atkeson.
\newblock Simultaneous tracking and activity recognition (star) using many
  anonymous, binary sensors.
\newblock In {\em International Conference on Pervasive Computing}, pages
  62--79. Springer, 2005.

\bibitem{wren2006toward}
Christopher~R Wren and Emmanuel~Munguia Tapia.
\newblock Toward scalable activity recognition for sensor networks.
\newblock In {\em International Symposium on Location-and Context-Awareness},
  pages 168--185. Springer, 2006.

\bibitem{wang2009sensor}
Liang Wang, Tao Gu, Xianping Tao, and Jian Lu.
\newblock Sensor-based human activity recognition in a multi-user scenario.
\newblock In {\em European Conference on Ambient Intelligence}, pages 78--87.
  Springer, 2009.

\bibitem{gordon2013towards}
Dawud Gordon, Jan-Hendrik Hanne, Martin Berchtold, Ali Asghar~Nazari
  Shirehjini, and Michael Beigl.
\newblock Towards collaborative group activity recognition using mobile
  devices.
\newblock {\em Mobile Networks and Applications}, 18(3):326--340, 2013.

\bibitem{chen2019framework}
Hao Chen, Seung~Hyun Cha, and Tae~Wan Kim.
\newblock A framework for group activity detection and recognition using
  smartphone sensors and beacons.
\newblock {\em Building and Environment}, 158:205--216, 2019.

\bibitem{bian2021github}
Sizhen Bian.
\newblock Toolkit-for-hbc-sensing.
\newblock \url{https://github.com/zhaxidele/Toolkit-for-HBC-sensing}, 2021.

\bibitem{bian2022using}
Sizhen Bian, Siyu Yuan, Vitor~Fortes Rey, and Paul Lukowicz.
\newblock Using human body capacitance sensing to monitor leg motion dominated
  activities with a wrist worn device.
\newblock In {\em Sensor-and Video-Based Activity and Behavior Computing},
  pages 81--94. Springer, 2022.

\bibitem{casale2011human}
Pierluigi Casale, Oriol Pujol, and Petia Radeva.
\newblock Human activity recognition from accelerometer data using a wearable
  device.
\newblock In {\em Iberian conference on pattern recognition and image
  analysis}, pages 289--296. Springer, 2011.

\bibitem{feng2015random}
Zengtao Feng, Lingfei Mo, and Meng Li.
\newblock A random forest-based ensemble method for activity recognition.
\newblock In {\em 2015 37th Annual International Conference of the IEEE
  Engineering in Medicine and Biology Society (EMBC)}, pages 5074--5077. IEEE,
  2015.

\bibitem{bayat2014study}
Akram Bayat, Marc Pomplun, and Duc~A Tran.
\newblock A study on human activity recognition using accelerometer data from
  smartphones.
\newblock {\em Procedia Computer Science}, 34:450--457, 2014.

\bibitem{nurwulan2020random}
Nurul~Retno Nurwulan and Gjergji Selamaj.
\newblock Random forest for human daily activity recognition.
\newblock In {\em Journal of Physics: Conference Series}, volume 1655, page
  012087. IOP Publishing, 2020.

\bibitem{hamalainen2011jerk}
Wilhelmiina Ham{\"a}l{\"a}inen, Mikko J{\"a}rvinen, Paula Martiskainen, and
  Jaakko Mononen.
\newblock Jerk-based feature extraction for robust activity recognition from
  acceleration data.
\newblock In {\em 2011 11th International Conference on Intelligent Systems
  Design and Applications}, pages 831--836. IEEE, 2011.

\bibitem{find_peaks}
SciPy.org.
\newblock Find peaks inside a signal based on peak properties.

\bibitem{chawla2002smote}
Nitesh~V Chawla, Kevin~W Bowyer, Lawrence~O Hall, and W~Philip Kegelmeyer.
\newblock Smote: synthetic minority over-sampling technique.
\newblock {\em Journal of artificial intelligence research}, 16:321--357, 2002.

\bibitem{haescher2015study}
Marian Haescher, Denys~JC Matthies, John Trimpop, and Bodo Urban.
\newblock A study on measuring heart-and respiration-rate via wrist-worn
  accelerometer-based seismocardiography (scg) in comparison to commonly
  applied technologies.
\newblock In {\em Proceedings of the 2nd international Workshop on Sensor-based
  Activity Recognition and Interaction}, pages 1--6, 2015.

\bibitem{toutiaee2020video}
Mohammadhossein Toutiaee, Abbas Keshavarzi, Abolfazl Farahani, and John~A
  Miller.
\newblock Video contents understanding using deep neural networks.
\newblock {\em arXiv preprint arXiv:2004.13959}, 2020.

\bibitem{amosov2020using}
Oleg~Semenovich Amosov, Svetlana~Gennadievna Amosova, Yuriy~Sergeevich Ivanov,
  and Sergey~Victorovich Zhiganov.
\newblock Using the deep neural networks for normal and abnormal situation
  recognition in the automatic access monitoring and control system of
  vehicles.
\newblock {\em Neural Computing and Applications}, pages 1--15, 2020.

\bibitem{torfi2020natural}
Amirsina Torfi, Rouzbeh~A Shirvani, Yaser Keneshloo, Nader Tavvaf, and Edward~A
  Fox.
\newblock Natural language processing advancements by deep learning: A survey.
\newblock {\em arXiv preprint arXiv:2003.01200}, 2020.

\bibitem{nassif2019speech}
Ali~Bou Nassif, Ismail Shahin, Imtinan Attili, Mohammad Azzeh, and Khaled
  Shaalan.
\newblock Speech recognition using deep neural networks: A systematic review.
\newblock {\em IEEE access}, 7:19143--19165, 2019.

\bibitem{wang2019deep}
Jindong Wang, Yiqiang Chen, Shuji Hao, Xiaohui Peng, and Lisha Hu.
\newblock Deep learning for sensor-based activity recognition: A survey.
\newblock {\em Pattern Recognition Letters}, 119:3--11, 2019.

\bibitem{chen2020deep}
Kaixuan Chen, Dalin Zhang, Lina Yao, Bin Guo, Zhiwen Yu, and Yunhao Liu.
\newblock Deep learning for sensor-based human activity recognition: overview,
  challenges and opportunities.
\newblock {\em arXiv preprint arXiv:2001.07416}, 2020.

\bibitem{wang2020summary}
Lin Wang, Hristijan Gjoreski, Mathias Ciliberto, Paula Lago, Kazuya Murao,
  Tsuyoshi Okita, and Daniel Roggen.
\newblock Summary of the sussex-huawei locomotion-transportation recognition
  challenge 2020.
\newblock In {\em Adjunct Proceedings of the 2020 ACM International Joint
  Conference on Pervasive and Ubiquitous Computing and Proceedings of the 2020
  ACM International Symposium on Wearable Computers}, pages 351--358, 2020.

\bibitem{ordonez2016deep}
Francisco Ord{\'o}{\~n}ez and Daniel Roggen.
\newblock Deep convolutional and lstm recurrent neural networks for multimodal
  wearable activity recognition.
\newblock {\em Sensors}, 16(1):115, 2016.

\bibitem{qin2020imaging}
Zhen Qin, Yibo Zhang, Shuyu Meng, Zhiguang Qin, and Kim-Kwang~Raymond Choo.
\newblock Imaging and fusing time series for wearable sensor-based human
  activity recognition.
\newblock {\em Information Fusion}, 53:80--87, 2020.

\bibitem{chollet2015keras}
Fran\c{c}ois Chollet et~al.
\newblock Keras.
\newblock \url{https://keras.io}, 2015.

\bibitem{cho2018divide}
Heeryon Cho and Sang Yoon.
\newblock Divide and conquer-based 1d cnn human activity recognition using test
  data sharpening.
\newblock {\em Sensors}, 18(4):1055, 2018.

\bibitem{cruciani2020feature}
Federico Cruciani, Anastasios Vafeiadis, Chris Nugent, Ian Cleland, Paul
  McCullagh, Konstantinos Votis, Dimitrios Giakoumis, Dimitrios Tzovaras,
  Liming Chen, and Raouf Hamzaoui.
\newblock Feature learning for human activity recognition using convolutional
  neural networks.
\newblock {\em CCF Transactions on Pervasive Computing and Interaction},
  2(1):18--32, 2020.

\bibitem{KingmaB14}
Diederik~P. Kingma and Jimmy Ba.
\newblock Adam: {A} method for stochastic optimization.
\newblock {\em CoRR}, abs/1412.6980, 2014.

\bibitem{wang2018summary}
Lin Wang, Hristijan Gjoreskia, Kazuya Murao, Tsuyoshi Okita, and Daniel Roggen.
\newblock Summary of the sussex-huawei locomotion-transportation recognition
  challenge.
\newblock In {\em Proceedings of the 2018 ACM international joint conference
  and 2018 international symposium on pervasive and ubiquitous computing and
  wearable computers}, pages 1521--1530, 2018.

\bibitem{wang2019summary}
Lin Wang, Hristijan Gjoreskia, Ciliberto Mathias, Lago Paula, Murao Kazuya,
  Okita Tsuyoshi, and Daniel Roggen.
\newblock Summary of the sussex-huawei locomotion-transportation recognition
  challenge 2019.
\newblock In {\em Adjunct Proceedings of the 2019 ACM International Joint
  Conference on Pervasive and Ubiquitous Computing and Proceedings of the 2019
  ACM International Symposium on Wearable Computers}, pages 849--856, 2019.

\bibitem{gjoreski2020classical}
Martin Gjoreski, Vito Janko, Ga{\v{s}}per Slapni{\v{c}}ar, Miha Mlakar, Nina
  Re{\v{s}}{\v{c}}i{\v{c}}, Jani Bizjak, Vid Drobni{\v{c}}, Matej Marinko, Nejc
  Mlakar, Mitja Lu{\v{s}}trek, et~al.
\newblock Classical and deep learning methods for recognizing human activities
  and modes of transportation with smartphone sensors.
\newblock {\em Information Fusion}, 62:47--62, 2020.

\bibitem{hoelzemann2020digging}
Alexander Hoelzemann and Kristof Van~Laerhoven.
\newblock Digging deeper: towards a better understanding of transfer learning
  for human activity recognition.
\newblock In {\em Proceedings of the 2020 International Symposium on Wearable
  Computers}, pages 50--54, 2020.

\bibitem{morales2016deep}
Francisco Javier~Ord{\'o}{\~n}ez Morales and Daniel Roggen.
\newblock Deep convolutional feature transfer across mobile activity
  recognition domains, sensor modalities and locations.
\newblock In {\em Proceedings of the 2016 ACM International Symposium on
  Wearable Computers}, pages 92--99, 2016.

\bibitem{lasagne}
J.; et~al. Dieleman, S.;~Schlüter.
\newblock Welcome to lasagne, 2014-2015.

\bibitem{castle1997contact}
GSP Castle.
\newblock Contact charging between insulators.
\newblock {\em Journal of Electrostatics}, 40:13--20, 1997.

\bibitem{zeng2014convolutional}
Ming Zeng, Le~T Nguyen, Bo~Yu, Ole~J Mengshoel, Jiang Zhu, Pang Wu, and Joy
  Zhang.
\newblock Convolutional neural networks for human activity recognition using
  mobile sensors.
\newblock In {\em 6th International Conference on Mobile Computing,
  Applications and Services}, pages 197--205. IEEE, 2014.

\bibitem{jiang2015human}
Wenchao Jiang and Zhaozheng Yin.
\newblock Human activity recognition using wearable sensors by deep
  convolutional neural networks.
\newblock In {\em Proceedings of the 23rd ACM international conference on
  Multimedia}, pages 1307--1310. Acm, 2015.

\bibitem{rana2015application}
Jitenkumar~Babubhai Rana, Rashmi Shetty, and Tanya Jha.
\newblock Application of machine learning techniques in human activity
  recognition.
\newblock {\em arXiv preprint arXiv:1510.05577}, 2015.

\bibitem{marinho2016new}
Leandro~B Marinho, Amauri~Holanda de~Souza~J{\'u}nior, and Pedro~Pedrosa
  Rebou{\c{c}}as~Filho.
\newblock A new approach to human activity recognition using machine learning
  techniques.
\newblock In {\em International Conference on Intelligent Systems Design and
  Applications}, pages 529--538. Springer, 2016.

\bibitem{ronao2016human}
Charissa~Ann Ronao and Sung-Bae Cho.
\newblock Human activity recognition with smartphone sensors using deep
  learning neural networks.
\newblock {\em Expert systems with applications}, 59:235--244, 2016.

\bibitem{murad2017deep}
Abdulmajid Murad and Jae-Young Pyun.
\newblock Deep recurrent neural networks for human activity recognition.
\newblock {\em Sensors}, 17(11):2556, 2017.

\bibitem{tensorflow2015-whitepaper}
Mart\'{\i}n Abadi, Ashish Agarwal, and Paul~Barham et~al.
\newblock {TensorFlow}: Large-scale machine learning on heterogeneous systems,
  2015.
\newblock Software available from tensorflow.org.

\bibitem{maaten2008visualizing}
Laurens van~der Maaten and Geoffrey Hinton.
\newblock Visualizing data using t-sne.
\newblock {\em Journal of machine learning research}, 9(Nov):2579--2605, 2008.

\bibitem{laguna2011dynamic}
Javier~Ortiz Laguna, Angel~Garc{\'\i}a Olaya, and Daniel Borrajo.
\newblock A dynamic sliding window approach for activity recognition.
\newblock In {\em International Conference on User Modeling, Adaptation, and
  Personalization}, pages 219--230. Springer, 2011.

\bibitem{banos2014window}
Oresti Banos, Juan-Manuel Galvez, Miguel Damas, Hector Pomares, and Ignacio
  Rojas.
\newblock Window size impact in human activity recognition.
\newblock {\em Sensors}, 14(4):6474--6499, 2014.

\bibitem{kuhn2019feature}
Max Kuhn and Kjell Johnson.
\newblock {\em Feature engineering and selection: A practical approach for
  predictive models}.
\newblock CRC Press, 2019.

\bibitem{van2014accelerating}
Laurens Van Der~Maaten.
\newblock Accelerating t-sne using tree-based algorithms.
\newblock {\em The Journal of Machine Learning Research}, 15(1):3221--3245,
  2014.

\bibitem{liu2013comparison}
Miao Liu, Mingjun Wang, Jun Wang, and Duo Li.
\newblock Comparison of random forest, support vector machine and back
  propagation neural network for electronic tongue data classification:
  Application to the recognition of orange beverage and chinese vinegar.
\newblock {\em Sensors and Actuators B: Chemical}, 177:970--980, 2013.

\bibitem{were2015comparative}
Kennedy Were, Dieu~Tien Bui, {\O}ystein~B Dick, and Bal~Ram Singh.
\newblock A comparative assessment of support vector regression, artificial
  neural networks, and random forests for predicting and mapping soil organic
  carbon stocks across an afromontane landscape.
\newblock {\em Ecological Indicators}, 52:394--403, 2015.

\bibitem{rodriguez2015machine}
V~Rodriguez-Galiano, M~Sanchez-Castillo, M~Chica-Olmo, and M~Chica-Rivas.
\newblock Machine learning predictive models for mineral prospectivity: An
  evaluation of neural networks, random forest, regression trees and support
  vector machines.
\newblock {\em Ore Geology Reviews}, 71:804--818, 2015.

\bibitem{sizhen2023exploring}
Sizhen Bian, Xiaying Wang, Tommaso Polonelli, and Michele Magno.
\newblock Exploring automatic gym workouts recognition locally on wearable
  resource-constrained devices.
\newblock {\em Sustainable Computing: Informatics and Systems}, 2023.

\bibitem{PeakUtils}
Peakutils.
\newblock \url{https://bitbucket.org/lucashnegri/peakutils}, 2019.

\end{thebibliography}






\end{document}